\documentclass[12pt,preprint]{aastex}
\usepackage{graphics}

\begin{document}
\slugcomment{Submitted to ApJ}

\title{Dynamical mass constraints on the ultraluminous X-ray Source NGC 1313 X-2}

\author{Jifeng Liu\altaffilmark{1,2}, Jerome Orosz\altaffilmark{3}, and Joel N. Bregman\altaffilmark{4}}
\altaffiltext{1}{National Astronomical Observatory of China, Beijing, China 100012}
\altaffiltext{2}{Harvard-Smithsonian center for Astrophysics, 60 Garden st. Cambridge, MA 02138}
\altaffiltext{3}{San Diego State University, San Diego, Ca 92182}
\altaffiltext{4}{University of Michigan, Ann Arbor, MI 40185}

\begin{abstract}

Dynamical mass measurements hold the key to answering whether ultraluminous
X-ray sources (ULXs) are intermediate mass black holes (IMBHs) or stellar mass
black holes with special radiation mechanisms. NGC 1313 X-2 is so far the only
ULX with HST light curves, the orbital period, and the black hole's radial
velocity amplitude based on the He II $\lambda4686$\AA\ disk emission line
shift of $\sim200$ km/s.  We constrain its black hole mass and other parameters
by fitting observations to a binary light curve code with accommodations for
X-ray heating of the accretion disk and the secondary.  Given the dynamical
constraints from the observed light curves and the black hole radial motion and
the observed stellar environment age, the only acceptable models are those with
40-50 Myrs old intermediate mass secondaries in their helium
core and hydrogen shell burning phase filling 40\%-80\% of their Roche lobes.
The black hole can be a massive black hole of a few tens of $M_\odot$ that can
be produced from stellar evolution of low metalicity stars, or an IMBH of a few
hundred to above 1000$M_\odot$ if its true radial velocity $2K^\prime<40$ km/s.
Further observations are required to better measure the black hole radial
motion and the light curves in order to determine whether NGC 1313 X-2 is a
stellar black hole or an IMBH.

\end{abstract}


\section{INTRODUCTION}

Recent X-ray observations have revealed many ultraluminous X-ray sources
(ULXs), i.e., non-nuclear point sources in nearby galaxies with X-ray
luminosities in the range of $2\times10^{39}$ - $10^{41}$ erg s$^{-1}$
(Fabbiano 1989; Colbert \& Mushotzky 1999). 
Some ULXs may well be foreground stars or background AGN/QSOs projected into
the host galaxies (e.g, Gutiérrez 2006), or recent supernovae (e.g., SN1993J in
M81). 
Most ULXs, however, should be intermediate mass black holes (IMBHs) of
100--1$0^5 M_\odot$ radiating at sub-Eddington levels, or stellar mass black
holes with special radiation mechanisms, such as relativistic beaming
(Georganopoulos et al. 2002), beaming effects (King et al. 2001) or truly
super-Eddington emission from magnetized accretion disks with photon bubbles
(Begelman et al.  2002). 
X-ray spectroscopy and X-ray quasi-periodic oscillation behaviors may support
the interpretation for IMBHs for some ULXs (e.g., 
Strohmayer \& Mushotzky 2003; Liu \& Di Stefano 2008).  However, the X-ray spectra might be explained
otherwise such as the very high state of stellar mass black holes under
super-critical accretion (Gladstone et al. 2009), and the QPO argument is also
inconclusive given our poor understanding of the QPO behaviors.

Whether ULXs are IMBHs can only be securely addressed by measuring their
dynamical masses. 
The dynamical mass measurement is a well-defined process and requires that we
(1) identify the secondary unambiguously in the optical and derive the
secondary properties, (2) obtain the orbital period $P$ and the inclination
angle $i$ from the light curve, and (3) measure the radial velocity
half-amplitude $K$ of the secondary with optical spectroscopy.
The resulting mass function $f(M)=PK^3/2{\pi}G = m_1^3sin^3i/(m_1+m_2)^2$ will
set the lower limit for the black hole mass $m_1$. The black hole mass can be
further determined if we know the inclination angle $i$ and the secondary mass
$m_2$.  
So far, about 20 ULXs were identified in the optical with unique or multiple
counterparts mostly through registering Chandra X-ray positions onto HST images
(e.g., Liu et al. 2002, 2004; Roberts et al. 2008). 
For ULX counterparts, it is impractical to measure the radial velocity with the
shallow narrow photospheric absorption lines from the secondary given their
optical faintness (V$\ge$22 mag) due to their location in distant galaxies.
Instead, we must resort to strong emission lines to determine the radial
motion.

Strong emission lines can come from the ULX secondary if the secondary is a
Wolf-Rayet (WR) star. 
Most recently, the secondary of M101-ULX1 is identified and spectroscopically
confirmed as a WR star with strong He II $\lambda4686$ and He I $\lambda5876$
emission lines (Liu 2009), and Keck/Gemini spectroscopic monitoring programs
(PI: Liu) are under way to determine the radial velocity half-amplitude $K$ of
the secondary and ultimately the black hole mass.
However, M101-ULX1 is the only ULX with a WR secondary, and may not be
representative of the class.
Strong emission lines are expected from the X-ray photoionization of the
accretion disk for ULXs 
%, especially those 
with high X-ray luminosities above a few $\times10^{39}$ erg/s.
% (e.g., NGC 1313 X-2) that are more likely to host IMBHs.
%
Indeed, emission lines are observed for low-mass X-ray binaries during
outbursts with X-ray luminosities above only a few $\times10^{38}$ erg/s.
The black hole mass can be derived from $f^\prime(M)=PK^{\prime3}/2{\pi}G =
m_2^3sin^3i/(m_2+m_1)^2$, where $K^\prime$ is the radial velocity
semi-amplitude of the black hole as measured from the disk emission lines.

The explanation of the ULX optical observations is complicated by the X-ray
irradiation of the accretion disk and the side of the secondary facing the
primary, which becomes very significant for ULXs.
Indeed, the X-ray irradiated accretion disk can overwhelm the secondary itself
by a factor of $\ge100$ in X-ray outbursts of Galactic low mass X-ray binaries
(van Paradijs \& McClintock 1994).
Kaaret \& Corbel (2009) claimed, based on the spectral  slopes and other
features of their VLT/FORS spectra, that the optical light from an ULX in
NGC 5408 mainly comes from an X-ray irradiated accretion disk.
If X-ray irradiation is unimportant, and periodic modulations come from the
orbital motion of the binary with its tidally distorted secondary,
%(nearly) filling its Roche lobe, 
one expects a smooth ellipsoidal light curve with two maxima and two unequal
minima.
The deeper minimum comes from the heavily gravitationally darkened side of the
secondary close to the L1 point (hereafter the L1 side), the shallower minimum
comes from its antipodal side, while the two maxima come from the other two
sides.
If X-ray heating of the L1 side of the secondary is so significant as to boost
it from the darkest to brightest side, then one expects a sinusoidal light
curve with one maximum from the X-ray heated side, and one minimum from its
antipodal side. 

In this work, we describe our efforts to constrain the black hole mass and the
secondary simultaneously for NGC 1313 X-2 through the detected light curves and
the radial velocity changes of the disk emission line.
The light curve for NGC 1313 X-2 is largely sinusoidal with an amplitude of
$\Delta {\rm F450W}$ = 0.2 mag (Liu et al. 2009), and the X-ray irradiation
must be significant.
The X-ray irradiated binary model used to fit the light curves is described
briefly in \S2, with details given in the appendix.
The constraints on the black hole mass and other parameters are derived based
on $\chi^2$ minimization as described in \S3. 
The results are discussed in \S4.

\section{AN X-RAY IRRADIATED BINARY MODEL}

A model with X-ray irradiation properly included should be developed to get
dynamical parameters of ULXs from the optical spectra and light curves.
In such a black hole binary model, the secondary star should fill a significant
part of its tidally distorted pear-like Roche lobe to provide the high
accretion rates required by the high X-ray luminosity (King et al. 2001).
Many models of close binaries have been developed to model the ellipsoidal
light curves without X-ray irradiation (see the review by Wilson 1994). 
One example is the eclipsing binary light curve code (ELC; Orosz \& Hauschildt 2000),
which fits the wide-band photometric light curve and the radial velocity curve
to get the dynamical parameters.
Copperwheat et al. (2005) developed an X-ray irradiated black hole binary model
with the secondary surface approximated as black body emitters, which predicts
the wide-band photometry but not the spectrum for ULX optical counterparts.
To fully utilize the optical spectroscopic and photometric monitoring
observations of ULXs, we have developed a computer code to model both the
spectrum and the photometry from the X-ray irradiated black hole binary with
the (X-ray heated) secondary described by Kurucz stellar atmospheres.

In constructing the X-ray irradiated close binary model, we described the
tidally distorted secondary following Kopal (1954) and Lucy (1967), treated the
accretion disk as a standard disk (Shakura \& Sunyaev 1973) that can be
fattened due to X-ray irradiation, modeled the X-ray heating of the disk and
the secondary surface following Frank, King \& Raine (2002), and described the
X-ray heated secondary following Kurucz stellar atmosphere models.
We tried not to just patch up existing code segments, but wrote the computer
code in {\tt C} from scratch.
The model construction is detailed in the appendix, so that one can rewrite the
code if one so wishes. 
This model has 10 parameters, including five parameters to set up the binary
model without X-ray irradiation: (1) the black hole mass $M_\bullet$, (2-3) the
secondary's iitial mass and age which determine its mass ($M_*$), radius ($R_*$) and temperature ($T_*$),
(4) the fraction ($f_R$) the secondary fills its Roche
lobe by volume,  and (5) the inclination angle $\Theta$.
Five extra parameters are required to incorporate the X-ray irradiation: (6)
the X-ray luminosity $L_X$, (7) the disk thickness, as described by a fattening
factor $X_h$ relative to a standard disk, (8) the disk size, as described by a
factor $X_d$ relative to its maximal tidal-disruption radius, (9) the disk
albedo $\beta$, and (10) the X-ray to thermal conversion factor $\gamma$ on the
X-ray heated secondary surface.

% illustrate the model output?

Given the parameters, the model first sets up the binary geometry, computes the
secondary and disk temperature profiles with and without X-ray irradiation,
then computes the emergent spectrum and the magnitudes for a binary phase
$\Phi$.
For illustration purposes, we put the model output in one figure for one binary
phase as shown in Figure~1 for a model with $M_\bullet = 30M_\odot$, a B0V
secondary ($M_* = 17.5 M_\odot, R_* = 7.4 R_\odot$, and $T_*$ = 30,000 K with
solar abundance) filling its Roche lobe ($f_R = 1$), $L_X = 10^{39}$ erg
s$^{-1}$, $X_h = 1$, $X_d = 0.5$, $\beta = 0.5$, $\gamma = 0.5$, and $\Theta =
60^\circ$.
The upper left panel shows the binary components projected on the plane normal
to the viewing vector, and the upper right panel shows the $UBVRIJHK$ light
curves for a whole period with the magnitudes marked for this phase.
The lower panel shows the emergent spectrum from the secondary with X-ray
irradiation and its spectrum without X-ray irradiation, the spectrum from the
disk itself and the spectrum with the X-ray irradiation contribution, and the
spectrum from the binary as a whole.

\section{DYNAMICAL PARAMETERS OF NGC 1313 X-2}

NGC 1313 X-2 is a ULX with an observed maximum luminosity of $3\times10^{40}$
erg/s (Feng \& Kaaret 2006) in a nearby spiral galaxy NGC  1313 at a distance of
3.7 Mpc (Tully 1988).
NGC  1313 X-2 has been studied extensively both in the
X-ray and the optical to understand its companion and environments.
The optical counterpart has been established as a B=23.5 mag blue star through
a train of efforts (Zampieri et al. 2004; Mucciarelli et al. 2005; Ramsey et
al. 2006; Liu et al. 2007).
Optical spectra of its counterpart have shown variable HeII $\lambda4686$
emission lines (FWHM=3$\sim$10\AA) from the accretion disk, and a line shift of
$380\pm30$ km/s was detected between two VLT spectroscopic observations (Pakull
et al. 2006; Grise et al. 2009).
Our HST Cycle 16 GO program monitored this counterpart with 20 WFPC2 B/V
observations, and revealed a sinusoid-like light curve with a half-amplitude of
$A_B = 0.102\pm0.016$ mag and a period of $P = 6.12 \pm 0.16$ days (Liu et al.
2009). A slightly shorter period of $\sim6$ days is obtained when about a dozen
of VLT observations over five months are included (Zampieri et al. 2011).
The above information leads to $f^\prime(M)\ge4.4/2.4/1.2M_\odot$ if we adopt
$2K^\prime=380/200/100$ km/s, which sets the lower limit for the secondary
mass.

%A follow-up spectrocopic monitoring program with Gemini-South was campaigned in
%attempt to obtain the radial velocity curve for NGC 1313 X-2 (PI: Roberts).
%
Ten Gemini-S/GMOS follow-up observations were obtained for NGC 1313 X-2 within
two months in semester 2009B, which detected the anticipated shift in the He II
$\lambda4686$\AA\ line with respect to its local vicinity (Roberts et al. 2010;
Gladstone et al.  2011).
The He II $\lambda4686$\AA\ emission line most probably comes from the X-ray
irradiated accretion disk, since
%, and cannot come from a Wolf-Rayet secondary, because the spectra also reveal
%strong H$_\beta$ and H$_\gamma$ emission lines, while the optical spectrum for
%Wolf-Rayet stars is free of hydrogen lines.
%
the underlying continuum of these GMOS spectra is a featureless power-law in
resemblance of X-ray irradiated accretion disk like in the case of VLT/FORS
spectra for NGC 5408-ULX (Kaaret \& Corbel 2009).
Unfortunately, there is no evidence for a sinusoidal radial velocity curve, and
the data does not provide a new, solid constraint on the mass function for NGC
1313 X-2.
Based on the detailed analysis by Gladstone et al. (2011), the $2\sigma$ rms
scatter of the velocity shift is $182\pm40$ km/s. 
This places an upper limit on the radial velocity change, which enables to
constrain the dynamical parameters for NGC 1313 X-2.
In \S3.1 we constrain the dynamical parameters by assuming $2K^\prime =
200\pm40$ km/s. 
When using the disk emission line to probe the black hole motion, the observed
radial velocity amplitude can be two times higher than the true radial velocity
amplitude (Cantrell \& Bailyn 2007); we thus run a second case in \S3.2
assuming $2K^\prime = 100\pm40$ km/s.
We also run a third case in \S3.3 assuming $2K^\prime = 25\pm15$ km/s to
illustrate how the dynamical parameters change for lower values of $2K^\prime$.

\subsection{In case of $2K^\prime=200\pm40$ km/s}

Based on the observed sinusoidal light curves and the black hole's radial
velocity amplitude for NGC 1313, the dynamical parameters, including the black
hole mass and the secondary properties, can be derived with the X-ray
irradiated binary model as described in \S2.
For this purpose, we define $\chi^2 = \sum_i{(m_i - m_d)^2/e_i^2} + (P_d -
P)^2/\sigma_P^2 + (K^\prime_d - K^\prime)^2/\sigma^2_{K^\prime}$.
Here $m_i$ and $e_i$ are the observed HST magnitudes and errors, $P=6.12$ days and
$\sigma_P=0.16$ days are the detected period and error (Liu et al. 2009),
$2K^\prime=200$ km/s and $\sigma_{K^\prime}=40$ km/s are the radial velocity
semi-amplitude of the black hole and error based on the Gemini observations
(Roberts et al. 2010; Gladstone et al. 2011), while $m_d$ is the predicted
magnitude at the observed phase, $P_d$ and $K_d$ are the corresponding
quantities predicted by the model.
Five F555W observations are excluded from the fit because they exhibited
abnormally red colors and abnormally large deviations from the sinusoidal light
curve (Liu et al. 2009).
The Padova stellar model grid (Girardi et al. 2002), with Z=0.1Z$_\odot$ as
appropriate for NGC 1313 (Ryder 1993), is used to determine the secondary
properties with two parameters, the age $t$ and the initial mass $m_{ini} =
m_f\cdot m_{ini,max}$ with $m_f$ defined as fraction of the maximum initial
mass allowed for that age.
In addition to the ten parameters to describe the X-ray irradiated binary
model, we assume an extinction E(B-V) with the standard galactic extinction law
to convert model magnitudes to observed magnitudes. 

%  // compute Mdot using de Jager 1988A&AS, 72, 259 p264, but 1X larger
%   double lglm = 2*log(staradius)/log(10.) + 4*log(teffstar/5700)/log(10);
%   lgmdot = 1.769*lglm - 1.676*log(teffstar)/log(10) - 8.158 + 0.0;
%   if(fillfrac<pflags[3][0] || fillfrac>pflags[3][3]) return 1.0e10;
%   if(lgmdot < -6 && imf < imfrgb && fillfrac < 0.98) return 1e10;

Acceptable models are searched over the 11-parameter phase space by minimizing
the above defined $\chi^2$ with 26 degrees of freedom (37 constraints and 11
free parameters).
To not miss any acceptable models, we start with very loose constraints on the
parameters, with $3 < M_\bullet/M_\odot < 3000$, $6.6< \lg(t) < 8.4$, $0.3 <
m_f < 1$, $0.05 < f_R < 1$, $0<\Theta<90$, $30 < L_X/10^{38} < 300$, $1 < X_h <
10$, $0.05 < X_d < 0.9$, $0.01 < \beta < 0.99$, $0.01 < \gamma < 0.99$, and
0.11 $<$ E(B-V) $<$ 0.44.
Here the extinction value is bounded within 0.11 mag (Galactic extinction) and
0.44 mag, i.e., $n_H = 2.7\times10^{21}$ cm$^{-2}$ as inferred from the fit to
the X-ray spectrum (Miller et al. 2003).
The X-ray luminosity is bounded within $3\times$ and $30\times10^{39}$ erg/s
with an average of $\sim4\times10^{39}$ erg/s based on previous X-ray
observations (Mucciarelli et al. 2005).
The average X-ray luminosity corresponds to an accretion rate of $\dot{M} =
2\times10^{-7} (L/10^{39}) (0.1/\eta) \sim 8\times10^{-7} M_\odot/{\rm yr}$
from $L_{acc} = \eta \dot{M} c^2$ regardless the black hole mass.  
Based on the accretion rate consideration, models are discarded unless the
secondary is filling ($>98$\% of) its Roche lobe, or the secondary has already
evolved off the main sequence into its red/blue giant or asymptotic giant
phases and fills a significant fraction of its Roche lobe, so that the heavy
stellar winds can be focused onto the primary to provide the required accretion
rate.

Given the vast parameter space, the $\chi^2$ minimization is an extremely
difficult problem that demands carefully designed intensive computations, which
are carried out using the Odyssey cluster at Harvard University.
As the first step, $10^8$ models are randomly drawn over the 11-dimension phase
space.
This is equivalent to an average of $10^{8/11} \sim 5.3$ sampled values over
the parameter range for each dimension, but with the advantage of continuous
coverage due to randomization.
Most of the $10^8$ models have extremely large $\chi^2$ values, including about
20 models with $\chi^2<260$ (i.e., $\chi^2_\nu < 10$), and about 25000 models
with $\chi^2 < 10000$, 
As shown in Figures~2a-f,  $M_\bullet$, $\Theta$, $\lg(t)$, $m_f$ and E(B-V)
are clearly constrained even at the $\chi^2 < 10000$ level, while $L_X$, $X_h$,
$X_d$, $f_R$, $\beta$ and $\gamma$ are not constrained.
At the $\chi^2 < 10000$ level, models with black hole masses above
1000$M_\odot$ or inclination angles below $20^\circ$ are not acceptable.
Only certain combinations of the secondary age and initial mass in the two thin
stripes are possibly acceptable.
Secondaries in the lower stripe are starting the helium core and hydrogen shell
burning phase, and secondaries in the upper stripe are starting the carbon core
and helium shell burning phase, while all have similar average densities
appropriate for the detected orbital period ($P = \sqrt{110/\rho}$ with $\rho$
as the mean matter density within the secondary's Roche lobe; Paczynski 1972).
While we start with a uniform distribution of E(B-V) within 0.11 mag and 0.44
mag, it is found that models with larger E(B-V) are more likely to be
unacceptable.

Local $\chi^2$ minima are then searched using the AMOEBA method as described in
{\it Numerical Recipe} (Press et al. 2003) adapted to computing with the
Odyssey cluster.
Selected as starting points are about 6000 models as shown in Figures~2a-f,
including all models with $\chi^2 < 2000$ and those with lowest $\chi^2$ within
fine intervals of each parameter.
This large number of initial guesses provides a fine coverage of the parameter
space, and yet is still doable with the available computing power.
For each AMOEBA search, we construct the initial simplex with a length scale of
20\% of the whole range for each parameter, in hope to have a complete coverage
of the whole parameter space.
The simulated annealing method is applied in the first 300 model evaluations of
an AMOEBA search in order to increase the chances to obtain a ``global''
$\chi^2$ minima in the large parameter space to be searched.
The AMOEBA search claims a $\chi^2$ minima if the fractional tolerance {\tt
ftol}$<10^{-5}$, and stops if the iteration number exceeds 1000.
To find the ``true'' $\chi^2$ minima, we restart another AMOEBA search from a
claimed $\chi^2$ minima, but with a $4\times$ smaller length scale.
In the end, two rounds of AMOEBA searches with about eight million model
evaluations lead to local $\chi^2$ minima from $\chi^2\sim52$ to $\ge1000$.
As shown in Figure~3, models with black hole masses above 100/300 $M_\odot$ are
excluded at the $\chi^2_\nu=3/4$ level.
Due to the large photometric fluctuations present in the observations, even the
``best'' model has $\chi^2_\nu > 2$.
Therefore, we do not claim one ``true'' model for NGC 1313 X-2, but instead
consider models with $\chi^2<78$ (i.e., $\chi^2_\nu \le 3$) as ``acceptable''
models.

The acceptable models are grouped into three categories based on the secondary
properties as shown in Figure~4a-e.
The secondaries in the first group of models are very young ($\le$ 5 Myrs) and
massive  in the carbon core and helium shell burning phase filling about 20\%
of their Roche lobes.
The black hole mass and the inclination angle cluster tightly in small ranges
of (30-40$M_\odot$, 80$^\circ$-90$^\circ$).
The X-ray luminosities are within 1-2 $\times10^{40}$ erg/s (about 2-6 times
the Eddington luminosity, or $\sim$2-6$\times L_E$), but the  accretion disks
are quite small ($<$15\% of the tidal-disruption radius) and as thin as the
standard thin disk.
The accretion disk albedo is in the range of 0-0.5, and the X-ray to thermal
conversion factor is above 0.8.
The best model in this group (Model~A1, $\chi^2=70.88$) contains a
$45.3M_\odot$ black hole and a young (5 Myrs) and massive (initially
44$M_\odot$, currently 17$M_\odot$, $5.1R_\odot$) secondary that fills 20\% of
its Roche lobe.
As shown in Figure~5a, the optical spectrum is dominated by the hot
($\sim70000K$) secondary, which is only slightly affected by the X-ray heating.
Although the X-ray luminosity is high ($L_X \sim1.6\times10^{40}$ erg/s
$\sim3\times L_E$), the optical light from the irradiated accretion disk is
only 1\% of that from the secondary in the B/V bands because the accretion disk
is small ($X_d\sim0.05$)  and thin ($X_h\sim1$).
The binary system is viewed almost edge-on ($\theta\sim82^\circ$), and the
predicted sinusoidal light curves have amplitudes of 0.04/0.04 mag in the B/V
bands (Figure~5d).

The secondaries in the second group are 40-50 Myrs old in the  helium core and
hydrogen shell burning phase filling about 40\%-80\% of their Roche lobes.
As compared to the first group, the secondaries are 10 times older and less
massive.
The black hole mass and the inclination angle occupy much larger parameter
ranges of (3-100$M_\odot$, 20$^\circ$-80$^\circ$).
The X-ray luminosities are $3-30 \times10^{39}$ erg/s ($3-20\times L_E$),
and the accretion disks can be rather large ($>$20\% of the tidal-disruption
radius) and about 2-6 times thicker than the standard thin disk.
The accretion disk albedo and the X-ray to thermal conversion factor on the
secondary surface range both from 0.01 to 0.99.
The best model in this group (model~B1,  $\chi^2=54.94$) contains a
$15.3M_\odot$ black hole and an old (40 Myrs) intermediate-mass (initially
8.08$M_\odot$, currently 8.07$M_\odot$, $9.0R_\odot$) secondary that fills
54\% of its Roche lobe.
The X-ray luminosity is $L_X \sim1.2\times10^{40}$ erg/s, or $\sim6\times L_E$.
The secondary has a surface temperature of $\sim18500K$ and is significantly
affected by the X-ray heating, leading to sinusoidal light curves with
amplitudes of 0.15/0.15 mag in the B/V bands for the inclination angle of
$\theta\sim38^\circ$ (Fig~5d).
The accretion disk is 55\% of the tidal-disruption radius and 3.5 times thicker
than the standard thin disk, and the X-ray heated disk contributes roughly
equally as the secondary in the B/V bands (Fig~5b).

The secondaries in the third group are 80-160 Myrs old in the  helium core and
hydrogen shell burning phase filling ($>$98\% of) their Roche lobes.
As compared to the second group, the secondaries are 2-3 times older and less
massive.
The black hole mass and the inclination angle cluster tightly in small ranges
of (3-10$M_\odot$, 50$^\circ$-70$^\circ$).
The X-ray luminosities are within $1-2\times10^{40}$ erg/s (or $7-20\times
L_E$), and the accretion disks can be rather large (40\%-70\% of the
tidal-disruption radius) and about 6-9 times thicker than the standard thin
disk.
The accretion disk albedo is constrained within 0.7-0.8, while the X-ray to
thermal conversion factor on the secondary surface ranges from 0.01 to 0.8.
The best model in this group (model~F1, $\chi^2=61.26$) contains a $6.7M_\odot$
black hole and a very old (150 Myrs) and rather low mass (initially 4.28$M_\odot$,
currently 4.27$M_\odot$, $10.6R_\odot$) secondary that fills its Roche lobe.
The X-ray luminosity is $L_X \sim1.9\times10^{40}$ erg/s ($\sim20\times L_E$),
the accretion disk is 40\% of the tidal-disruption radius and 7 times thicker
than the standard thin disk.
The X-ray heated disk contributes about 4 times more light than the secondary
in the B/V bands (Fig~5c).
The secondary has a surface temperature of $\sim9500K$ and is significantly affected by the X-ray
heating, leading to sinusoidal light curves with amplitudes of 0.17/0.17 mag in
the B/V bands for the inclination angle of $\theta\sim56^\circ$ (Fig~5d).

\subsection{In case of $2K^\prime=100\pm40$ km/s}

The model $\chi^2$ for assumed $2K^\prime=100\pm40$ km/s is defined in the same
way as described in \S3.1, and $\chi^2$ minima are sought in a large phase
space with two rounds of AMOEBA searches following the same procedures.
The resulted models are grouped into three categories based on the secondary
properties as in \S3.1, and the ``acceptable'' models with $\chi^2<78$ are
plotted in Figure~6a-f.
As compared to the case of $2K^\prime=200\pm40$ km/s, the first group of models
with secondaries in their carbon core and helium shell burning phase are no
longer ``acceptable''. Even the best model (model~A2) for the first
group has $\chi^2=84.0$, an apparently bad fit to the observed light curve as
shown in Figure~7.

Models in the second group occupy similar phase space as compared to
the case of $2K^\prime=200\pm40$ km/s, while the best model is quite different.
As shown in Figure~6-7, the best model (model~B2, $\chi^2=51.9$) contains a
$39M_\odot$ black hole and a 50Myrs old intermediate mass secondary (initially
7.18$M_\odot$, currently 7.16$M_\odot$, 8.2$R_\odot$) in its helium core and
hydrogen shell burning phase filling 51\% of its Roche lobe. 
The X-ray luminosity is $L_X \sim1.2\times10^{40}$ erg/s, or $\sim2.4\times
L_E$.
The accretion disk is 15\% of the tidal-disruption radius and 5 times thicker
than the standard thin disk, and the X-ray heated disk contributes two times that of
the secondary in the B/V bands.
The secondary has a surface temperature of $\sim17500K$  and is significantly affected by the
X-ray heating, leading to sinusoidal light curves with amplitudes of
0.155/0.155 mag in the B/V bands for the inclination angle of
$\theta\sim45^\circ$ (Fig~7).

Models in the third group occupy roughly similar phase space as compared to the
case of $2K^\prime=200\pm40$ km/s, while the secondaries are older and less
massive.
The best model (model~F2, $\chi^2=55.8$) contains a $8.2M_\odot$ black hole and
a 200 Myrs old secondary (initially 3.75$M_\odot$, currently 3.74$M_\odot$,
9.9$R_\odot$) in its helium core and hydrogen shell burning phase filling its
Roche lobe.
The X-ray luminosity is $L_X \sim9\times10^{39}$ erg/s ($\sim8.5\times L_E$),
the accretion disk is 30\% of the tidal-disruption radius and 6 times thicker
than the standard thin disk.
The X-ray heated disk contributes about 4 times that of the secondary
in the B/V bands.
The secondary has a surface temperature of $\sim8700K$ and is significantly affected by the X-ray
heating, leading to sinusoidal light curves with amplitudes of 0.156/0.156 mag
in the B/V bands for the inclination angle of $\theta\sim38^\circ$ (Fig~7).

\subsection{In case of $2K^\prime=25\pm15$ km/s}

The model $\chi^2$ for assumed $2K^\prime=25\pm15$ km/s is defined in the same
way as described in \S3.1, and $\chi^2$ minima are sought in a large phase
space with two rounds of AMOEBA searches following the same procedures.
The resulted models are grouped into three categories based on the secondary
properties as in \S3.1, and the ``acceptable'' models with $\chi^2<78$ are
plotted in Figure~8a-f.
As compared to the case of $2K^\prime=200\pm40$ km/s, the first group of models
with secondaries in their carbon core and helium shell burning phase are no
longer ``acceptable'' like the case of $2K^\prime=100\pm40$ km/s. Even the best
model (model~A3) for the first group has $\chi^2=96.5$, an apparently bad fit
to the observed light curve as shown in Figure~9.

Models in the second group occupy similar phase space as compared to the case
of $2K^\prime=200\pm40$ km/s, except that the black hole mass can be above
1000$M_\odot$.
The best model (model~B3, $\chi^2=52.43$) contains a $275M_\odot$ black hole
and a 40Myrs old intermediate mass secondary (initially 8.09$M_\odot$,
currently 8.08$M_\odot$, 7.9$R_\odot$) in its helium core and hydrogen shell
burning phase filling 44\% of its Roche lobe. 
The X-ray luminosity is $L_X \sim1.3\times10^{40}$ erg/s, or $\sim0.4\times
L_E$.
The accretion disk is 12\% of the tidal-disruption radius and 4 times thicker
than the standard thin disk, and the X-ray heated disk contributes 1.7 times
that of the secondary in the B/V bands.
The secondary has a surface temperature of $\sim20000K$ and is significantly
affected by the X-ray heating, leading to sinusoidal light curves with
amplitudes of 0.157/0.157 mag in the B/V bands for the inclination angle of
$\theta\sim73^\circ$ (Fig~9).

Models in the third group occupy roughly similar phase space as compared to the
case of $2K^\prime=200\pm40$ km/s, while the secondaries are older and less
massive.
The best model (model~F3, $\chi^2=53.89$) contains a $6.1M_\odot$ black hole
and a 220 Myrs old secondary (initially 3.58$M_\odot$, currently 3.57$M_\odot$,
9.8$R_\odot$) in its helium core and hydrogen shell burning phase filling its
Roche lobe.
The X-ray luminosity is $L_X \sim1.6\times10^{40}$ erg/s ($\sim20\times L_E$).
The secondary has a surface temperature of $\sim8200K$  and is significantly
affected by the X-ray heating, leading to sinusoidal light curves with
amplitudes of 0.156/0.156 mag in the B/V bands for the inclination angle of
$\theta\sim15^\circ$ (Fig~9).
The accretion disk is 28\% of the tidal-disruption radius and 6 times thicker
than the standard thin disk, and the X-ray heated disk contributes about 2
times that of the secondary in the B/V bands.

The best models for the three cases are listed in Table~1.
To summarize, there are three groups of acceptable models based on the
secondary properties depending on whether the secondary is in the carbon core
and helium shell burning phase or in the helium core and hydrogen shell burning
phase, and whether the secondary is filling the Roche lobe or just part of the
Roche lobe.
The first group of models with secondaries in the carbon core and helium shell
burning phase, while acceptable for the highest $2K^\prime=200\pm40$ km/s,
become unacceptable when $2K^\prime$ drops. 
Models of the other two groups, with secondaries in the helium core and
hydrogen shell burning phase, are acceptable for all cases of assumed
$2K^\prime$, but the black hole mass can become as massive as 1000$M_\odot$ as
the assumed $2K^\prime$ becomes lower as shown in Figure~10.
We note that, while the extinction value is started to range from 0.11 mag to
0.44 mag, it is always close to E(B-V)=0.11 mag for all three groups of
``acceptable'' models regardless the assumed $2K^\prime$ as shown in
Figure~4f/6f/8f.

\section{DISCUSSION}

In this work, we constrain the binary masses and other parameters for ULX
NGC 1313 X-2 by fitting observations to an X-ray irradiated binary model.
The model is based on the eclipsing binary light curve code by Jerome Orosz
(Orosz \& Hauschildt 2000; Orosz et al. 2007) with accommodations for the X-ray
irradiation of the accretion disk and the secondary in the ULX systems.
This model treats the accretion disk as a standard disk that can be fattened
due to X-ray irradiation. The disk fattening factor should be linked to the
X-ray luminosity; however, the exact relation is not well studied, and for now
these two parameters are allowed to vary independently in our model.
The X-ray heating of the disk and the secondary surface is treated the same way
as in the cases of Galactic X-ray binaries (Vrtilek et al. 1990; FKR2002;
Copperwheat et al. 2005) with two parameters, disk albedo $\beta$ and the
X-ray-to-thermal conversion factor $\gamma$, respectively.
While $\beta$ and $\gamma$ can be calculated given the disk/secondary surface
structure and incident flux, such calculations are beyond the scope of this
work, and we simply allow them to vary between 0 and 1 exclusive.
The model describes the (X-ray heated) secondary surface with Kurucz stellar
atmosphere model, allowing to calculate the emergent spectrum with absorption
lines from the secondary.

NGC 1313 X-2 is chosen for this study because it is so far the only ULX with the
orbital period, B/V light curves and the black hole's radial velocity
amplitude.
Our HST GO program has obtained the B/V light curves and an orbital period of
$P = 6.12\pm0.16$ days for NGC 1313 X-2 (Liu et al. 2009), while the black
hole's radial velocity amplitude is measured from the disk emission line
HeII $\lambda4686$\AA\ (Pakull et al. 2006; Roberts et al. 2010; Gladstone et al.
2011).
The observed HeII $\lambda4686$\AA\ shift may deviate from the true amplitude
for the black hole (e.g.,  due to the emission line genesis location change
under different X-ray luminosities), as implied by the phase shift between the
radial velocity curve and the light curve often seen in Galactic accreting
binaries.
Indeed, a comprehensive study shows that the semi-amplitude measured with disk
emission lines can differ from the true value by a factor of 0.9$\sim$2.5
(Cantrell \& Bailyn 2007).
The black hole mass measurements based on the disk emission lines are thus
considered less reliable as compared to those based on absorption/emission
lines from the secondary, and not commonly used for Galactic black hole
binaries given the detectability of absorption or Bowen fluorescent lines from
their secondaries.
For most ULXs with their faint optical counterparts, however, the disk emission
line method is the only way to estimate the black hole masses, with an accuracy
that is not superb but enough to determine whether they are stellar mass black
holes or IMBHs.

Dynamical models for NGC 1313 X-2 are sought with the Monte Carlo sampling of
the parameter space followed by two rounds of AMOEBA searches of $\chi^2$
minima.
Given that the observed He II $\lambda4686$\AA\ emission line shift
($\sim180\pm40$ km/s) can be 0.9-2.5 times that of the true radial velocity
amplitude of the black hole, the true amplitude can be $\sim$200-70 km/s.
We thus have tested cases of $2K^\prime=200\pm40$ km/s and $2K^\prime=100\pm40$
km/s, and additional case of $2K^\prime=25\pm15$ km/s.
While the model parameters are allowed to vary in ranges as large as possible
to not miss any acceptable models, the available observations are able to
exclude most of the parameter space.
For example, the extinction value is found to be E(B-V) = 0.11 mag for all
acceptable models with $\chi^2_\nu<3$ regardless the assumed $2K^\prime$
values, although it is allowed to vary between 0.11 and 0.44 mag.
Interestingly, independent measurements with nebular lines showed that E(B-V) =
0.11 mag (Pakull et al. 2006), justifying our model fits.
When X-ray irradiation parameters $\beta$, $\gamma$ and the dependency of the
disk thickness on the X-ray luminosity are further constrained by physical
considerations, some acceptable models may become unacceptable.  
However, no unacceptable models can become acceptable given such additional
constraints.

While $\chi^2_\nu\sim1$ for the true model is expected when there are only
Poisson errors for the photometric measurements, all models obtained from the
fitting have $\chi^2_\nu>2$. 
This is because the HST light curves of NGC 1313 X-2 are afflicted with
photometric fluctuations caused by, e.g., X-ray variations between
observations.
Such photometric fluctuations are clearly present as shown in Figure~5d, even
after we exclude five F555W observations with abnormally red colors that
exhibit extremely large deviations from the fitted sinusoidal light curve.
We thus do not claim one ``true'' model, but instead consider models with
$\chi^2_\nu<3$ as ``acceptable'' models, which fall into three groups based on
the secondary properties.
The first group of models have 4-5 Myrs old secondaries in the carbon core
and helium shell burning phase filling only 20\% of its Roche lobe, 
the second group of models have 40-50 Myrs old secondaries in the helium core
and hydrogen shell burning phase filling 40\%-80\% of its Roche lobe, 
while the third group have 150-220 Myrs old secondaries in the helium core and
hydrogen shell burning phase filling its Roche lobe.

The acceptable models can be further assessed with additional constraints.
The first group of models have X-ray luminosities of 1-2 $\times10^{40}$ erg/s,
or 2-6 times the Eddinton luminosity, yet the accretion disks are as thin as
the standard thin disk. This contradicts the understanding that a standard disk
forms under a few tenths of the Eddington luminosity based on observations of
Galactic black hole X-ray binaries (Remillard \& McClintock 2006).
ESO/VLT and SUBARU studies show that the stellar environment around NGC 1313
X-2 is about 40-70 Myrs old (Pakull et al. 2006). This is quite consistent with
the secondary age of 40-50 Myrs for the second group, but significantly
different from the secondary age of 150-220 Myrs for the third group and the
secondary age of 4-5 Myrs for the first group.
Thus, the second group of models are the only possible models given the
dynamical constraints, the reasonableness of the accretion disk, and the
observed stellar environment age.

Concerns arise whether the secondaries in the second group can provide the
required accretion rate to power the observed X-ray luminosity.
Unlike the third group where secondaries are filling the Roche lobe, the black
holes in the second group must accrete via capturing the focused stellar winds
from the secondaries that are only filling 40\%-80\% of the Roche lobe.
The stellar wind mass loss rate for the these secondaries is usually around
$10^{-8} M_\odot/{\rm yr}$ based on the de Jager et al. (1988) prescription,
but the observed average X-ray luminosity of $4\times10^{39}$ erg/s requires an
accretion rate of $\dot{M} \sim 8\times10^{-7} M_\odot/{\rm yr}$ regardless the
black hole mass.
However, the de Jager et al. prescription describes stars in isolation, while
the secondaries in the second group are in a close binary configuration, where
the equivalent surface gravity is reduced by the gravitational pull exerted by
the companion black hole.
Numerical calculations have shown that the mass loss rate due to this reduced surface gravity
can be boosted by more than an order of magnitude if the secondary is almost filling 
its Roche lobe (Frankowski \& Tylenda 2001).
In addition, the strong X-ray radiation from the accretion disk will heat up
the secondary surface and increase the kinetic energy of the surface materials,
and possibly enhence the mass loss further.

The black hole mass for NGC 1313 X-2 ranges from a few to above 1000 $M_\odot$
depending on the true radial velocity amplitude of the black hole.
Figure~10 shows the black hole mass versus the black hole radial velocity
semi-amplitude for all acceptable models from this work, with larger symbols
for more likely models of lower $\chi^2$ values.
The black hole mass can be 3-10 $M_\odot$ (e.g., model B1) if $K^\prime$ =
120-140 km/s, or 20-30 $M_\odot$ if $K^\prime$ = 65-75 km/s, or 30-50 $M_\odot$
(e.g., model B2) if $K^\prime$ = 40-60 km/s.
If the true radial velocity semi-amplitude is lower than 40 km/s, the black
hole mass can be a few $M_\odot$ (e.g., model B3b as listed in Table~1), a few
$\times 10 M_\odot$, a few $\times 100 M_\odot$ (e.g., model B3), or above 1000
$M_\odot$.
While the death of stars with solar metalicity can not produce black holes more
massive than 10$M_\odot$ (Fryer \& Kalogera, 2001), NGC 1313 X-2 is located in
an environment of 0.1-0.2$Z_\odot$ (Ryder 1993), and a black hole up to
$50M_\odot$ may be produced.
Indeed, massive black holes have been discovered in environments of similarly
low metalicities, e.g., the black hole of $\sim16M_\odot$  for M33 X-7 (Orosz
et al. 2007), and the black hole of 23-35 $M_\odot$ for IC 10 X-1 (Silverman \&
Filippenko 2009).
Therefore, NGC 1313 X-2 can be a black hole of up to a few $\times 10 M_\odot$
that may be explained by the stellar evolution of low metalicity stars, or an
IMBH of a few $\times100 M_\odot$ or above $1000M_\odot$ if the true black hole
radial velocity semi-amplitude is more than 5 times lower than the observed
velocity shift. 
Such IMBHs can not be produced through the death of stars, but may form in the
centers of dense stellar clusters via the merging of stellar mass black holes
(e.g., Miller \& Hamilton 2002), or from the direct collapse of merged
supermassive stars in very dense star clusters (e.g., Portegies Zwart \&
McMillan 2002).

Further observations are required to determine whether NGC 1313 X-2 is a massive
stellar black hole or an IMBH.
Current observations use the HeII $\lambda4686$\AA\ disk emission line to
constrain the radial motion of the black hole, which is difficult due to the
weakness of the line, the line genesis location change and the contamination
from the surrounding nebula.
Such complications can be circumvented if UV emission lines are used to measure
the radial velocity shift, because the UV emission lines are usually two orders
of magnitudes more luminous than the HeII $\lambda4686$\AA\ line.  
For example, X-ray illuminated disk and corona model predicts that the line
luminosities for N V $\lambda1240$\AA, Si IV $\lambda1396$\AA, and C IV
$\lambda1550$\AA\ are 390, 186, 373 times that of HeII $\lambda4686$\AA, quite
consistent with observations of several X-ray binaries (Raymond 1993).
The model also shows that the HeII $\lambda4686$\AA\ line is produced mostly in
the inner disk region, a small region that is easily affected by small changes
in the accretion rate or X-ray luminosity, resulting in line genesis location
changes and apparent line centroid changes.
In contrast, the above UV emission lines are produced across the whole disk,
and are less affected by small changes in the accretion rate or X-ray
luminosities. This reduces the line centroid changes due to X-ray luminosity
changes as compared to the HeII $\lambda4686$\AA\ line.
In addition, these UV emission lines cannot form in the surrounding nebula,
thus are free of nebular contamination.
Observations of these UV emission lines can be carried out with Cosmic Origin
Spectrograph (COS) aboard Hubble Space Telescope.

Better light curves are also required to determine whether NGC 1313 X-2 is a
massive stellar black hole or an IMBH.
Given the better determined radial motion of the black hole, the black hole
mass can still range from a few to a few $\times10M_\odot$ to a few
$\times100M_\odot$ to above $1000M_\odot$.
Although all models have similar light curves in the HST F450W/F555W bands as
resulted from the $\chi^2$ minimization process, the predicted light curves can
be quite different in other bands given the difference in the secondaries, the
inclination angles and the X-ray heating effects.
As shown in Figure~11 as examples, the predicted light curves of model B3 and
B3b for $K^\prime \sim20$ km/s, while similar in WFPC2 F450W/F555W bands, are
offset by 0.04 in the WFC3/UVIS F275W band and by 0.02 mag in the WFC3/UVIS
F814W band.
Such difference can be easily detectable with HST with reasonable exposure
time, and can be used to distinguish between a stellar black hole (model B3b)
and an IMBH (model B3).
Because the X-ray heating of the accretion disk and the secondary surface are
important in these models, the photometric monitoring observations should be
accompanied by quasi-simultaneous X-ray observations with, e.g., swift/XRT, in
order to assess the X-ray heating effects of the light curves.

\acknowledgements

We would like to thank Dr. Jeffery E. MaClintock and Rosanne Di Stefano for
helpful discussions. The computations in this paper were run on the Odyssey
cluster supported by the FAS Science Division Research Computing Group at
Harvard University.

\setlength{\tabcolsep}{3pt}

\begin{deluxetable}{lccccccccccccccccr}
\tablewidth{6.7in} 
\tabletypesize{\scriptsize}
\tablecaption{Model parameters for best cases\tablenotemark{a}}
\tablehead{
   \colhead{model} & \colhead{$\chi^2$} & \colhead{$M_{bh}$} & \colhead{age} & \colhead{$M_{*,ini}$} & \colhead{$M_{*}$} & \colhead{$R_*$} & \colhead{$T_*$} & \colhead{$f_R$} & \colhead{$\theta$} & \colhead{$L_{38}$} & \colhead{$X_h$} & \colhead{$X_d$} & \colhead{$\beta$} & \colhead{$\gamma$} & \colhead{E(B-V)} & \colhead{$P$} & \colhead{$K^\prime$} \\
}

\startdata

A1 & 70.88 & 45.3 & 5.0 & 44.10 & 16.99 & 5.1 & 71099 & 0.22 & 82 & 163 & 1.0 & 0.05 & 1.00 & 0.86 & 0.11 & 150.1 & 124 \\
B1 & 54.94 & 15.3 & 39.8 & 8.08 & 8.07 & 9.0 & 18462 & 0.54 & 38 & 120 & 6.3 & 0.18 & 0.48 & 0.74 & 0.11 & 145.8 & 71 \\
F1 & 61.26 & 6.7 & 146.8 & 4.28 & 4.27 & 10.6 & 9505 & 1.00 & 56 & 187 & 7.3 & 0.40 & 0.73 & 0.65 & 0.11 & 146.3 & 84 \\
A2 & 84.01 & 126.4 & 5.6 & 37.22 & 15.99 & 5.3 & 65950 & 0.24 & 84 & 113 & 1.0 & 0.06 & 0.72 & 0.67 & 0.11 & 143.1 & 68 \\
B2 & 51.91 & 39.1 & 50.0 & 7.18 & 7.16 & 8.2 & 17500 & 0.51 & 46 & 124 & 4.9 & 0.14 & 0.02 & 0.32 & 0.11 & 143.0 & 47 \\
F2 & 55.84 & 8.2 & 196.0 & 3.75 & 3.74 & 9.9 & 8695 & 0.99 & 38 & 91 & 6.1 & 0.30 & 0.27 & 0.14 & 0.11 & 143.0 & 52 \\
A3 & 96.53 & 4.5 & 4.1 & 57.94 & 25.66 & 3.1 & 98108 & 0.10 & 3 & 277 & 5.3 & 0.18 & 0.07 & 0.79 & 0.11 & 149.0 & 18 \\
B3 & 52.43 & 275.1 & 39.7 & 8.09 & 8.08 & 7.9 & 20259 & 0.44 & 74 & 131 & 4.4 & 0.12 & 0.51 & 0.82 & 0.11 & 147.4 & 21 \\
B3b & 53.01 & 5.9 & 39.8 & 8.08 & 8.07 & 8.0 & 19864 & 0.45 & 9 & 148 & 2.4 & 0.18 & 0.59 & 0.96 & 0.11 & 147.6 & 26 \\
F3 & 53.89 & 6.1 & 217.0 & 3.58 & 3.57 & 9.8 & 8190 & 1.00 & 15 & 162 & 6.0 & 0.28 & 0.88 & 0.89 & 0.11 & 143.0 & 24 \\

\enddata

\tablenotetext{a}{The columns are (1) model as named in text, (2) $\chi^2$, (3) black hole mass in $M_\odot$, (4) secondary age in Myrs, (5) initial mass in $M_\odot$, (6) current mass, (7) current radius in $R_\odot$, (8) stellar temperature, (9) filling fraction the secondary fills its Roche lobe, (10) viewing angle, (11) X-ray luminosity in $10^{38}$ erg/s, (12) disk fattening factor, (13) disk size in unit of tidal disruption radius, (14) disk albedo, (15) X-ray to thermal conversion factor on secondary surface, (16) E(B-V), (17) period in hours, and (18) black hole radial velocity semi-amplitude when viewed edge-on.}

\end{deluxetable}

\begin{figure}

\plottwo{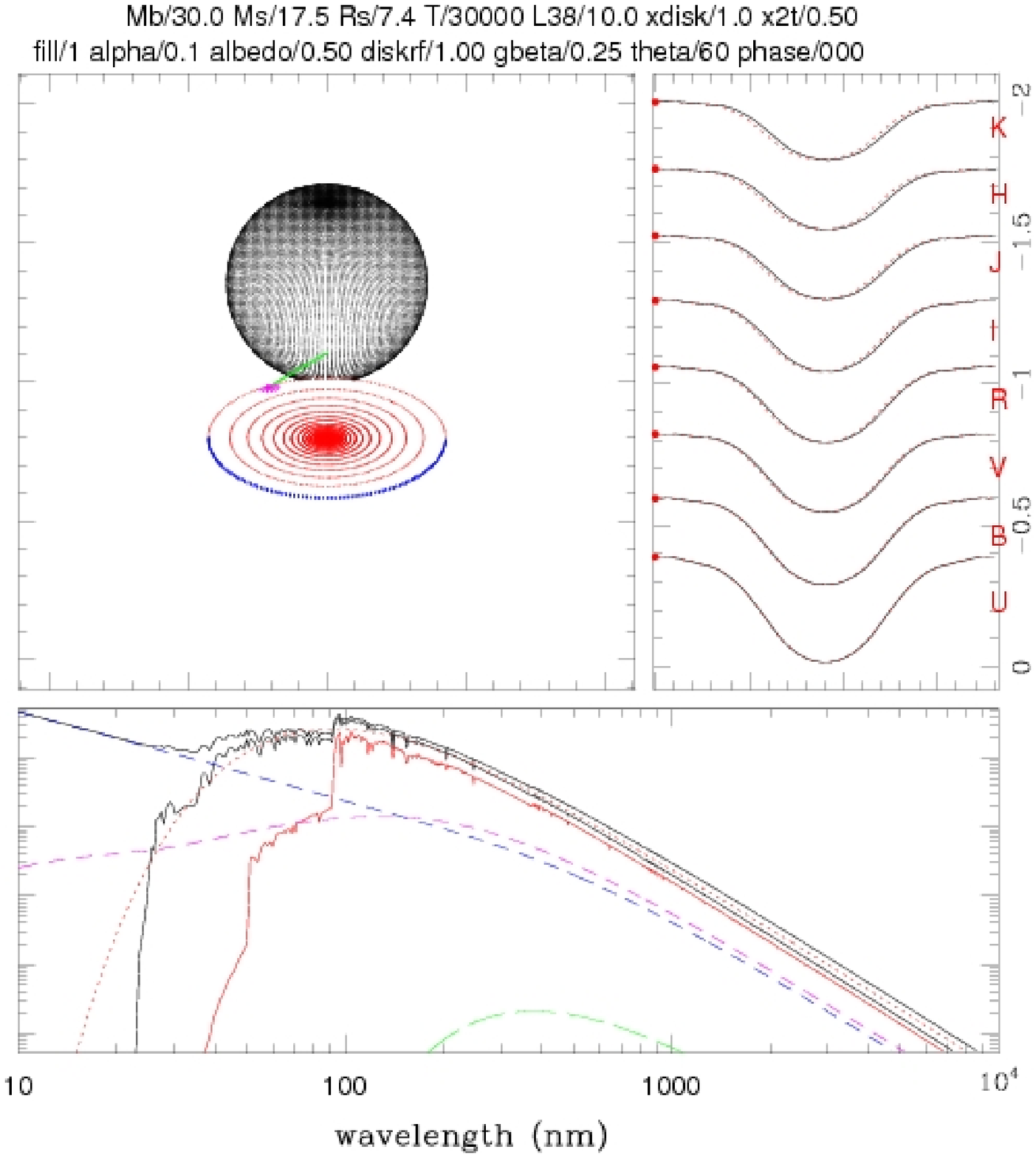}{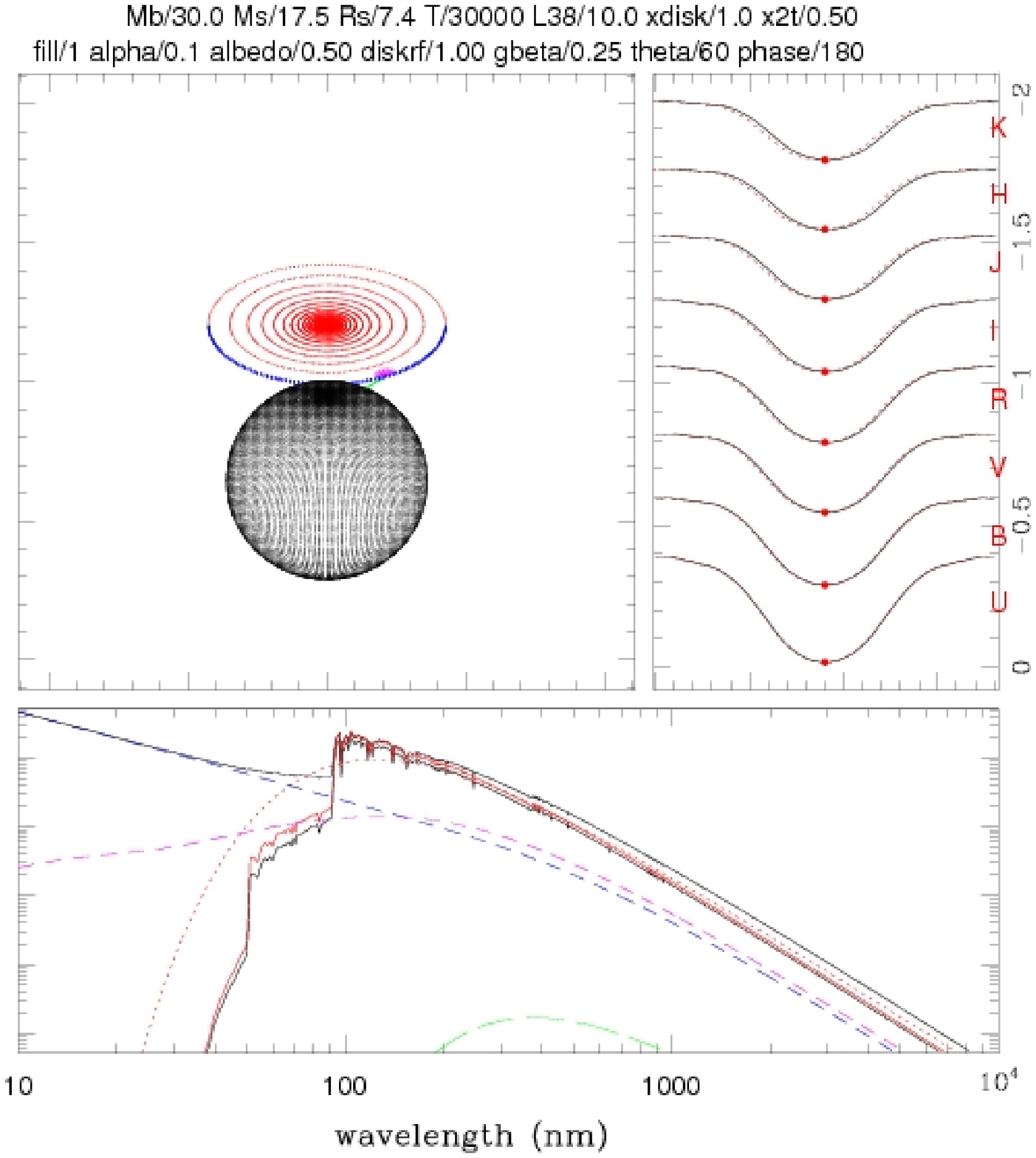}

\caption{Model output for (a) phase 0 and (b) phase 180 for the base model with
$M_\bullet = 30M_\odot$, a B0V secondary ($M_* = 17.5 M_\odot, R_* = 7.4
R_\odot$, and $T_*$ = 30,000 K with solar abundance), $L_X = 10^{39}$ erg
s$^{-1}$, $X_h = 1$, $X_d = 0.5$, $\beta = 0.5$, $\gamma = 0.5$, and $\Theta =
60^\circ$. In each figure, the upper left panel shows the binary components
projected on the plane normal to $\hat{v}$, and the upper right panel shows the
$UBVRIJHK$ light curves for a whole period with the magnitudes marked for this
phase.  The lower panel shows the emergent spectra from the secondary with
(black solid) and without (red solid) X-ray irradiation, from the disk itself
(blue long-dash) and the X-ray irradiation contribution (purple long-dash), and
from the binary as a whole (the uppermost black solid). The red dotted is the
blackbody approximation of the heated star spectrum. The wavelength unit is
nanometer (nm). }

\end{figure}

\begin{figure}

{\includegraphics[width=3.1in,height=2.1in]{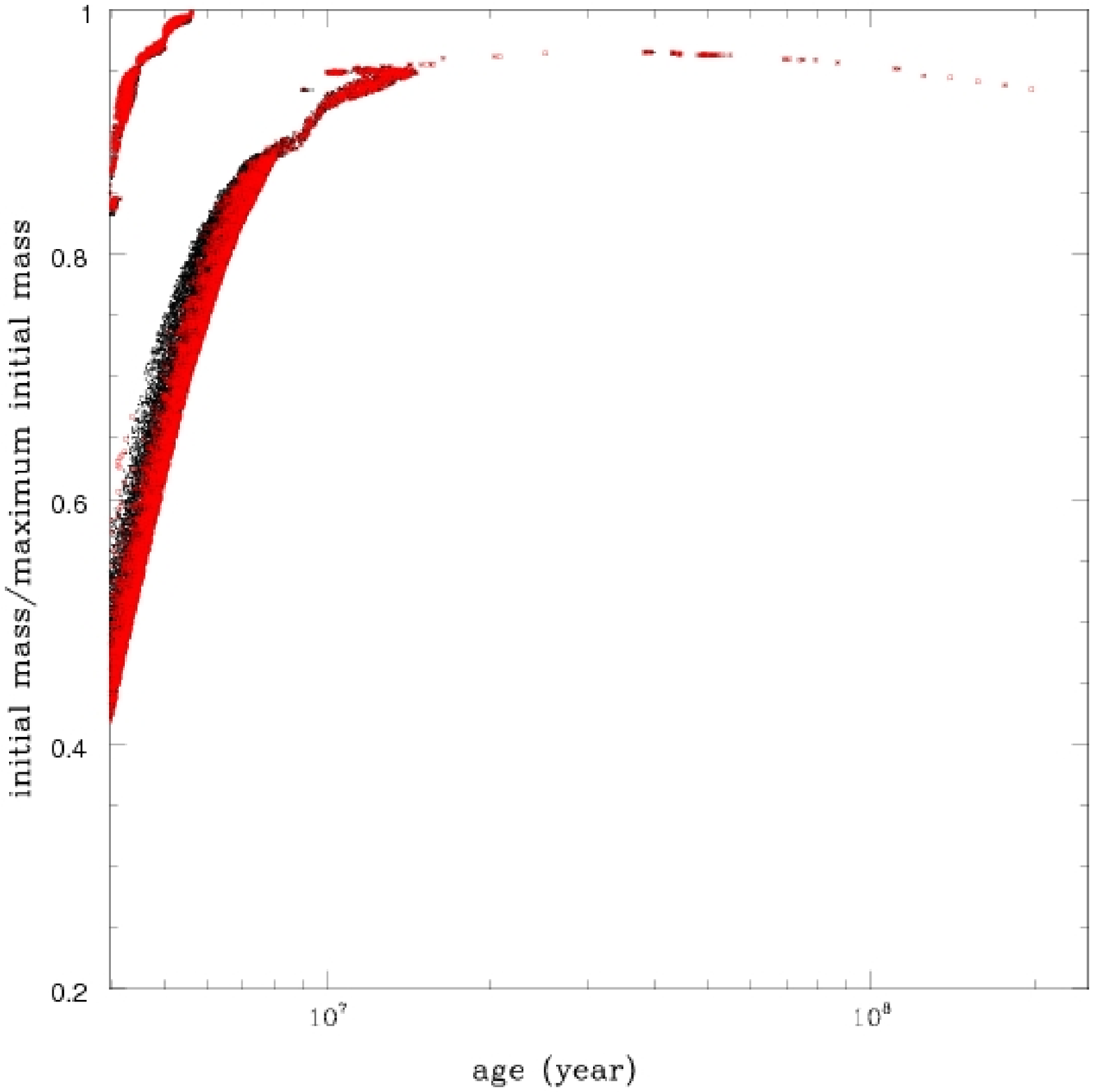}\includegraphics[width=3.1in,height=2.1in]{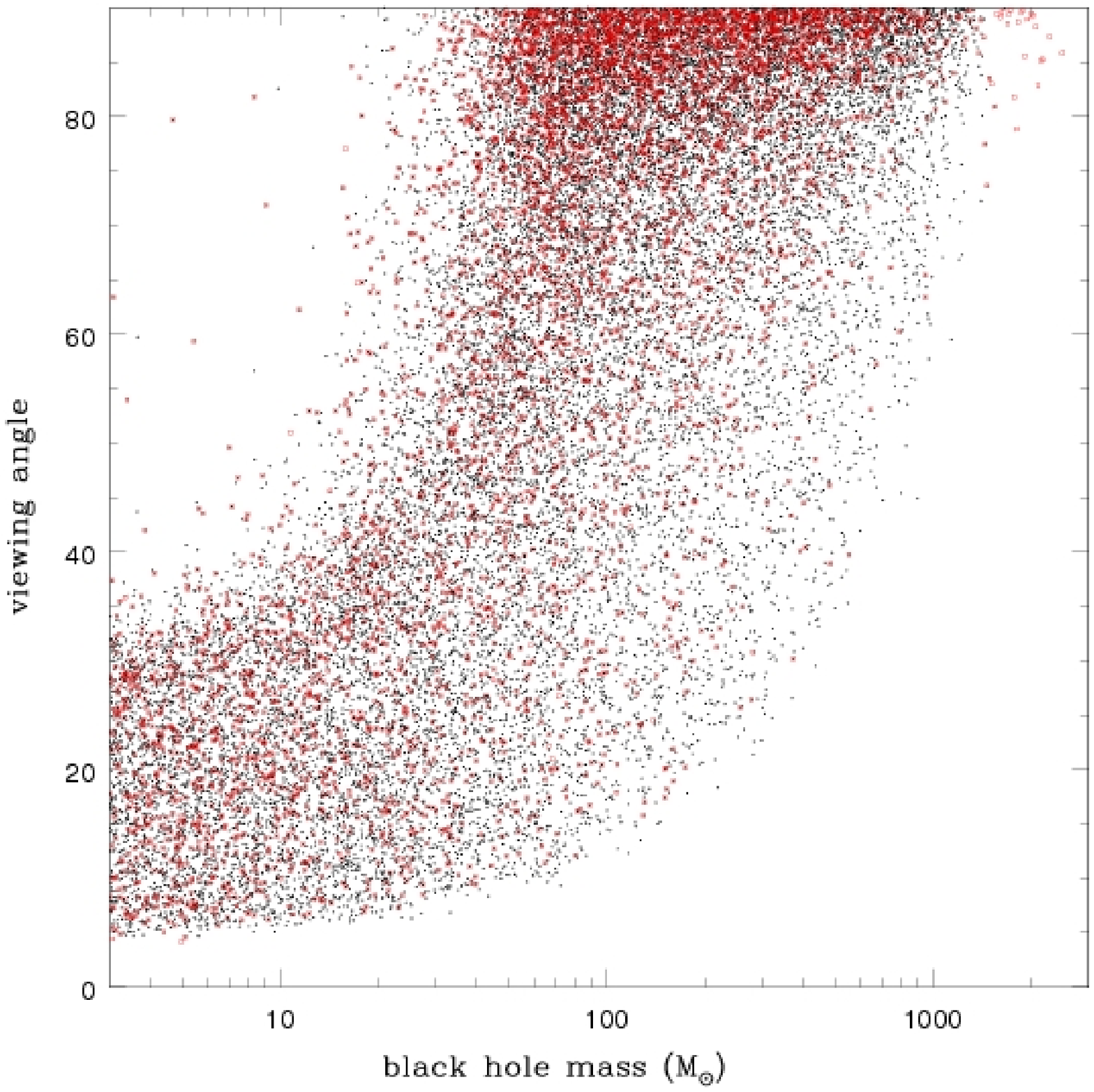}}
{\includegraphics[width=3.1in,height=2.1in]{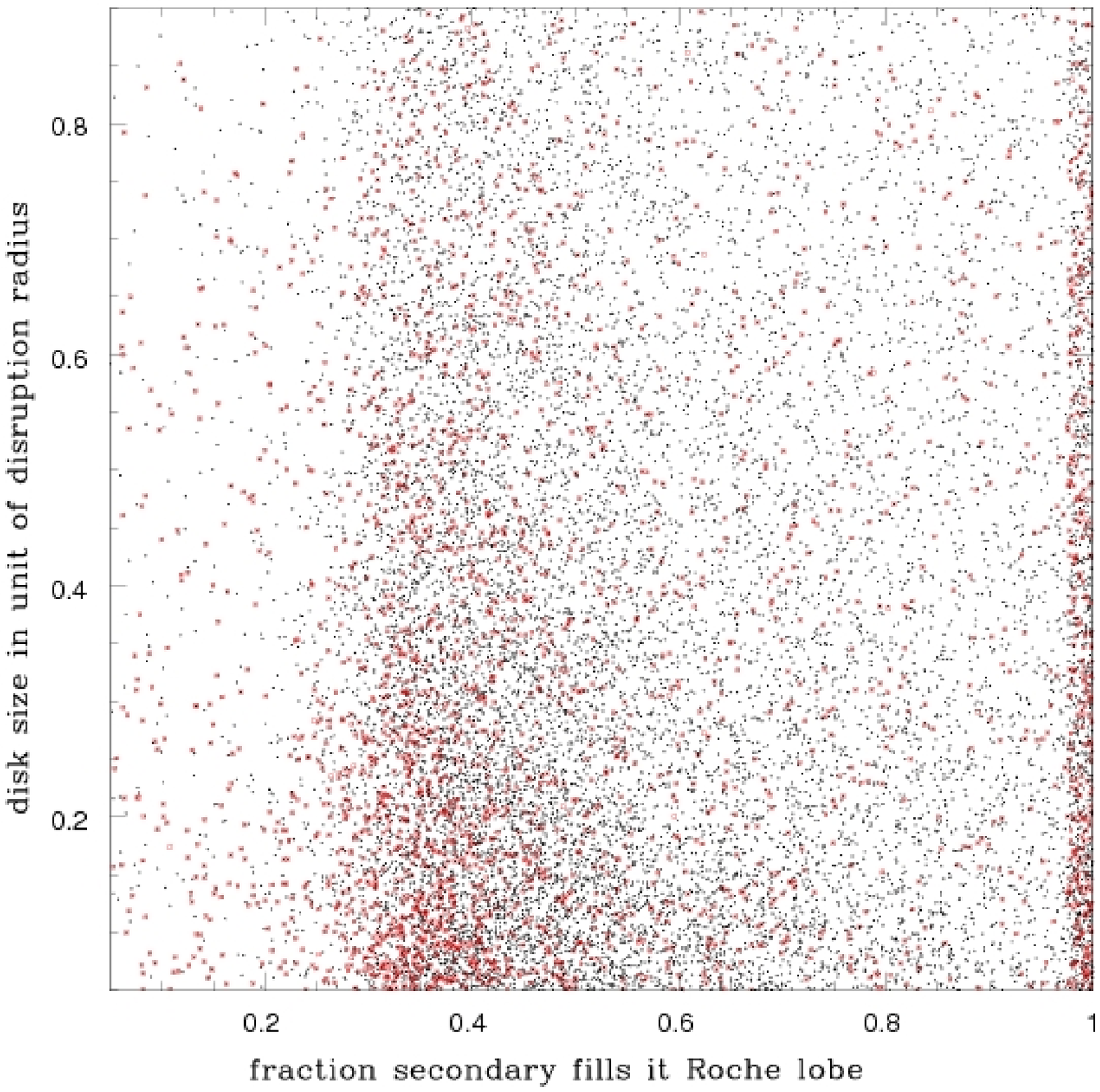}\includegraphics[width=3.1in,height=2.1in]{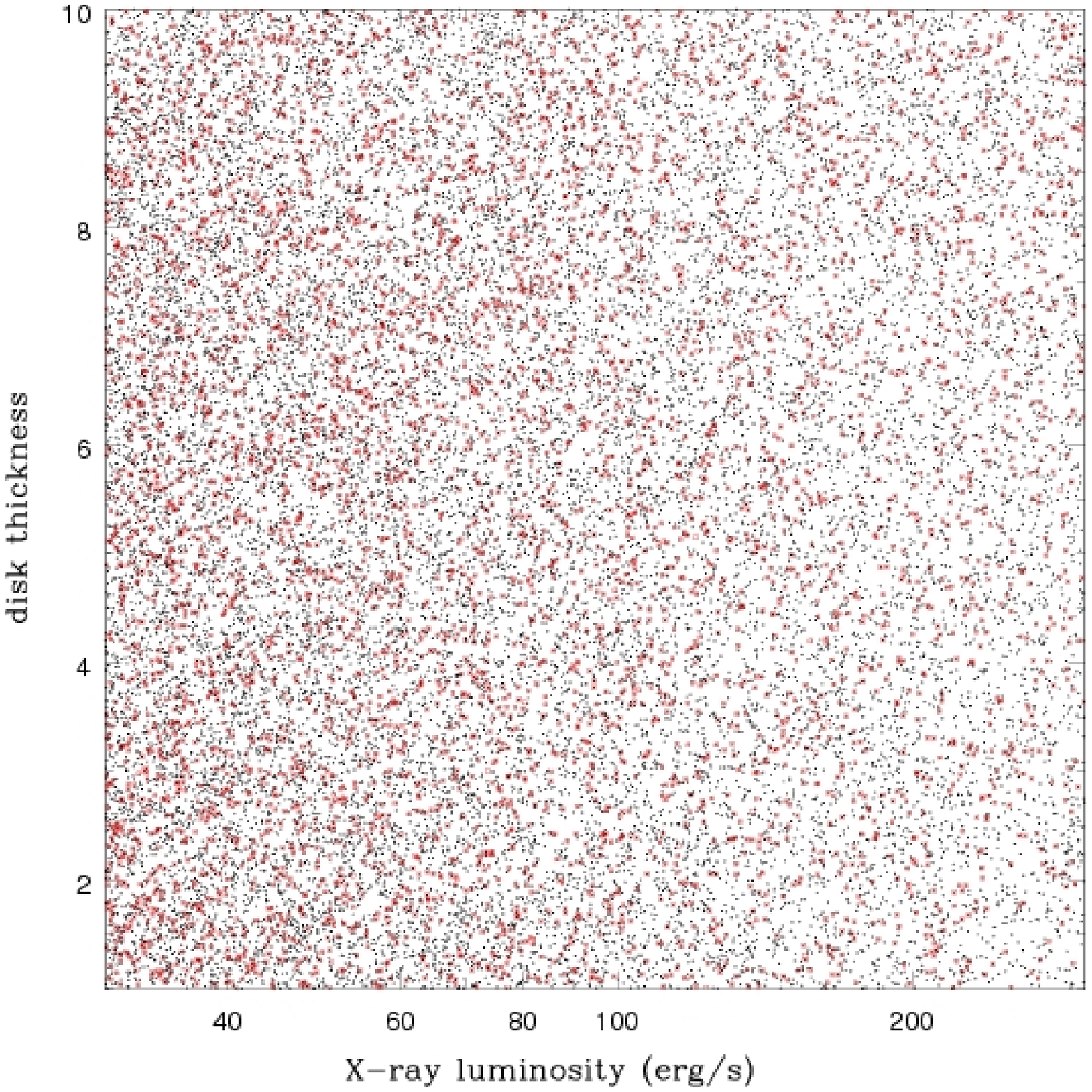}}
{\includegraphics[width=3.1in,height=2.1in]{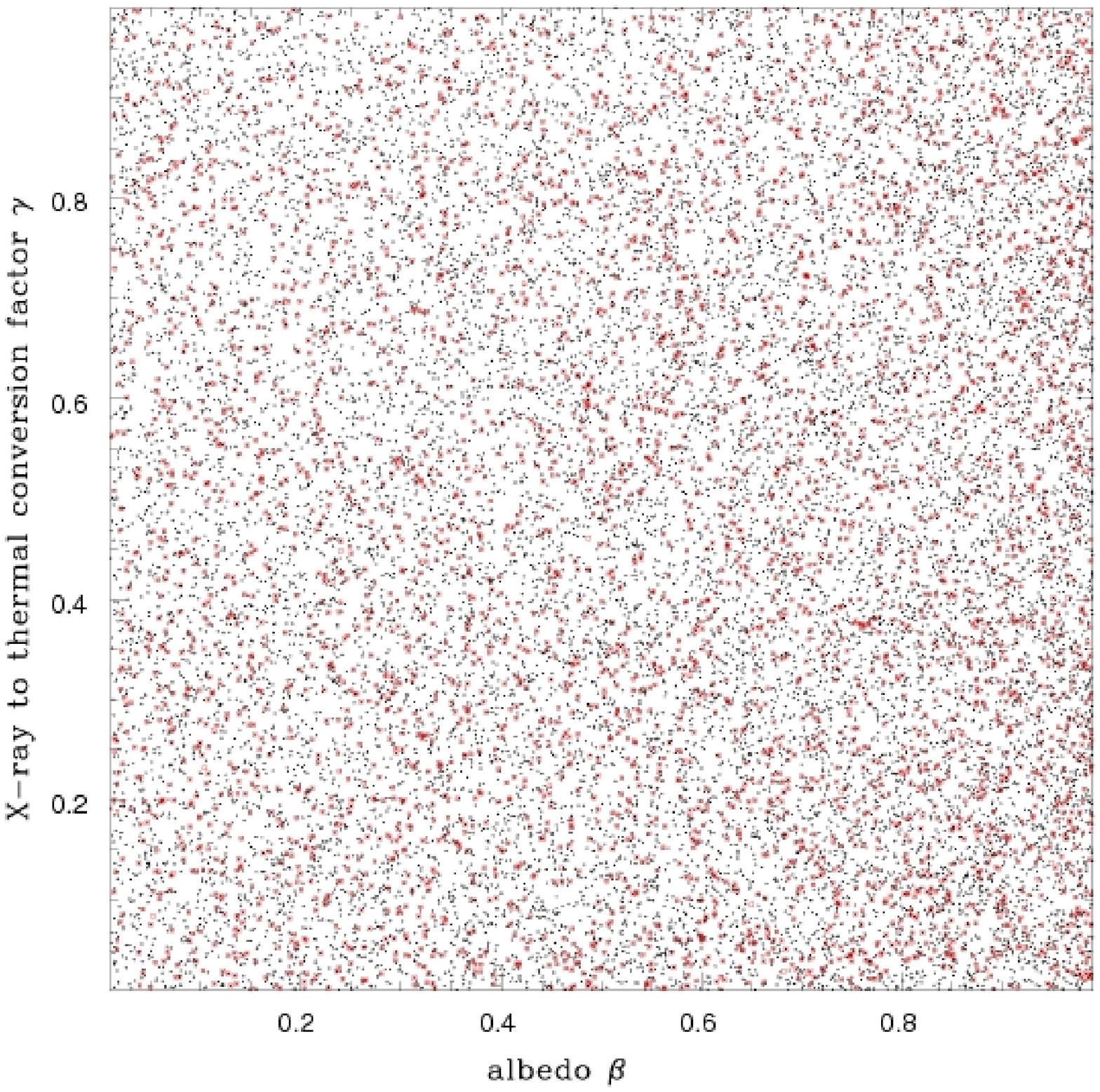}\includegraphics[width=3.1in,height=2.1in]{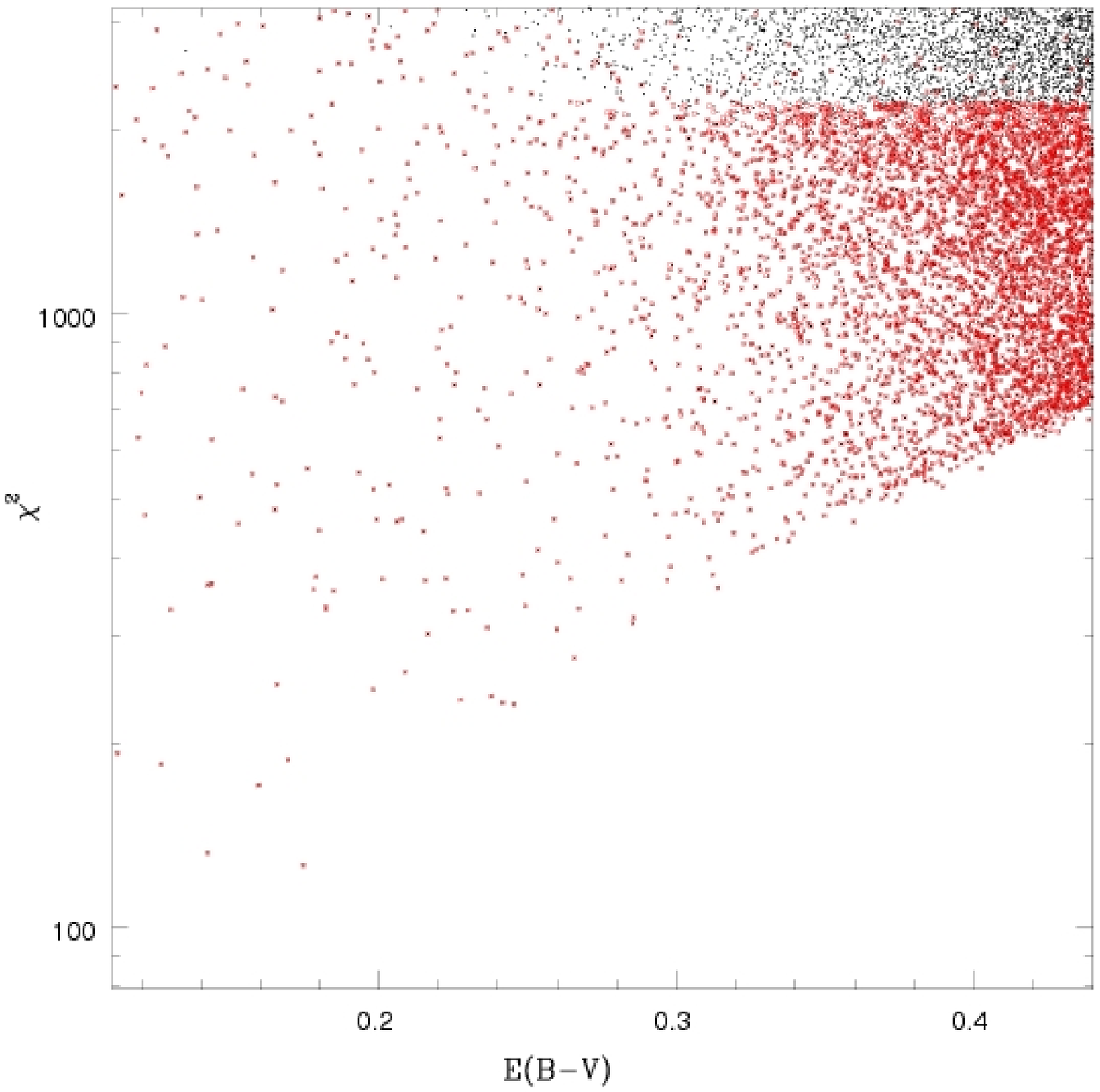}}

\caption{Models with $\chi^2<10000$ (black dots) for assumed $2K^\prime=200\pm40$ km/s. The six panels show (a) the
secondary age vs initial mass, (b) the black hole mass vs inclination angle,
(c) the fraction the secondary fills its Roche lobe vs the disk size in unit of
disruption radius, (d) the X-ray luminosity vs the disk thickness, (e) the disk
surface albedo vs the X-ray-to-thermal conversion factor on the secondary
surface, and (f) the extinction E(B-V) vs the model $\chi^2$. The red dots show
the models selected as starting points for AMOEBA searches. Note that the
considerations in the accretion rate lead to the concentration of models with
secondaries filling ($>$98\% of) the Roche lobe in (c). }

\end{figure}

\begin{figure}[t]

\center{\includegraphics[width=3.1in,height=3.1in]{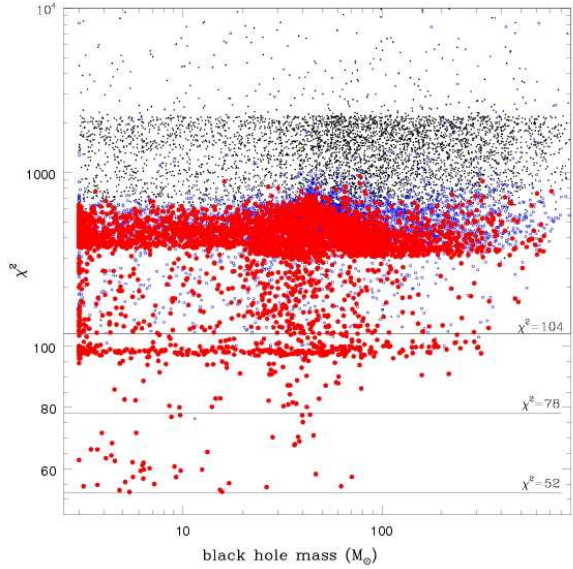}}

\caption{The black hole mass versus $\chi^2$ for models selected from the monte
carlo sampling of the phase space as starting points for AMOEBA searches (black
thin dots), the results for the first round of AMOEBA searches (blue open
circles) and for the second round of AMOEBA searches (red filled circles). The
horizontal lines denote $\chi^2$=52/78/104 (i.e., $\chi^2_\nu$=2/3/4). }

\end{figure}

\begin{figure}

{\includegraphics[width=3.1in,height=2.1in]{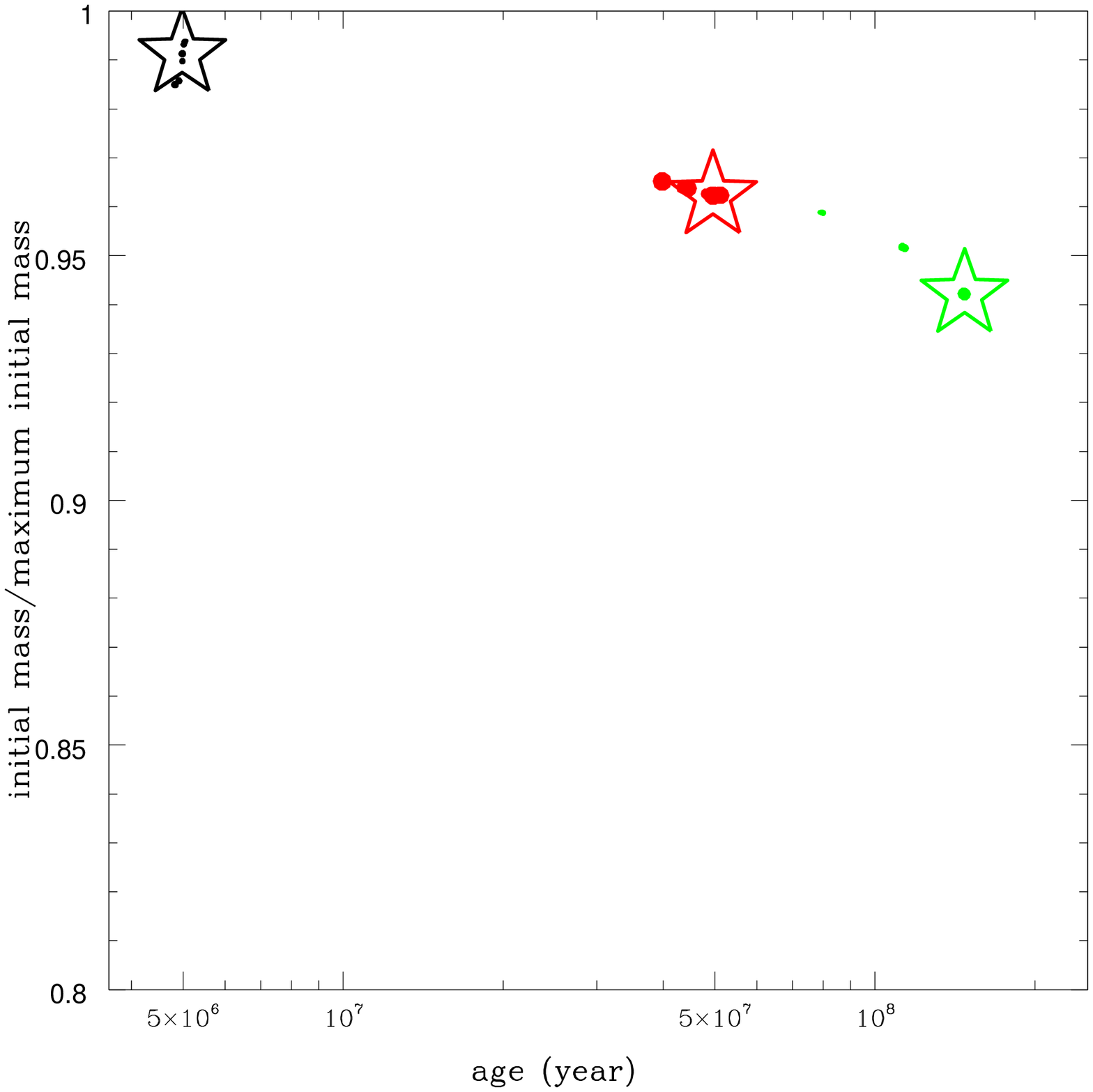}\includegraphics[width=3.1in,height=2.1in]{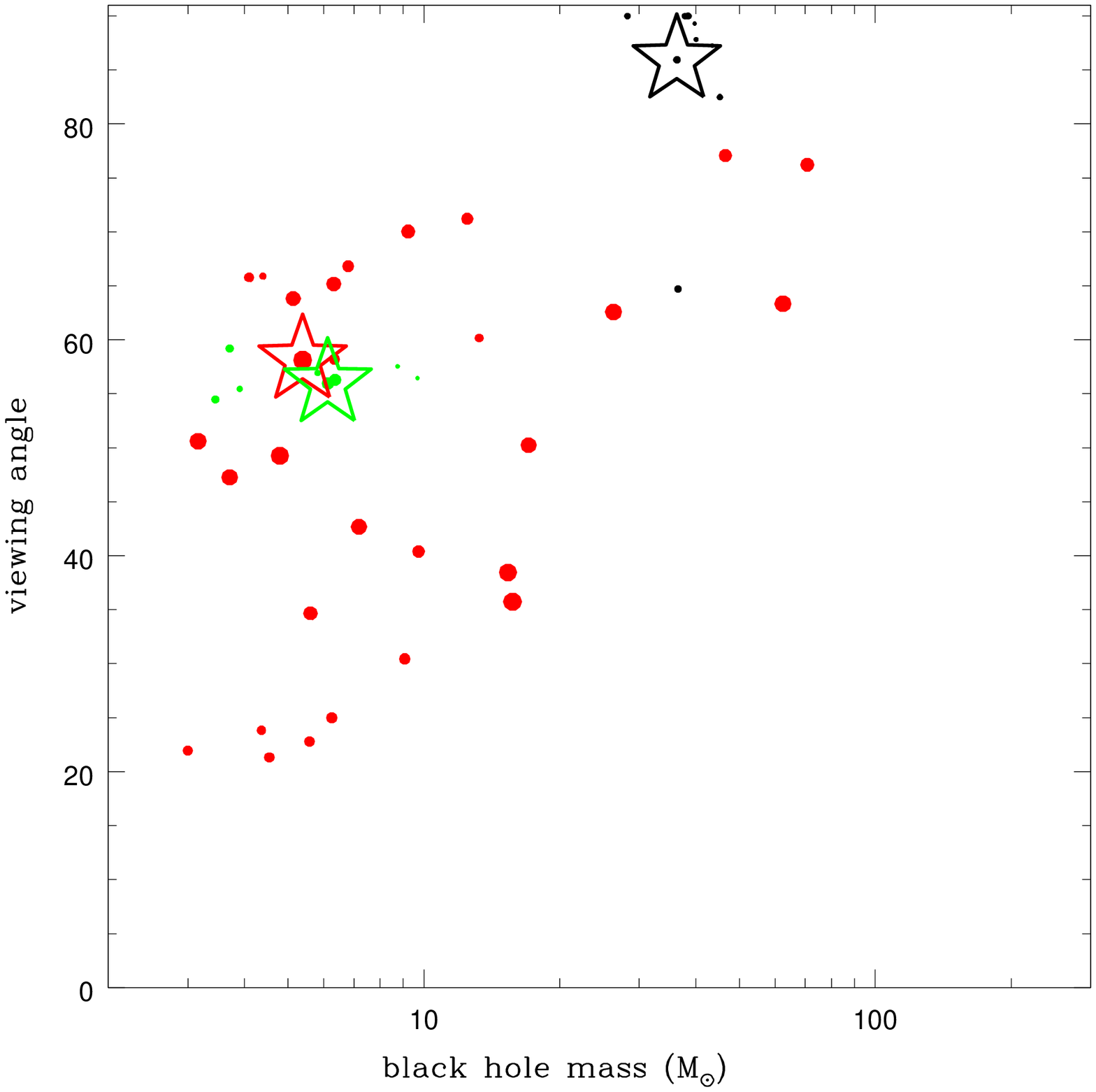}}
{\includegraphics[width=3.1in,height=2.1in]{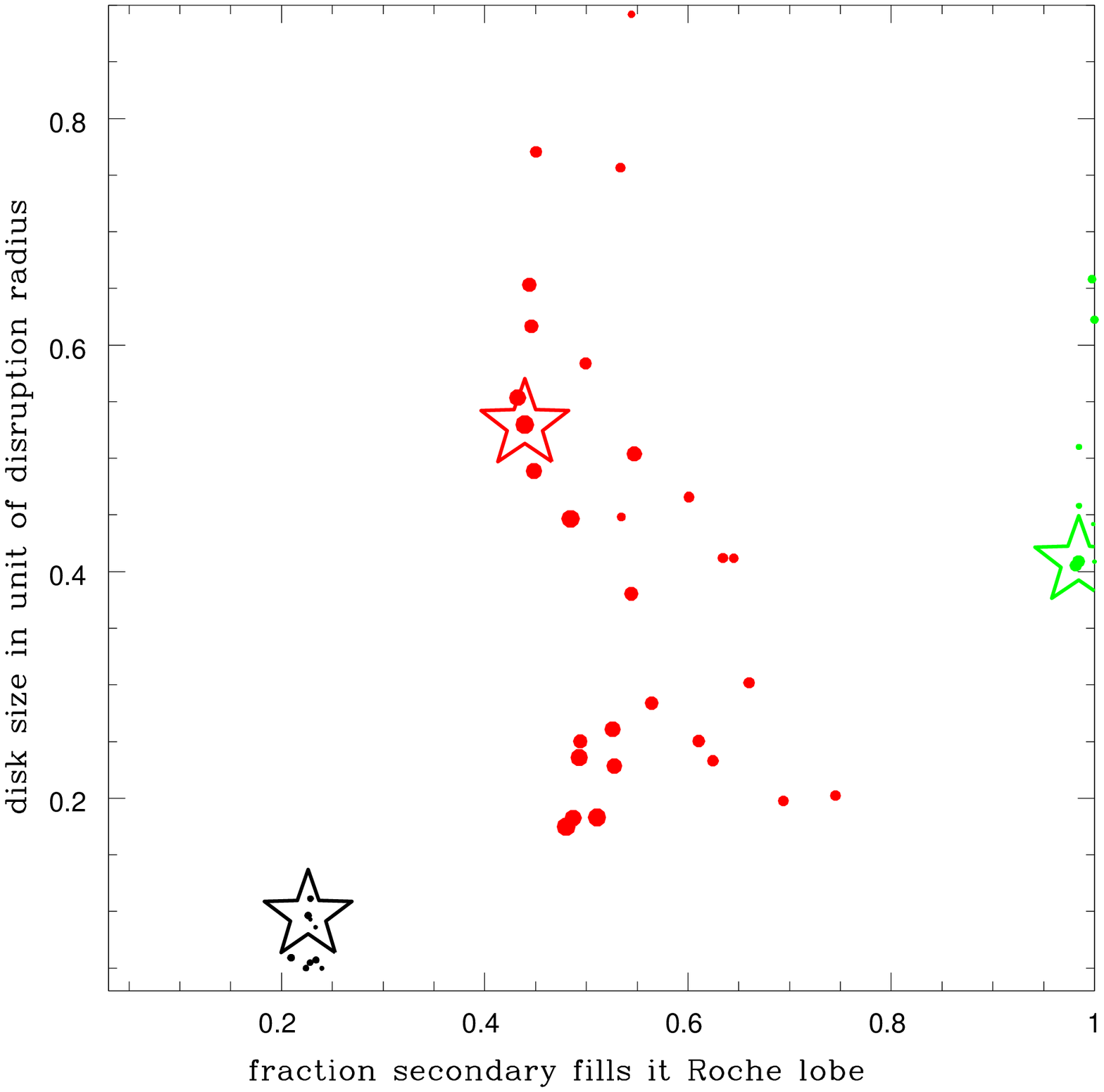}\includegraphics[width=3.1in,height=2.1in]{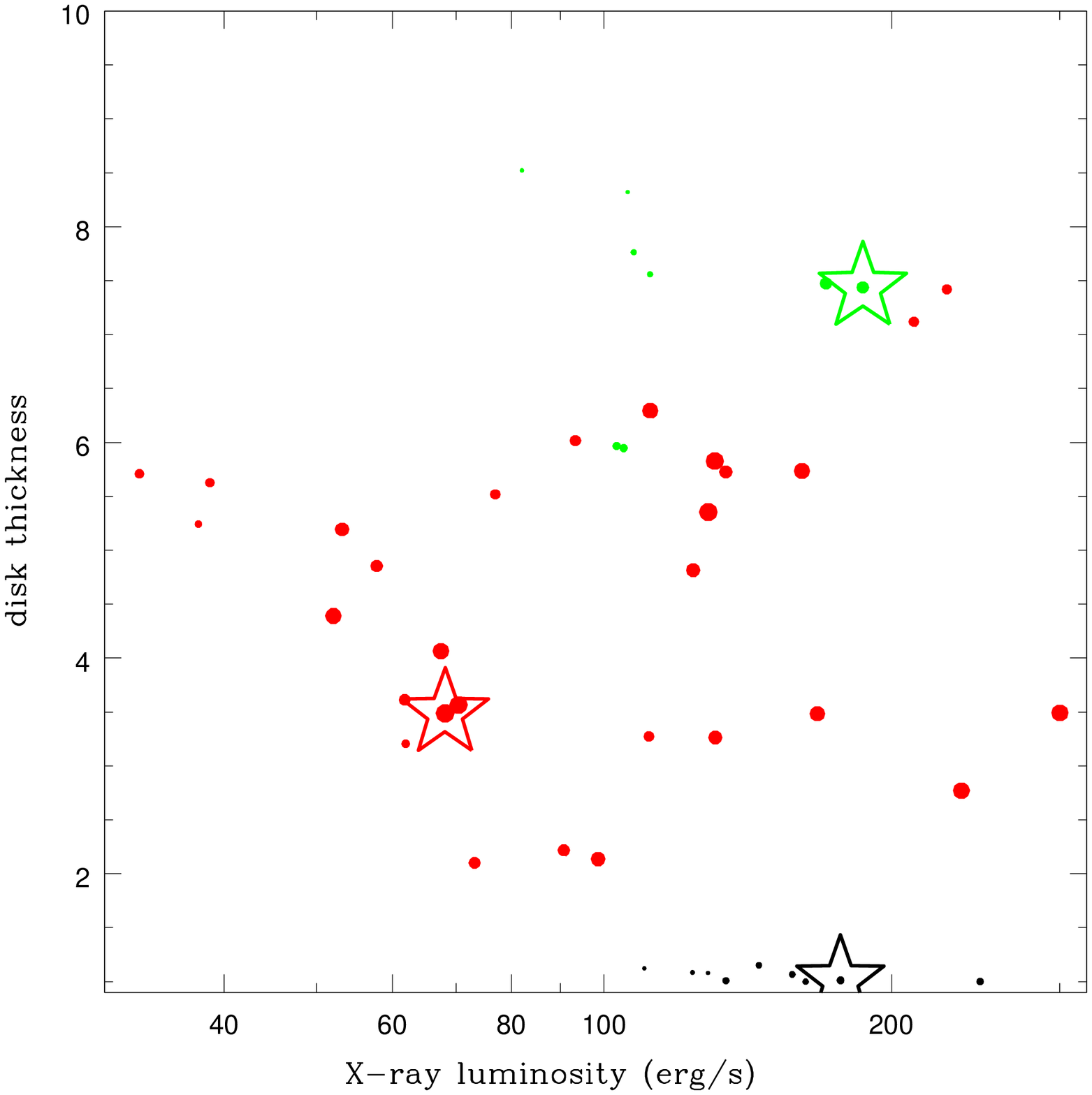}}
{\includegraphics[width=3.1in,height=2.1in]{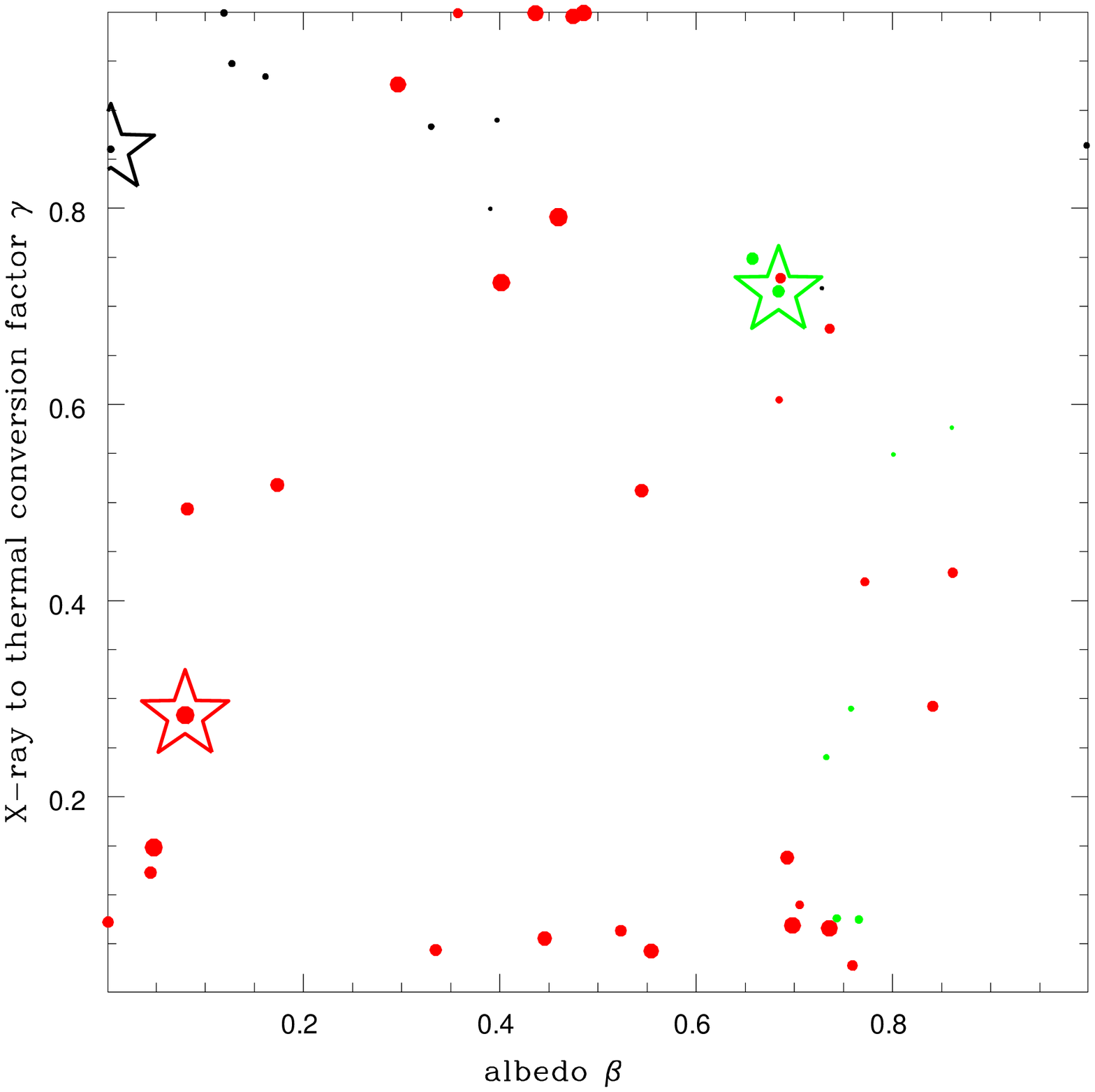}\includegraphics[width=3.1in,height=2.1in]{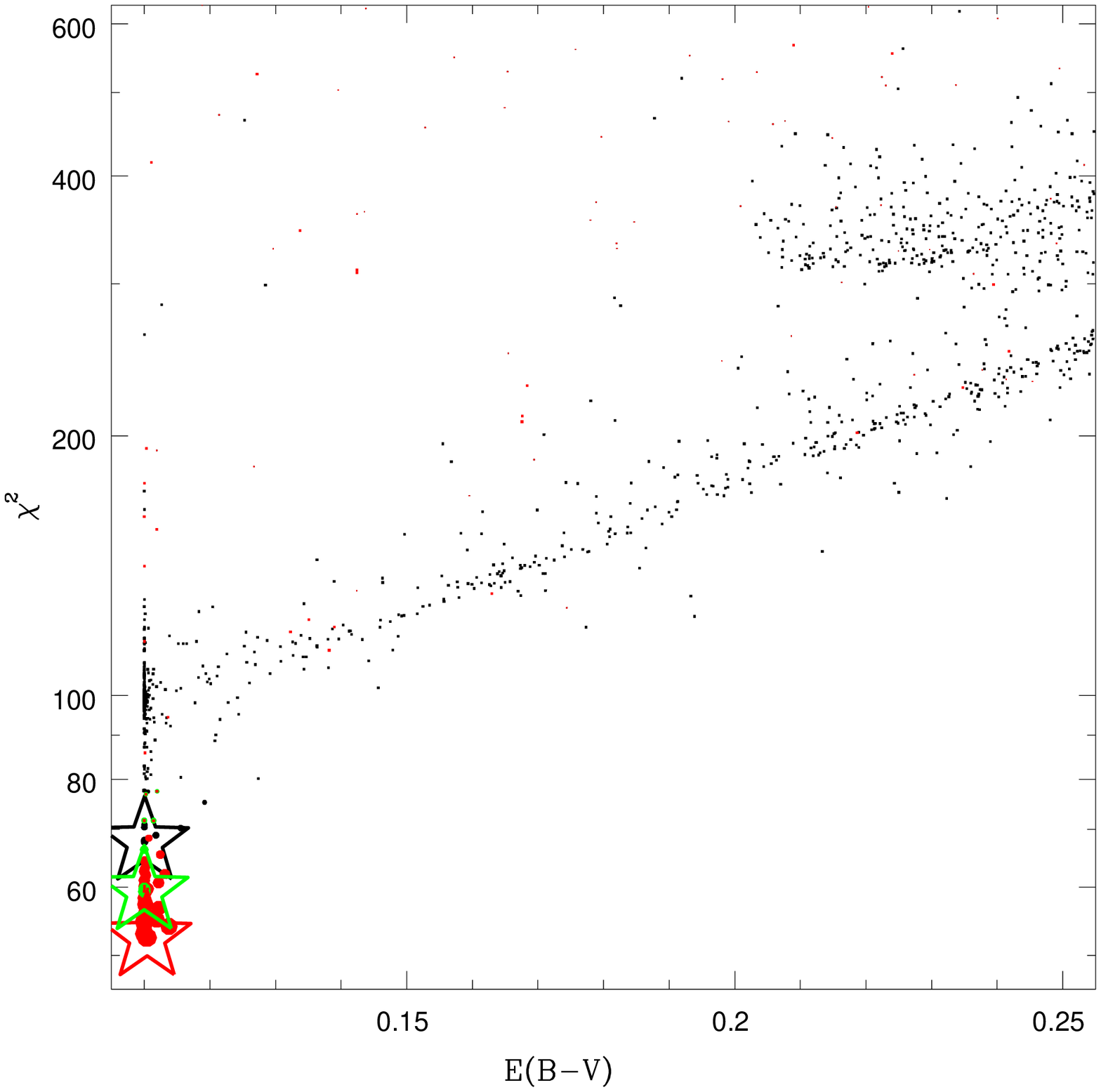}}

\caption{Models with $\chi^2<78$ obtained with two rounds of AMOEBA searches.
Models are divided into groups one (black), two (red) and three (green). The
symbol sizes are inversely scaled with the $\chi^2$ values of the models.  The
three asterisks are model A1 (black), model B1 (red) and model F1 (green) as
described in the text.  The six panels show (a) the secondary age vs initial
mass, (b) the black hole mass vs inclination angle, (c) the fraction the
secondary fills its Roche lobe vs the disk size in unit of disruption radius,
(d) the X-ray luminosity vs the disk thickness, (e) the disk surface albedo vs
the X-ray-to-thermal conversion factor on the secondary surface, and (f) the
extinction E(B-V) vs the model $\chi^2$ for models with $\chi^2<600$. }

\end{figure}

\begin{figure}[t]

{\includegraphics[width=3.1in,height=3.1in]{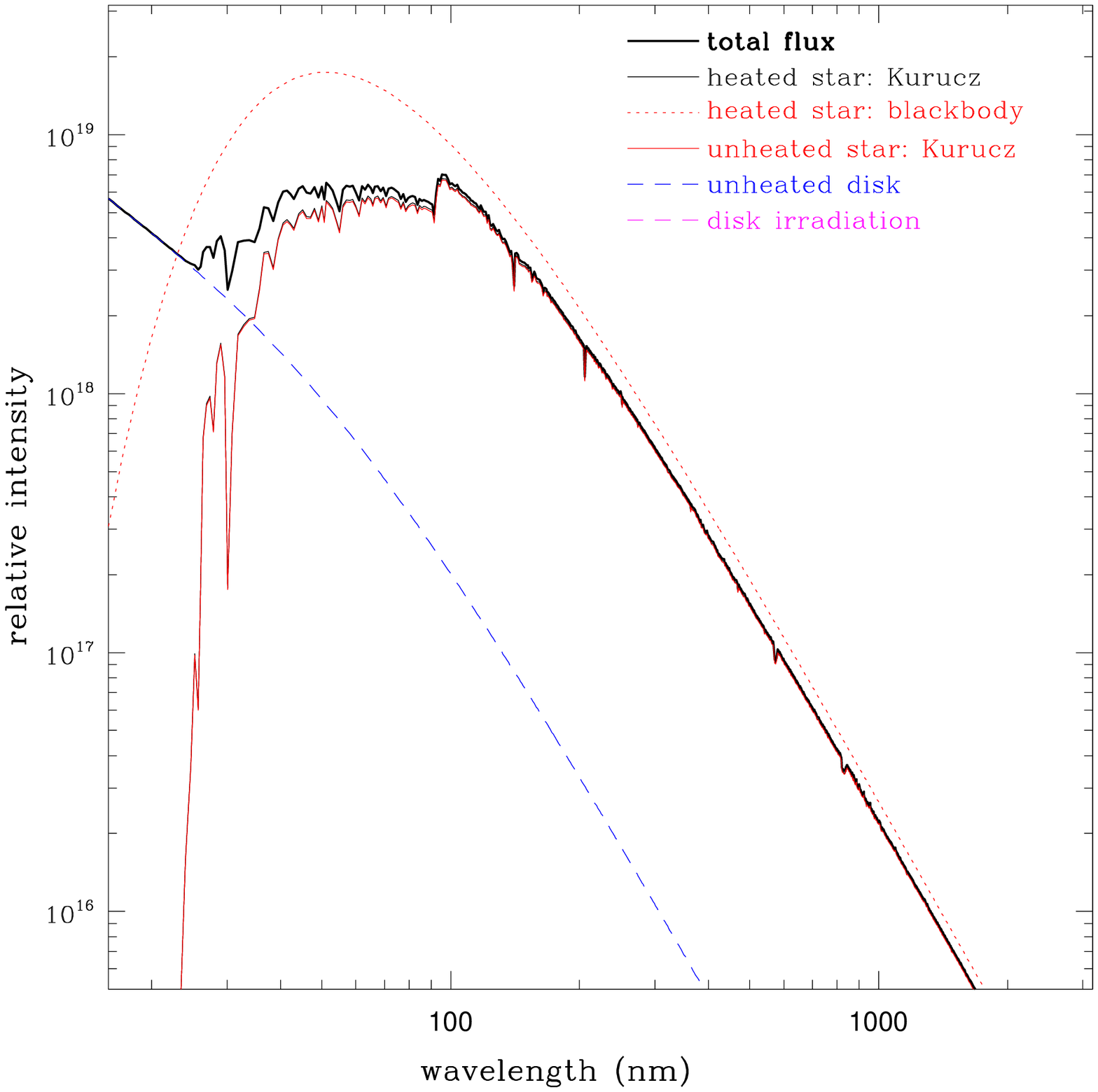}\includegraphics[width=3.1in,height=3.1in]{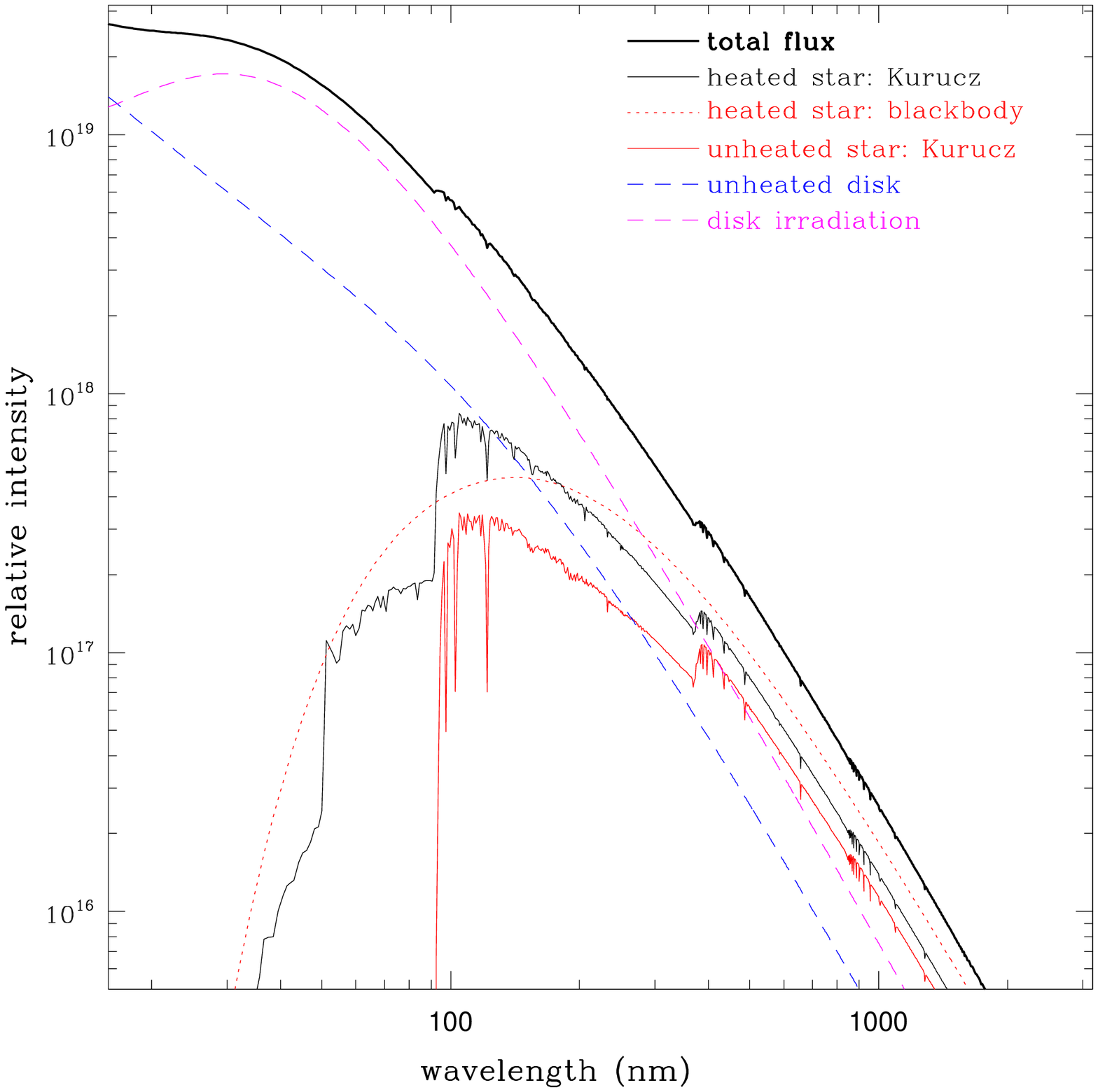}}
{\includegraphics[width=3.1in,height=3.1in]{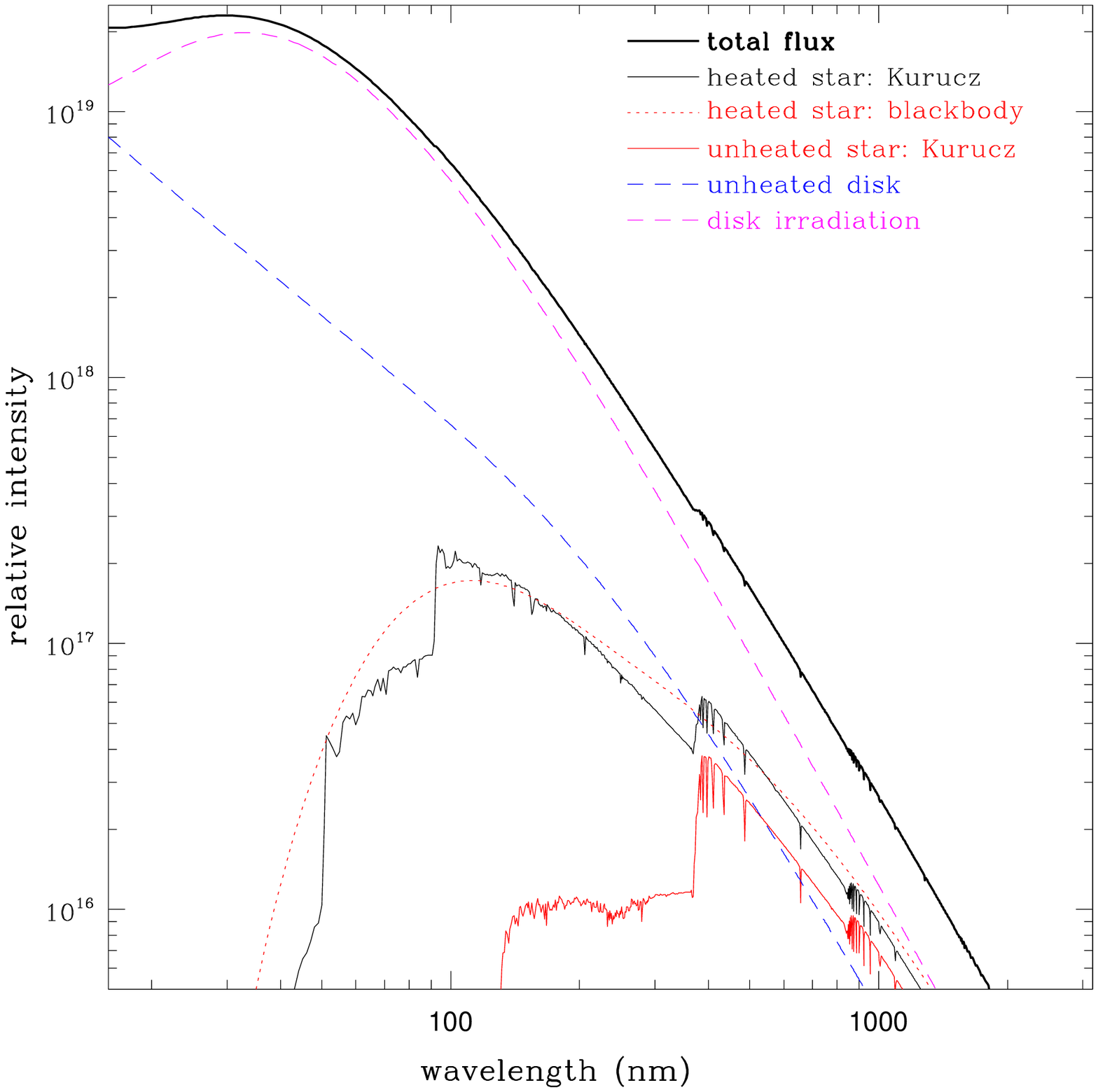}\includegraphics[width=3.1in,height=3.1in]{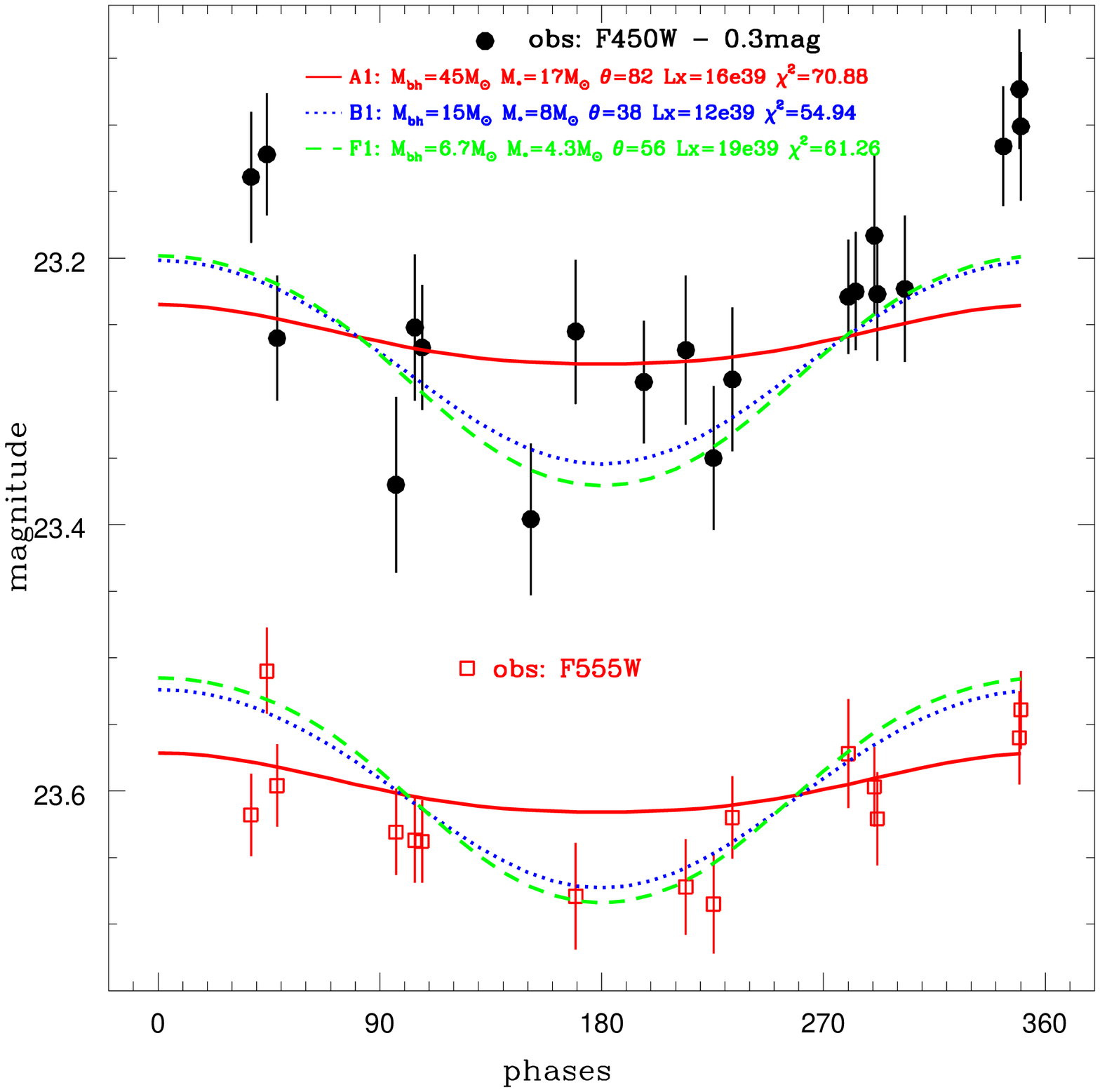}}

\caption{(a) The emergent spectrum from model A1 at binary phase $90^\circ$. The
spectrum is dominated by the hot secondary that is hardly affected by the X-ray
heating. (b) The emergent spectrum from model B1 at binary phase 90$^\circ$. The
secondary and the disk contribute roughly equally to the emergent spectrum at
4000\AA, while the secondary/disk contributes slightly more at longer/shorter
wavelengths. The X-ray heating boosts the secondary light by $>$30\%. (c) The
emergent spectrum from model F1 at binary phase 90$^\circ$.  The spectrum is
dominated by the X-ray heated accretion disk, which contributes four times more
light than the X-ray heated secondary at 4000\AA. The X-ray heating boosts the
secondary light by $>$60\%. (d)
The predicted light curves for model A1 (red solid), model B1 (blue
dotted line) and model F1 (green dashed line) with HST/WFPC2 F450W/F555W
observations overplotted. }

\end{figure}

\begin{figure}

{\includegraphics[width=3.1in,height=2.1in]{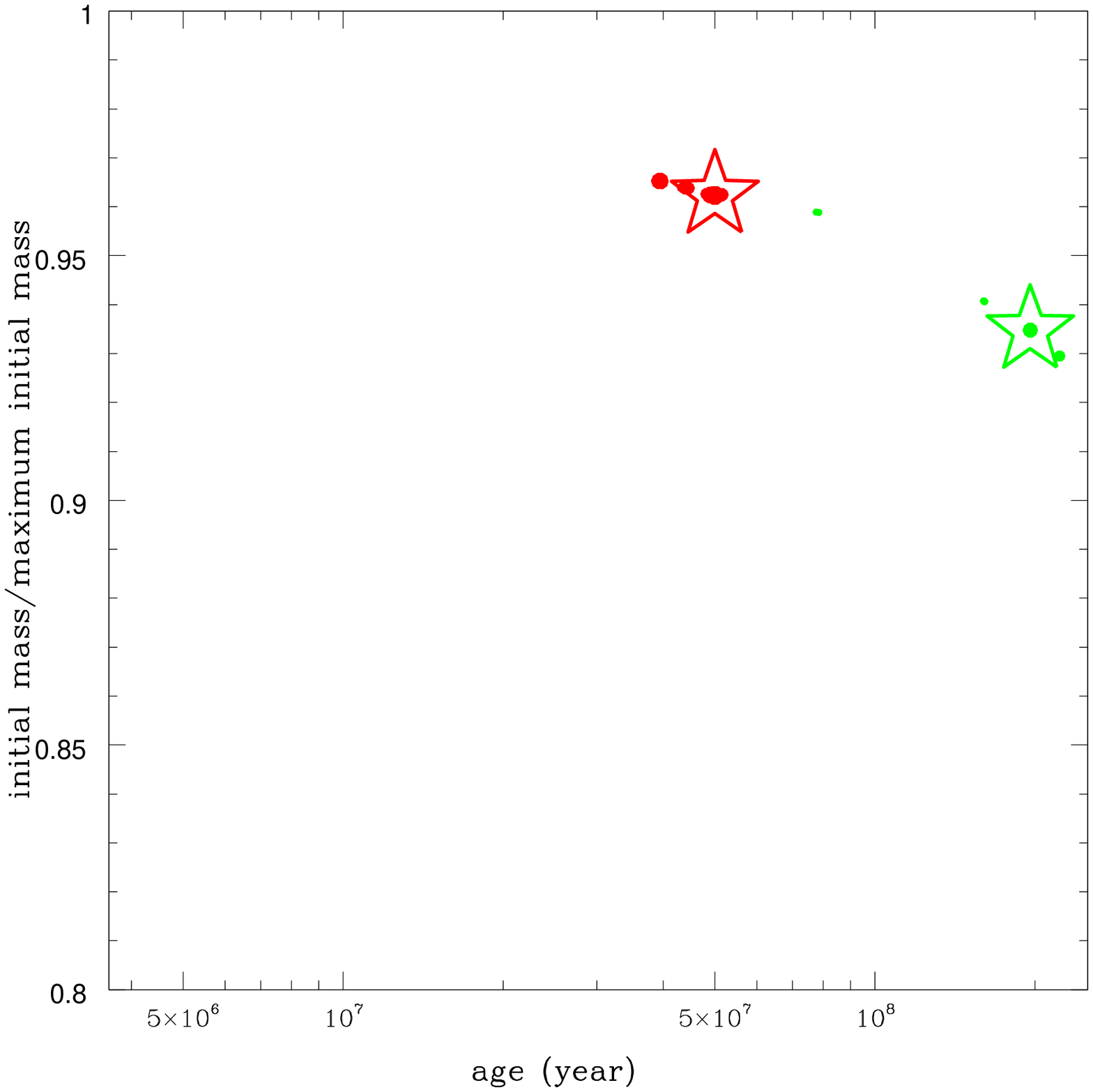}\includegraphics[width=3.1in,height=2.1in]{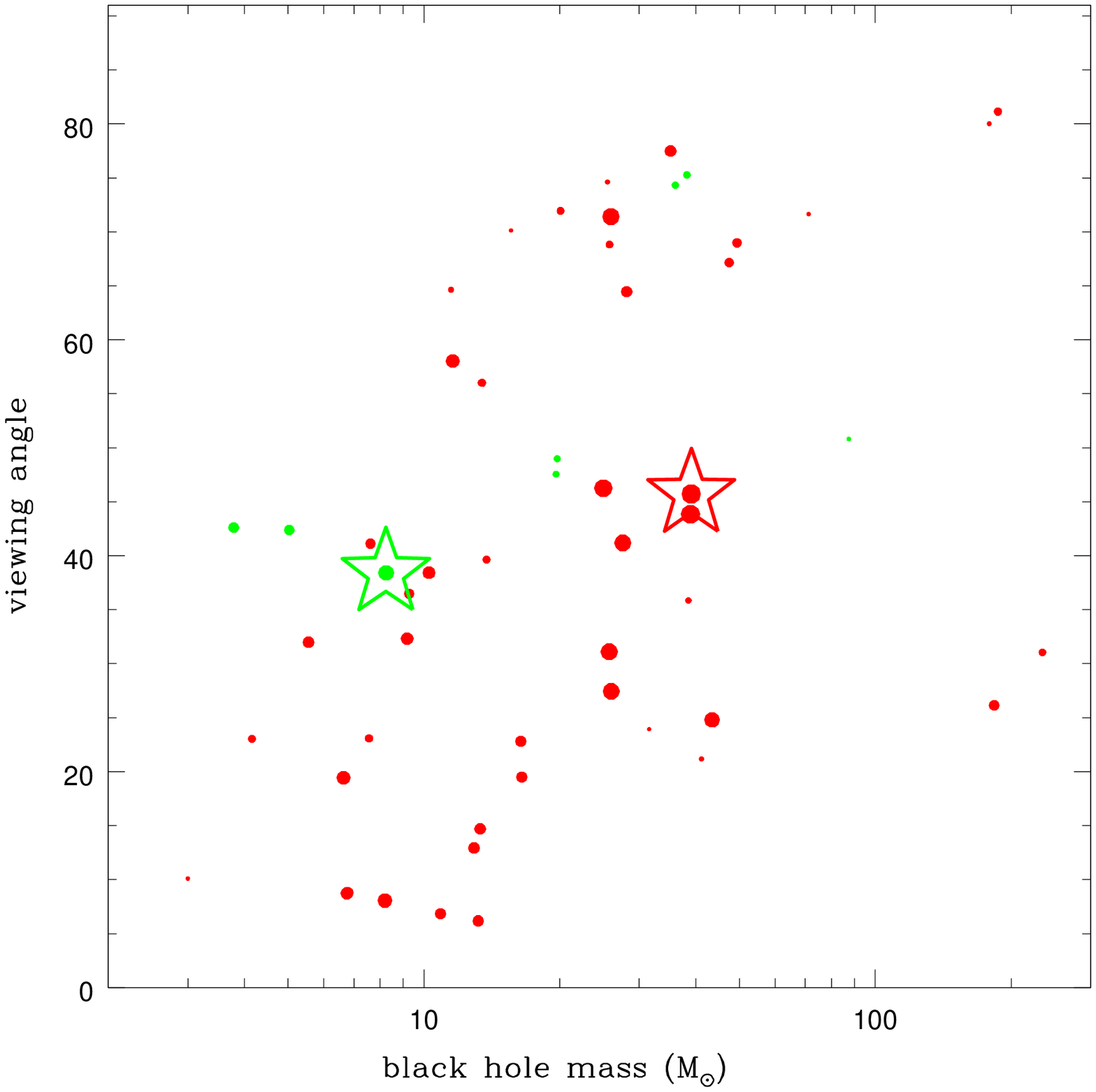}}
{\includegraphics[width=3.1in,height=2.1in]{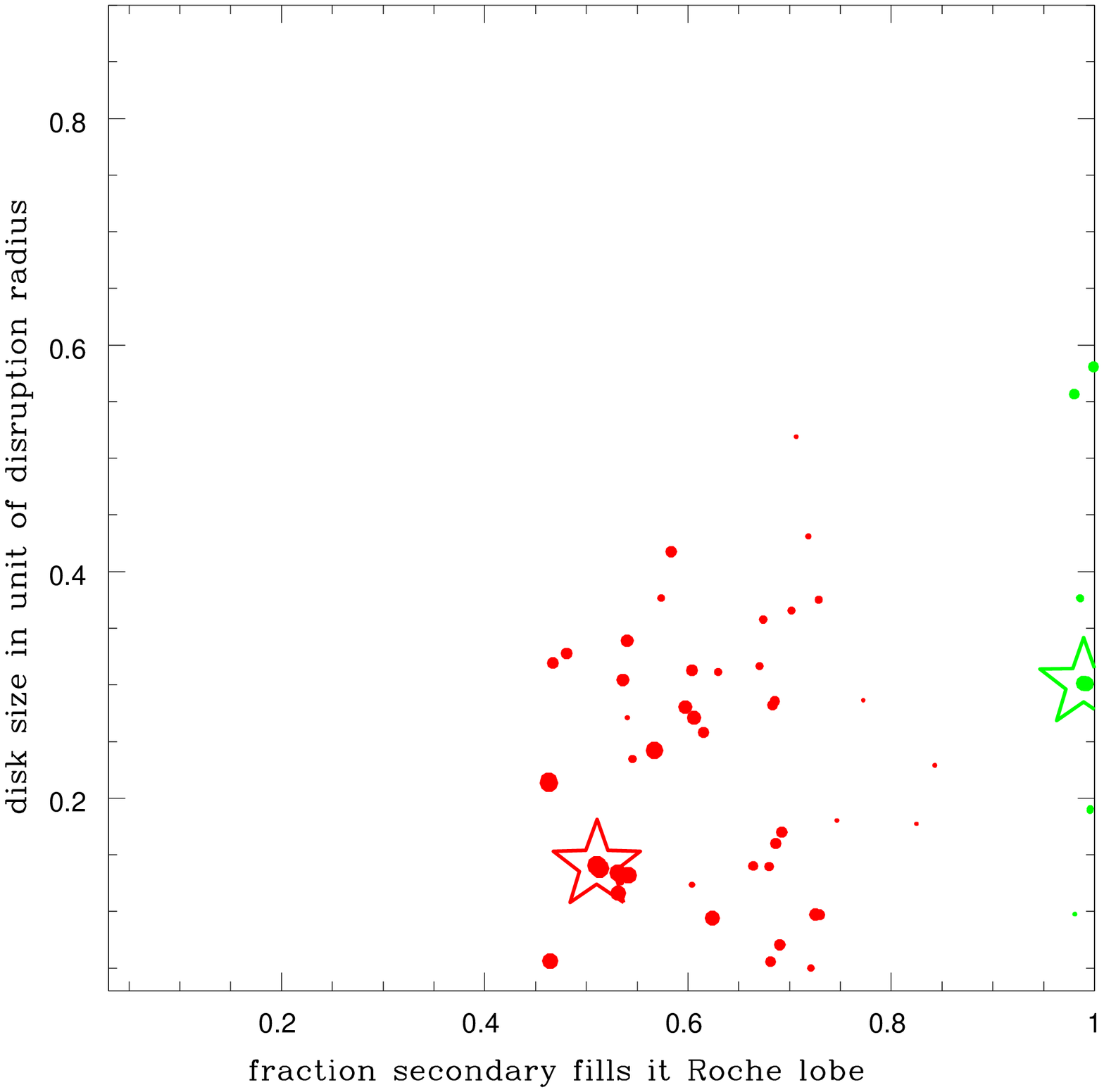}\includegraphics[width=3.1in,height=2.1in]{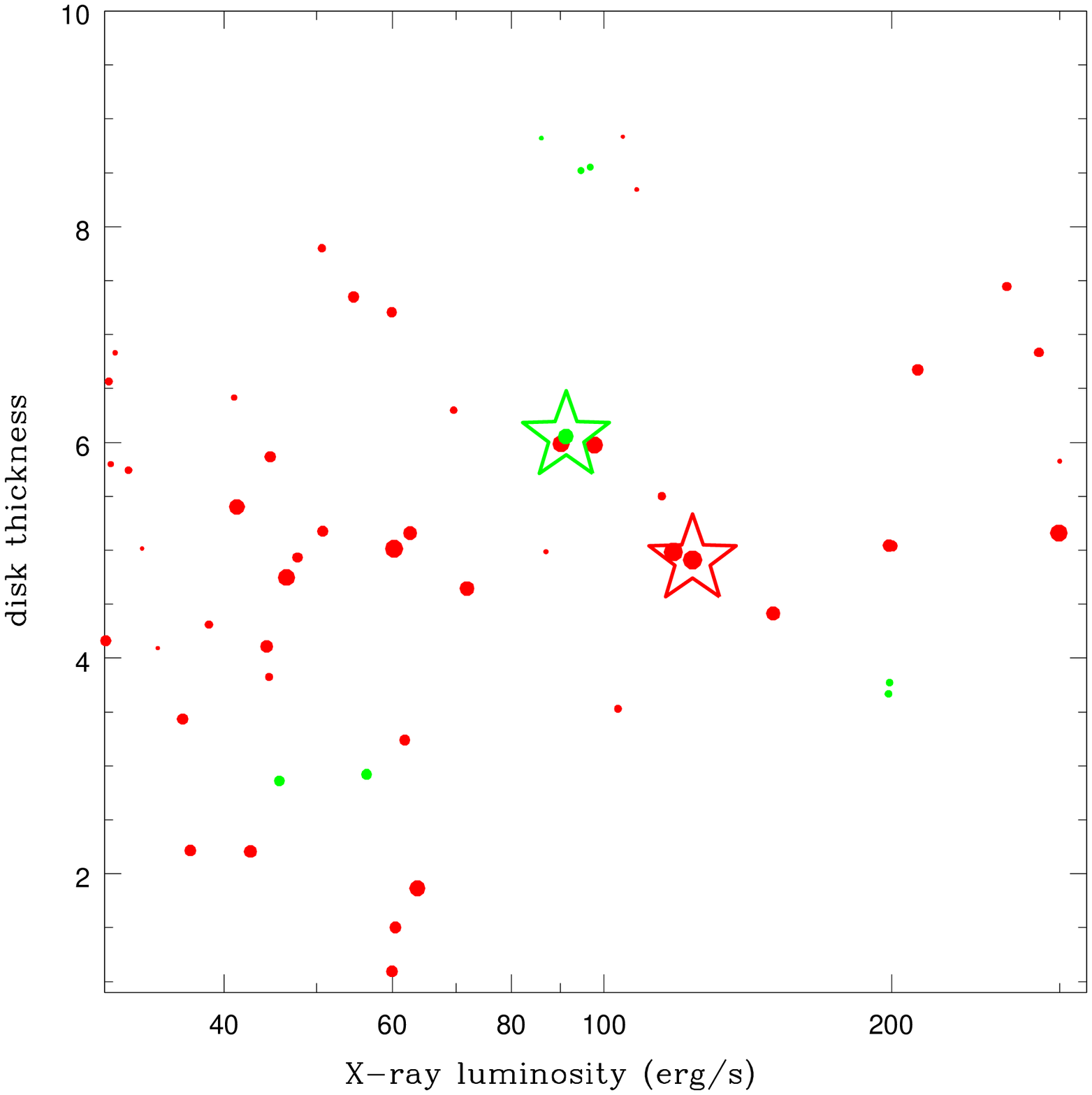}}
{\includegraphics[width=3.1in,height=2.1in]{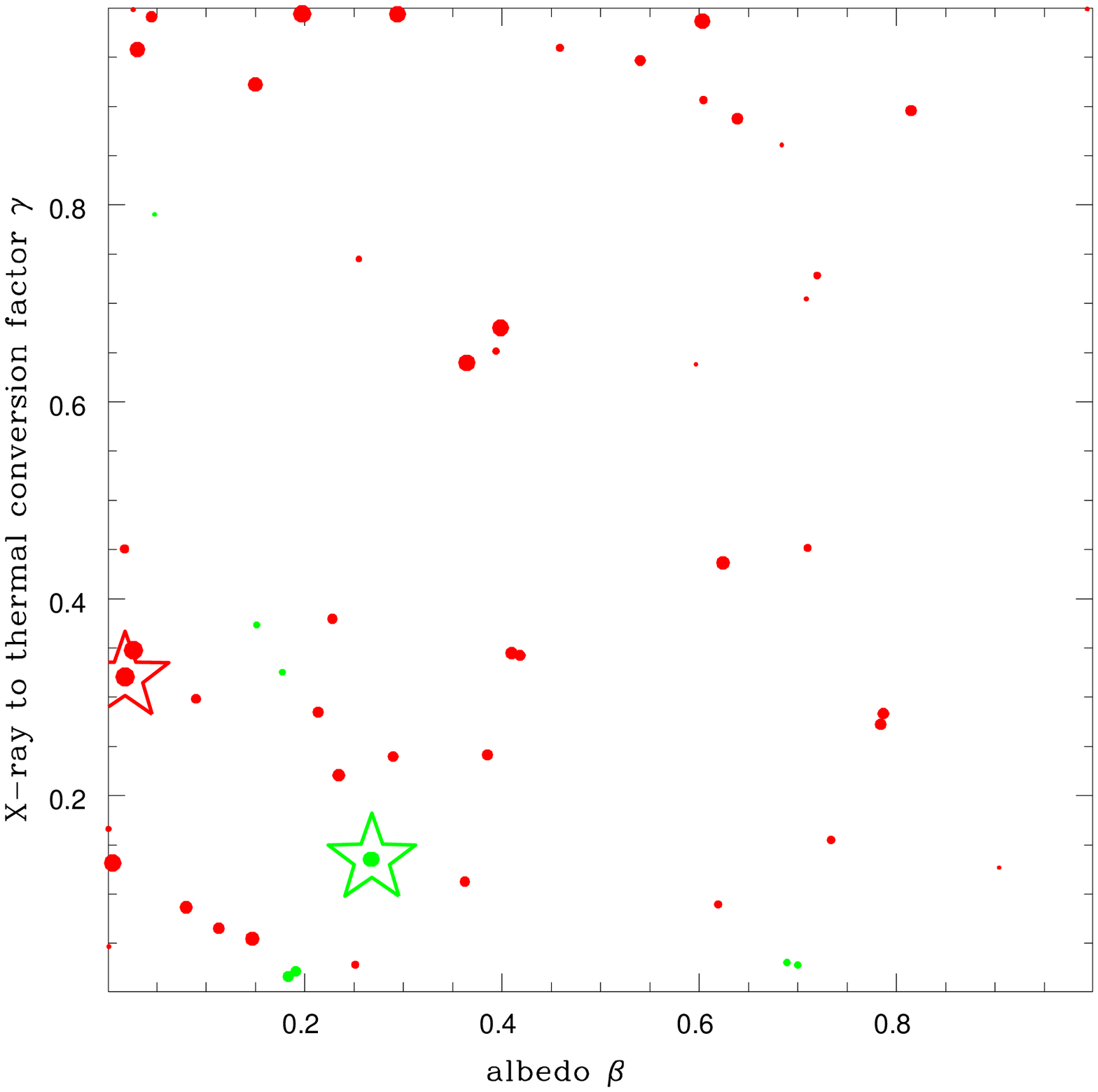}\includegraphics[width=3.1in,height=2.1in]{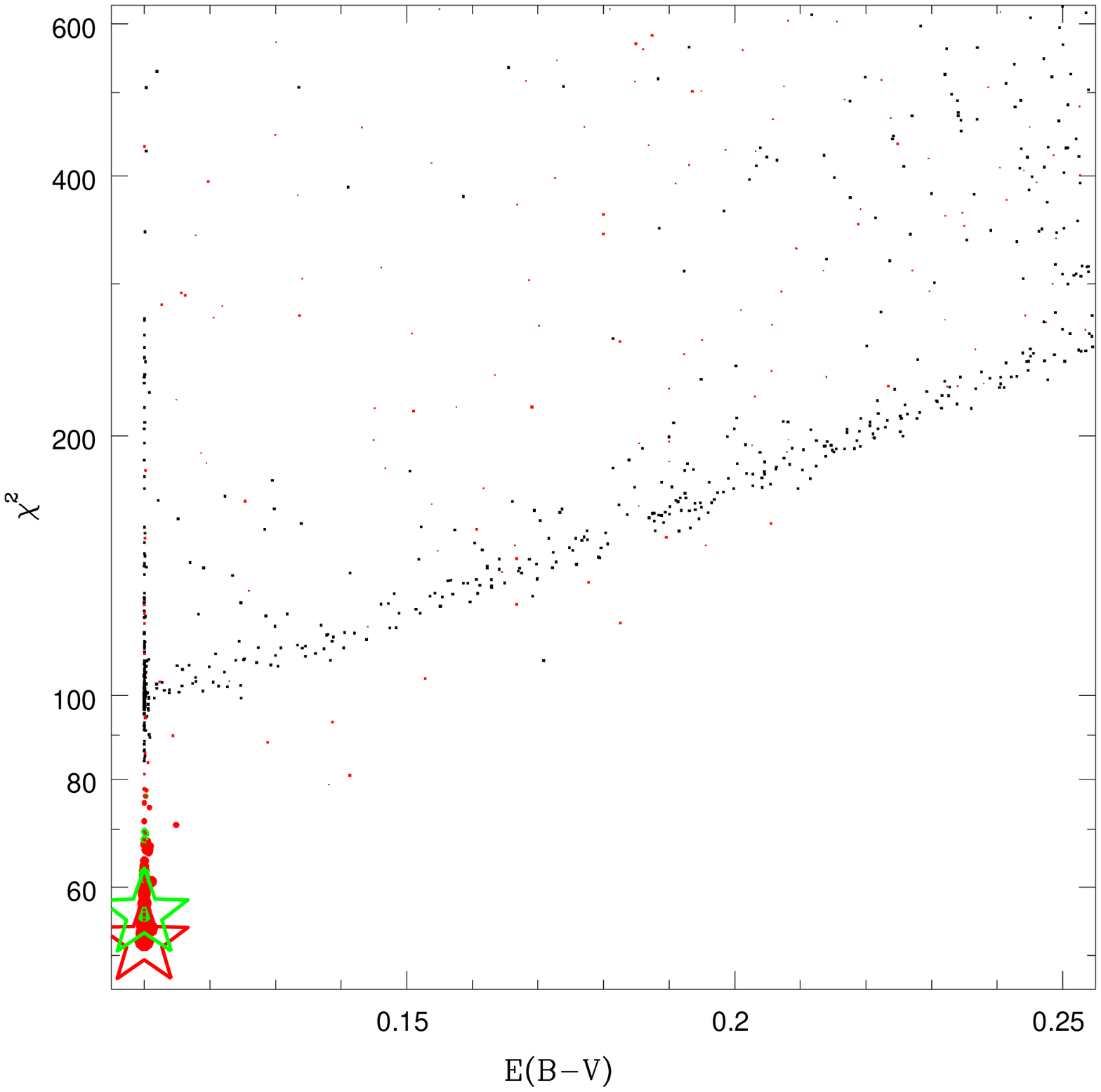}}

\caption{Models with $\chi^2<78$ obtained with two rounds of AMOEBA searches
for assumed $2K^\prime=100\pm40$ km/s.  The sysmbols are the same as in
Figure~4.  The asterisks are for model B2 (red) and model F2 (green) as described
in the text.  The six panels show (a) the secondary age vs initial mass, (b)
the black hole mass vs inclination angle, (c) the fraction the secondary fills
its Roche lobe vs the disk size in unit of disruption radius, (d) the X-ray
luminosity vs the disk thickness, (e) the disk surface albedo vs the
X-ray-to-thermal conversion factor on the secondary surface, and (f) the
extinction E(B-V) vs the model $\chi^2$ for models with $\chi^2<600$. }

\end{figure}

\begin{figure}[t]

\center{\includegraphics[width=3.1in,height=3.1in]{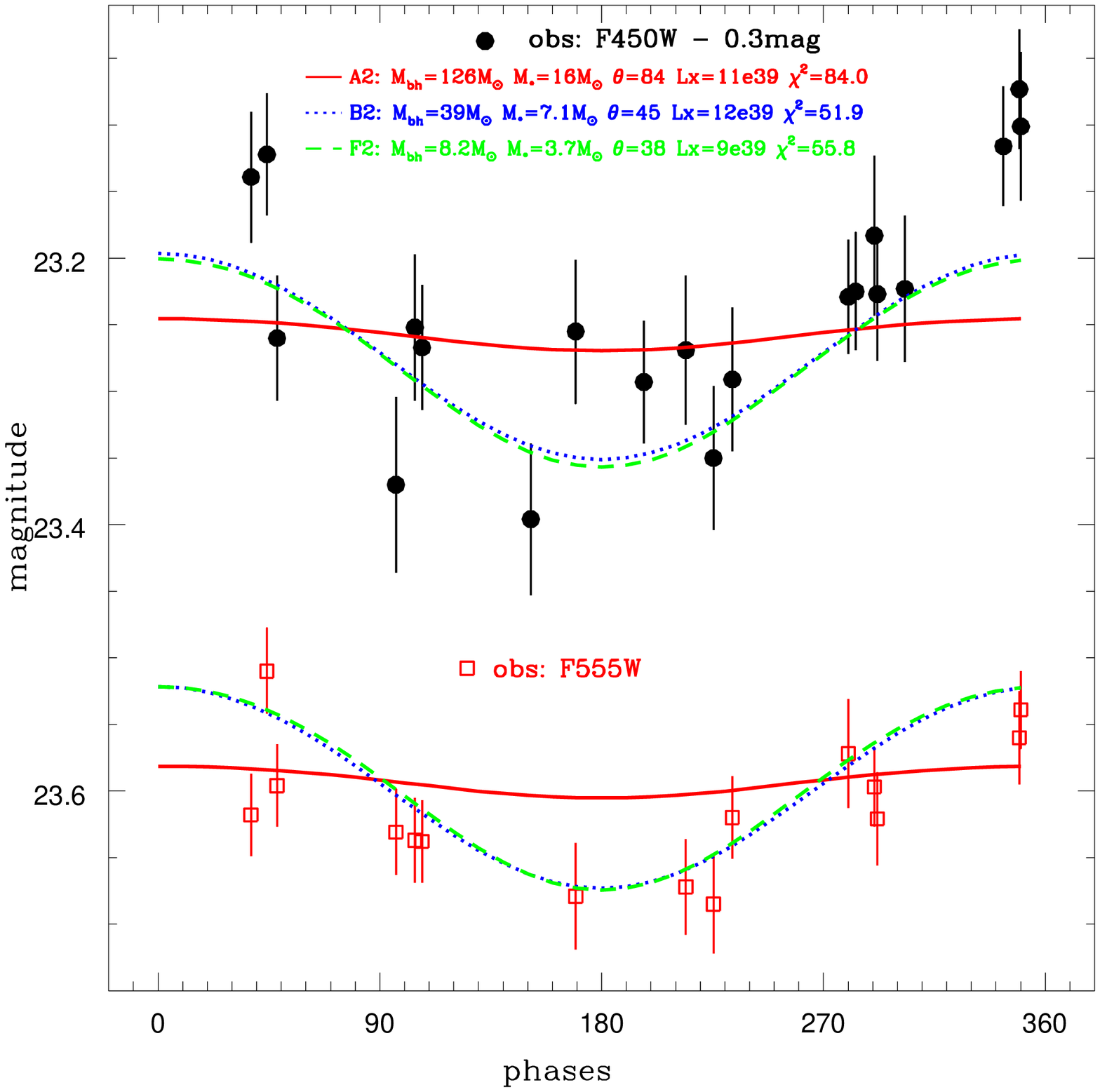}}

\caption{The predicted light curves for model B2 (blue dotted line) and model
F2 (green dashed line) with HST/WFPC2 F450W/F555W observations overplotted. }

\end{figure}

\begin{figure}

{\includegraphics[width=3.1in,height=2.1in]{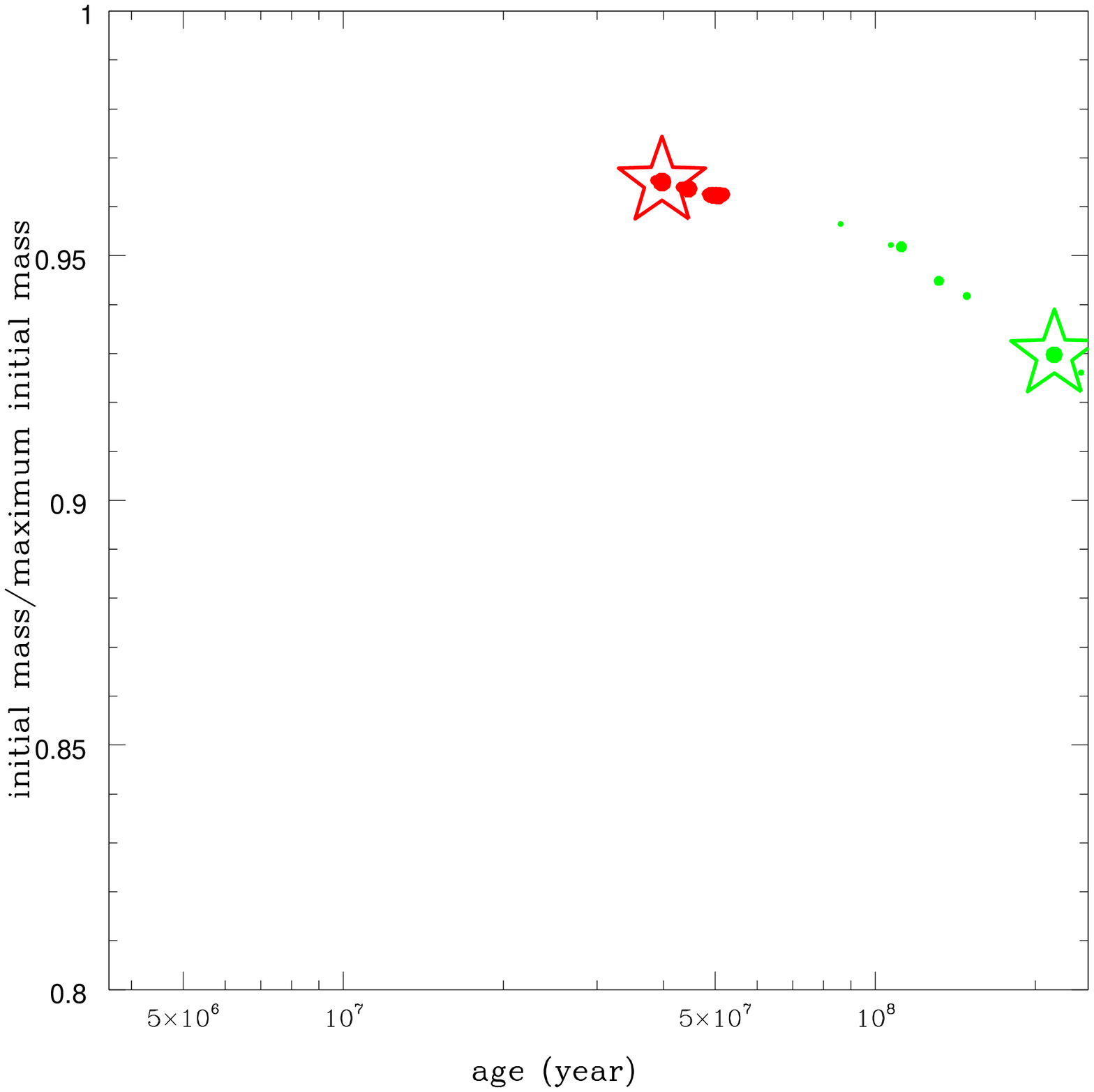}\includegraphics[width=3.1in,height=2.1in]{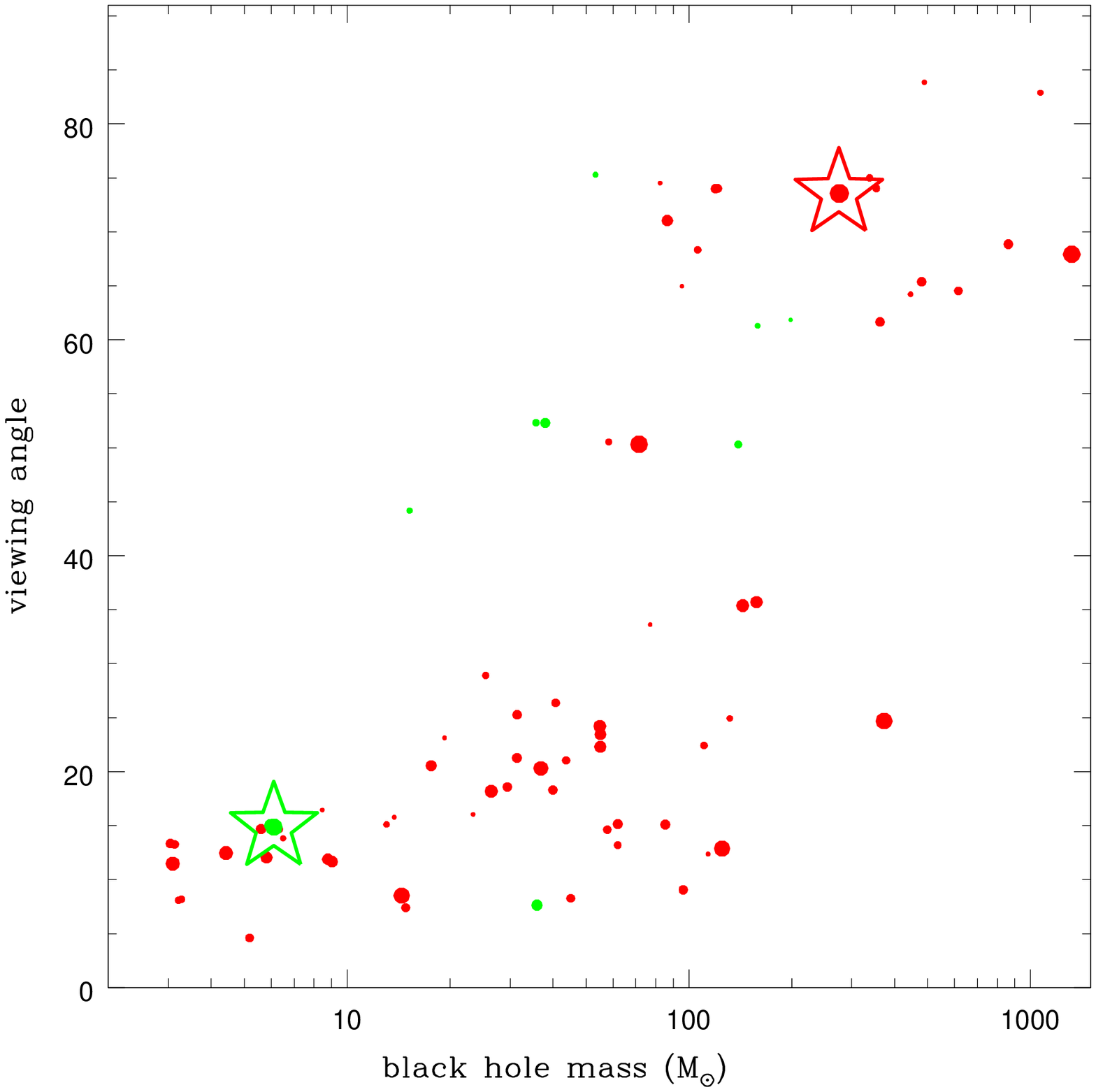}}
{\includegraphics[width=3.1in,height=2.1in]{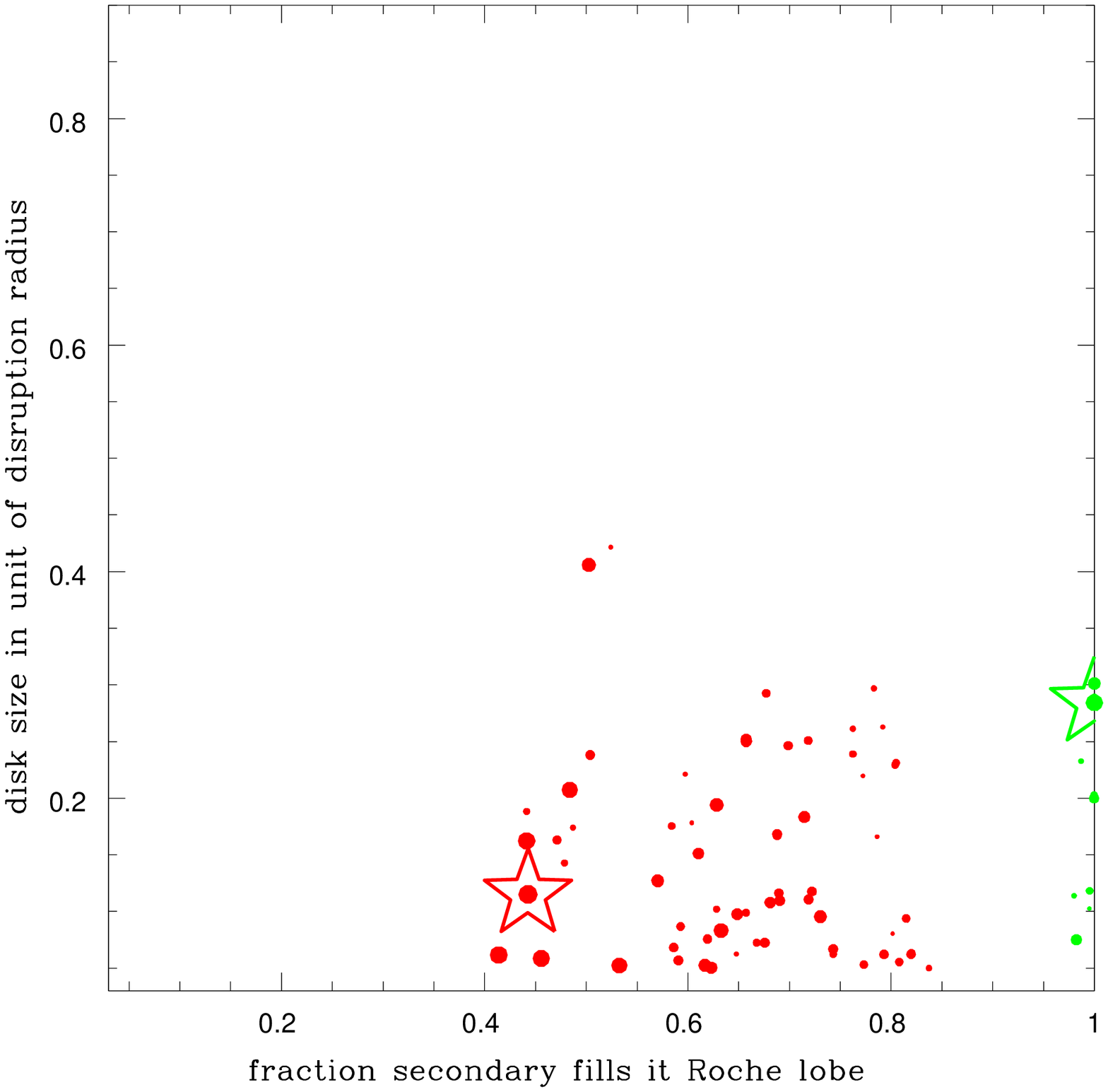}\includegraphics[width=3.1in,height=2.1in]{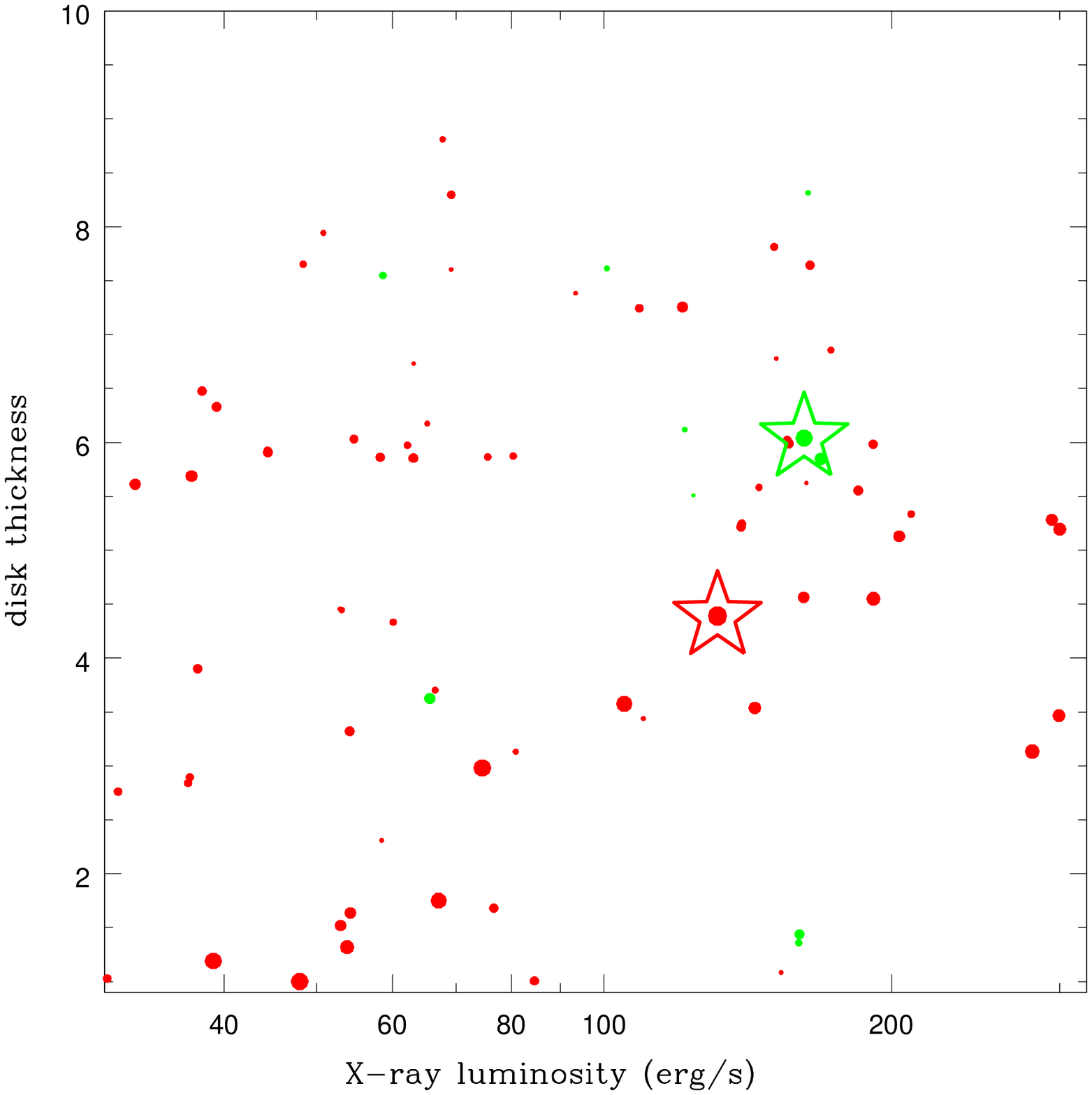}}
{\includegraphics[width=3.1in,height=2.1in]{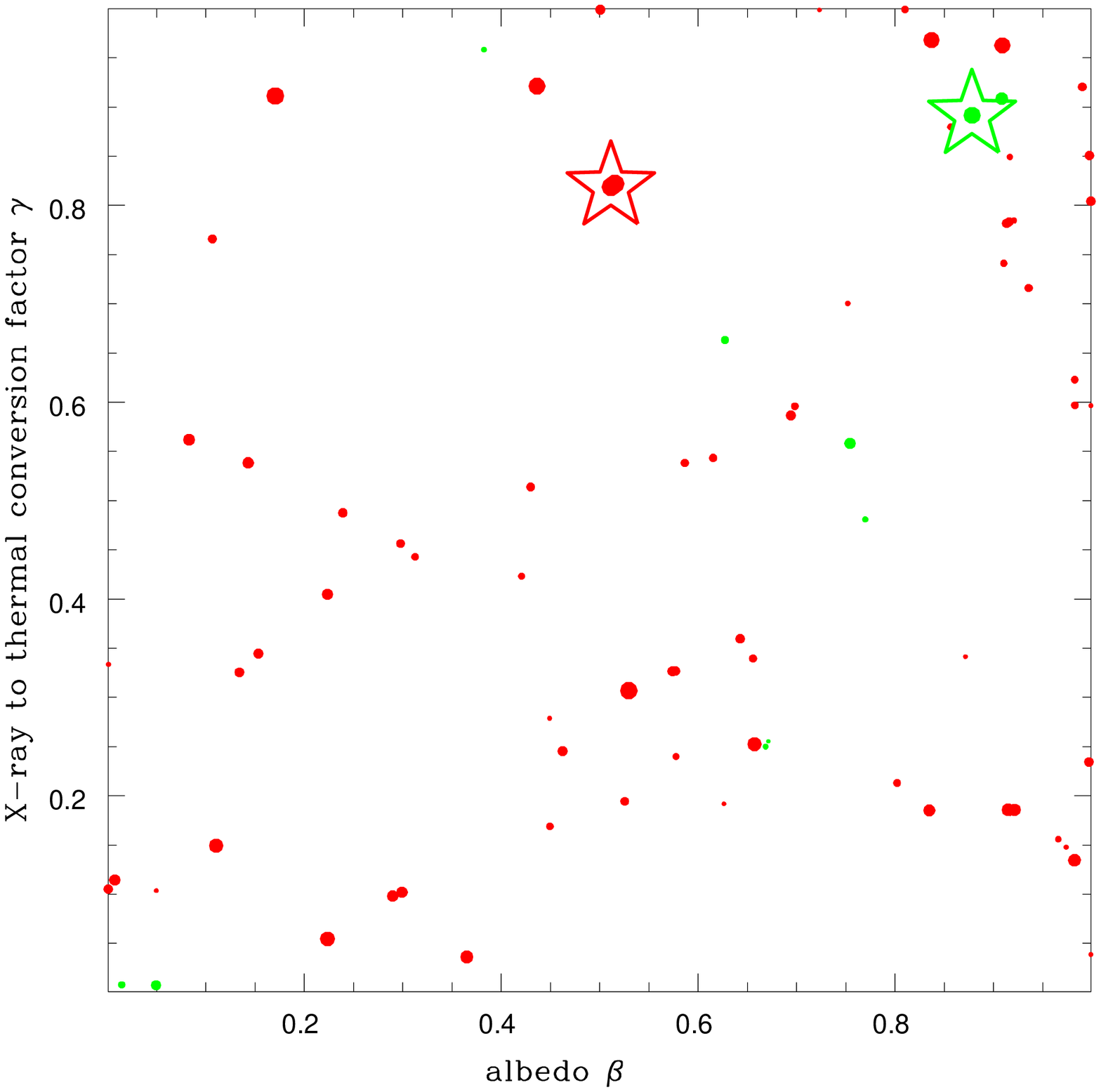}\includegraphics[width=3.1in,height=2.1in]{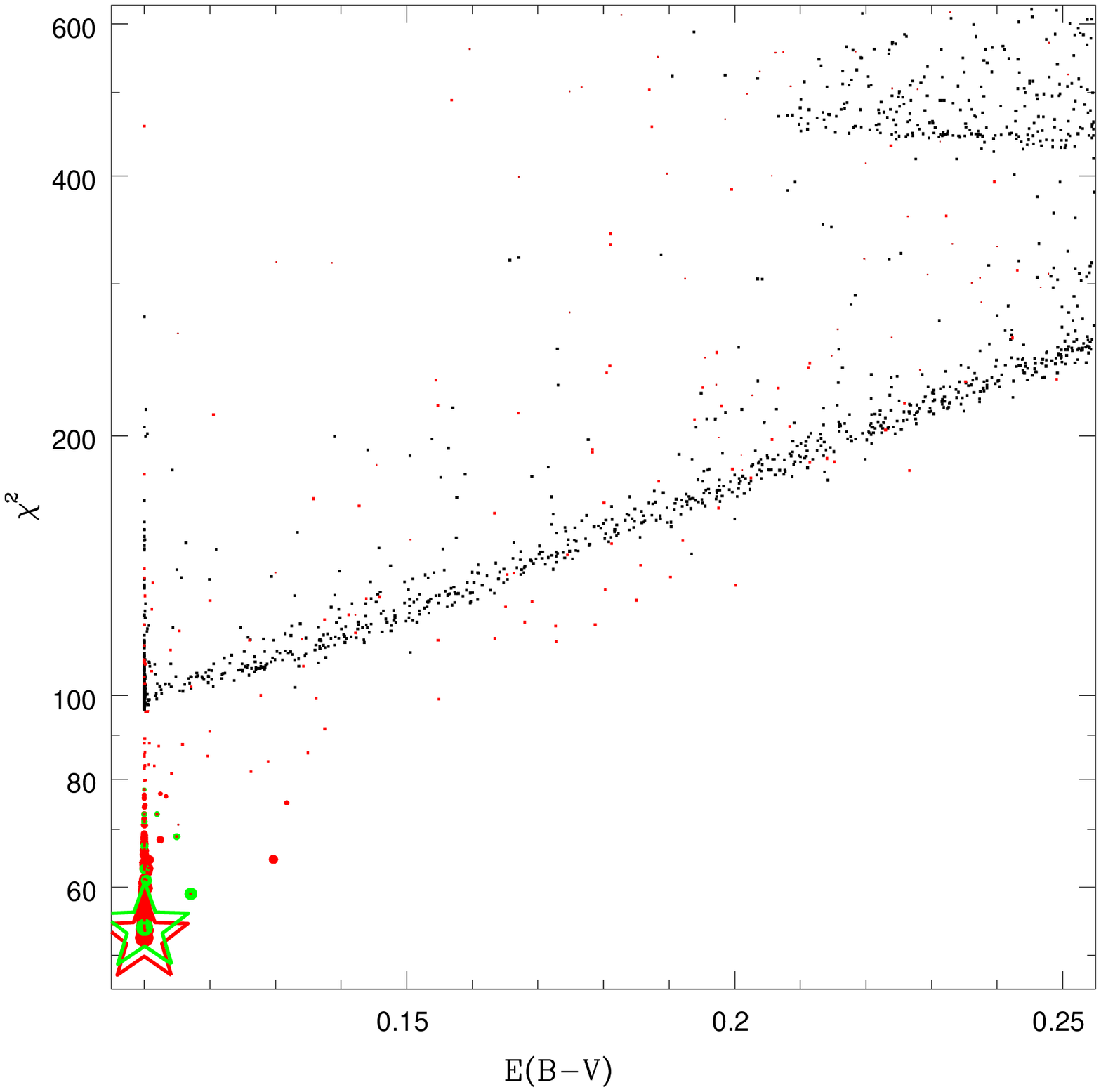}}

\caption{Models with $\chi^2<78$ obtained with two rounds of AMOEBA searches
for assumed $2K^\prime=25\pm15$ km/s.  The sysmbols are the same as in
Figure~4.  The asterisks are for model B3 (red) and model F3 (green) as described
in the text.  The six panels show (a) the secondary age vs initial mass, (b)
the black hole mass vs inclination angle, (c) the fraction the secondary fills
its Roche lobe vs the disk size in unit of disruption radius, (d) the X-ray
luminosity vs the disk thickness, (e) the disk surface albedo vs the
X-ray-to-thermal conversion factor on the secondary surface, and (f) the
extinction E(B-V) vs the model $\chi^2$ for models with $\chi^2<600$. }

\end{figure}

\begin{figure}[t]

\center{\includegraphics[width=3.1in,height=3.1in]{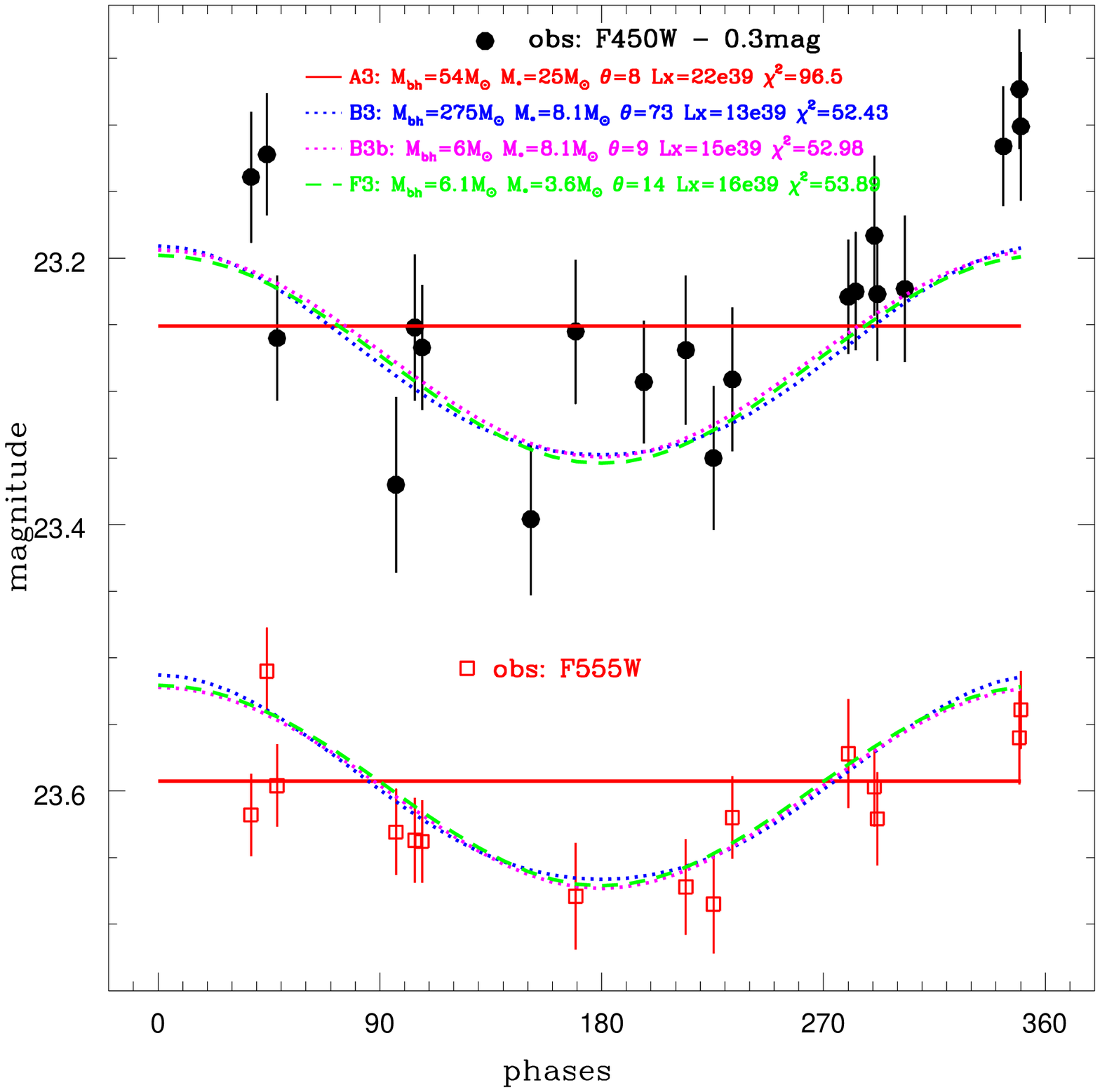}}

\caption{The predicted light curves for model B3 (blue dotted line) and model
F3 (green dashed line) with HST/WFPC2 F450W/F555W observations overplotted. }

\end{figure}

\begin{figure}[t]

\center{\includegraphics[width=3.1in,height=3.1in]{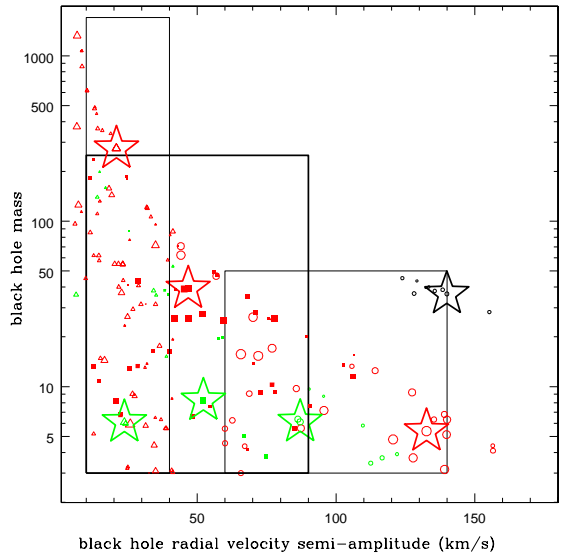}}

\caption{The black hole radial velocity semi-amplitude versus black hole mass
for acceptible models with $\chi^2<78$ for the cases of $K^\prime=100\pm40$
km/s (open circles), $K^\prime=50\pm40$ km/s (filled squares) and
$K^\prime=25\pm15$ km/s (open triangles).  The symbol sizes are inversely
scaled with the $\chi^2$ values of the models. The asterisks are the best
models for groups one (black), two (red) and three (green) as described in the
text. The three boxes indicate the rough ranges of ``acceptible'' models for three cases.}

\end{figure}

\begin{figure}[t]

\plottwo{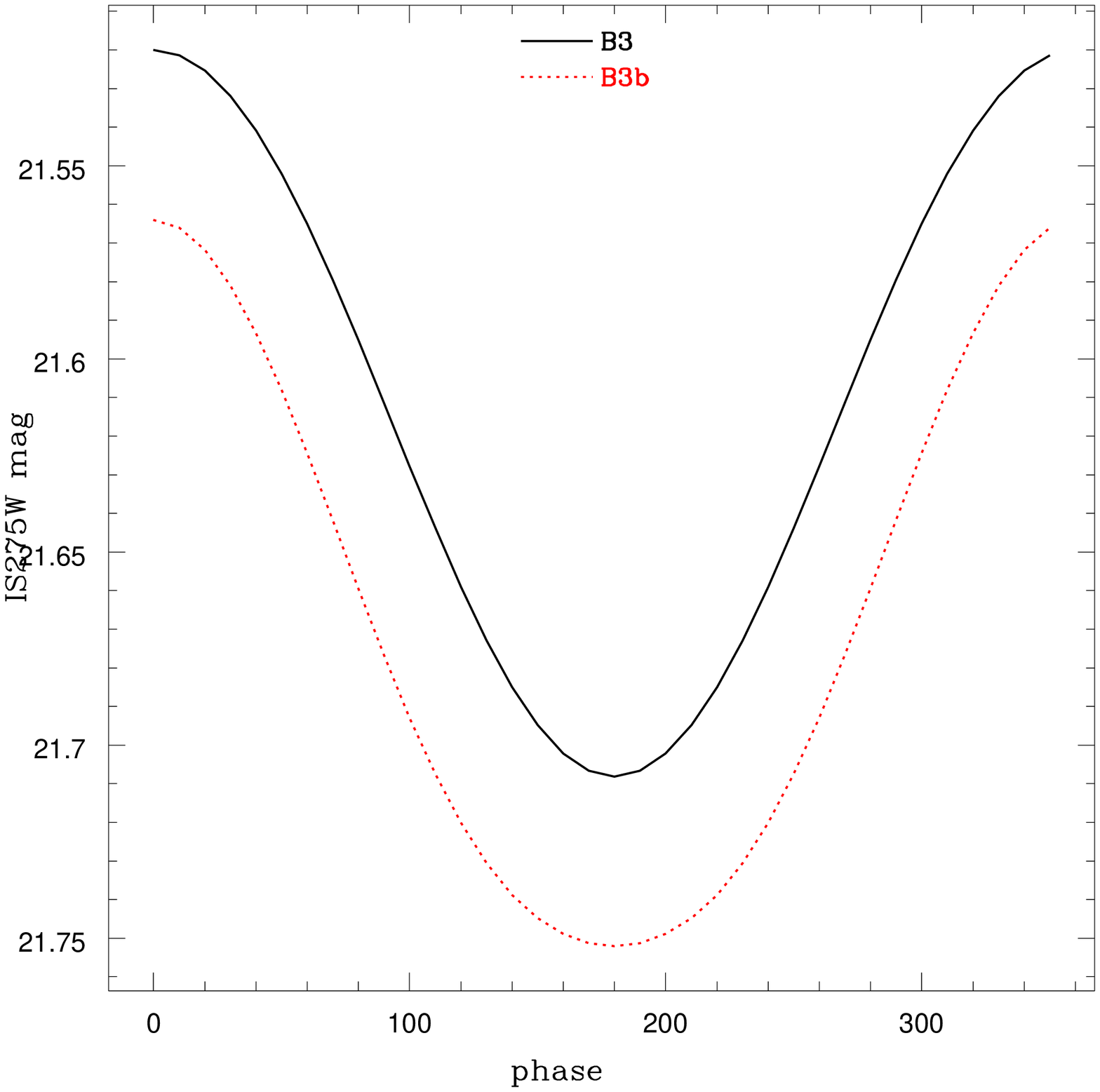}{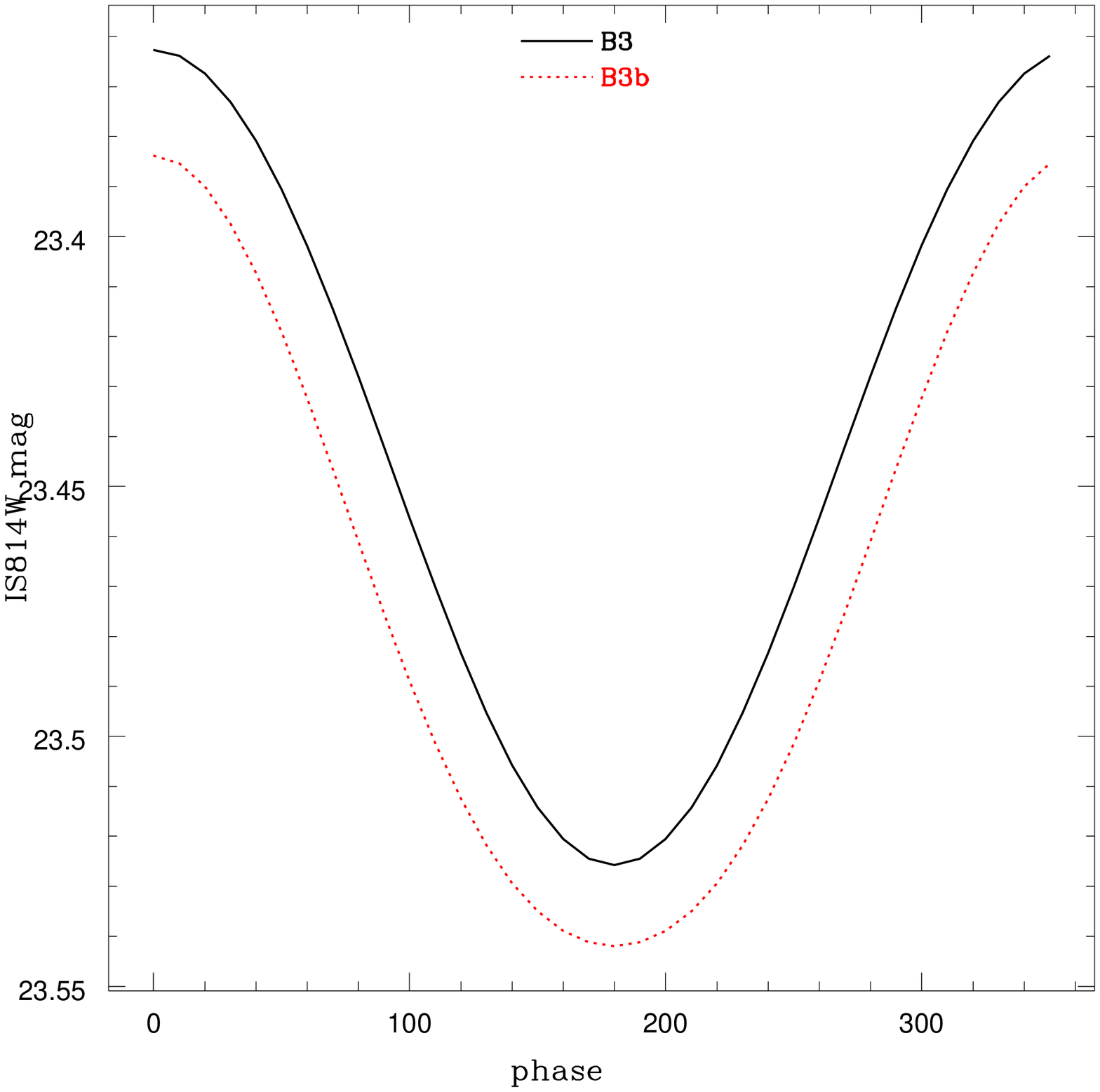}

\caption{The predicted HST WFC3/UVIS F275W and F814W light curves for the two
models B3 and B3b for $K^\prime\sim20$ km/s. }

\end{figure}

% Comparison of emergent spectrum between the
%three models. All spectra are scaled by the same power-law spectrum to clearly
%show the differences. Model A predicts a much bluer spectrum at ultraviolet
%without Balmer jump as compared to models B and F, while model B is slightly
%bluer than model F above 4000\AA.   }

\appendix

\setcounter{table}{0}
\renewcommand\thetable{\Alph{section}\arabic{table}}

\setcounter{figure}{0}
\renewcommand\thefigure{\Alph{section}\arabic{figure}}

\section{A Brief Model Description}

To fully model the X-ray irradiated binaries like ULXs, one needs to consider 
the gravitational distortion of the secondary, the accretion disk, and the X-ray irradiation effects.
Below we describe our treatments of these problems, and how we calculate the emergent spectra and light curves,
followed by examples of model outputs to illustrate the model dependence on key parameters.

\subsection{Roche geometry}

We follow the formulation of Kopal (1959) to compute the Roche lobe for given
$M_\bullet, M_*$, and $a$ (i.e., primary and secondary masses and separation).
In our computation, we define a Cartesian coordinate system with $a$ as the
unit length to place the secondary at (0, 0, 0) and the primary at (1, 0, 0).
The Roche potential at each point ($x,y,z$) is 
$$\Phi(x,y,z) = -GM_*/r_1 - GM_\bullet/r_2 - G(M_\bullet+M_*)r^2/2$$
where $r_1 = \sqrt{x^2+y^2+z^2}$, $r_2 = \sqrt{(x-1)^2+y^2+z^2}$ and $r^2 =
(x-M_\bullet/(M_\bullet+M_*))^2 + y^2$. 
The L1 point is located at ($x_1, 0, 0$), with $x_1$ solved numerically from
$x_1^{-1} - x_1 = M_\bullet/M_* [ (1-x_1)^{-2} - (1-x_1) ]$.
To compute the Roche lobe, we set up a polar coordinate system
($r,\theta,\phi$), and solve numerically for each of the $\theta,\phi$ grid the
radius $r$ at which the Roche potential is the same as at the L1 point.
The effective radius $R_{L,*}$ (in unit of $a$) for the Roche lobe around the
secondary is computed from the volume ($V = 4\pi R_{L,*}^3/3$) enclosed by the
equipotential surface.  The effective radius $R_{L,\bullet}$ is also computed
for the Roche lobe around the primary black hole.

A secondary with mass $M_*$, radius $R_*$, surface temperature $T_*$ and
gravity $g_*$ ($ = GM_*/R_*^2$) is filled into its Roche lobe by equating
$a\cdot R_{L,*} f_R = R_*$, with $f_R$ as the fraction the secondary fills
its Roche lobe. 
This immediately sets the physical dimension for $a$ and the orbital period
with $$P^2 = 2\pi a^3/G(M_\bullet+M_*)$$
The gravity at each surface element of the Roche lobe can be computed as
$$\vec{g} \equiv g\hat{g} = \nabla\Phi(x,y,z) = ({\partial/\partial
x},{\partial/\partial y},{\partial/\partial z})\Phi(x,y,z)$$ and
$\mbox{-}\hat{g}$ is the surface normal vector.
This gravity varies over the stellar surface. It is higher near the poles and
lower near the equators, and the lowest at the parts of the star nearest to the
L1 point.
In hydrostatic equilibrium, the gravity variation modifies the local effective
temperature over the stellar surface. 
To describe this gravitational darkening (brightening) effect, we adopt the
scaling relation $T_g = T_* (g/g_*)^b$ with $b$=1/4 for radiative stellar
atmospheres (von Zeipel, 1924) and $b$=0.08 for convective atmospheres (Lucy,
1967).

\subsection{Accretion disk}

The accretion disk around the black hole is described as a standard accretion
disk with the $\gamma$ prescription (Shakura \& Sunyaev 1973; Frank, King \&
Raine 2002, hereafter FKR2002). 
For such an accretion disk with radius $r$, the inner radius $r_i =
6GM_\bullet/c^2$ assuming a non-spinning black hole, 
the luminosity $L_d = GM_\bullet\dot{M_\bullet}/2r_i$,
and the temperature profile $T_a(r) = T_0 (r_i/r)^{3/4} f$ with $f^4 = 1 -
(r_i/r)^{1/2}$ and $T_0^4 = 3L_d/4\pi\sigma r_i^2$. 
To avoid tidal disruption, the disk must be within the tidal radius $R_{Tides}$
$ \approx 0.9R_{L,\bullet}$.  We assume the outer radius $r_o = X_r \cdot
R_{Tides}$ with $X_r \le 1$.
The disk can be divided into the inner disk, the middle disk, and the outer
disk, with the transitions at $r_{im} = 3.6\times10^2 \gamma^{2/21}
\dot{m}^{16/21} m^{2/21} f^{64/21}$ and at $r_{mo} = 2.6\times10^4
\dot{m}^{2/3}f^{8/3}$. Here $m = M_\bullet/M_\odot$, $\dot{m} =
\dot{M}_\bullet/\dot{M}_{Edd}$, $\dot{M}_{Edd} = L_{Edd}/0.1c^2$, and $L_{Edd}
= 4\pi G M_\bullet c/\kappa = 1.3\times10^{38}m$ erg s$^{-1}$.
The height of the disk $h(r) = 7.5\dot{m} f^4$ for the inner disk, $h(r) =
1.6\times10^{-2}\gamma^{-1/10}\dot{m}^{1/5}m^{-1/10}r^{21/20}f^{4/5}$ for the
middle disk, and $h(r) =
7.2\times10^{-3}\gamma^{-7/10}\dot{m}^{3/20}m^{-1/10}r^{9/8}f^{3/5}$ for the
outer disk.
In our model, we assume the ``real'' disk height $H(r) = X_h \cdot h(r)$ to
account for the fattened disk at high X-ray luminosities.
With the above formulae, we compute the surface temperature $T(r)$, height
$H(r)$ and surface normal vector $\hat{s}(\vec{r})$ for the disk under the
reasonable assumption that $L_X = L_d$, i.e., X-ray dominates the disk
radiation in ULXs.

\subsection{X-ray irradiation}

X-ray irradiation of the accretion disk is treated following FKR2002 (\S5.10).
The X-ray flux crossing the disk surface at $\vec{r}$ is $$F = {L_X\over4\pi
d^2} (1-\beta) \cos\psi$$ with $\vec{d} = d\hat{d} = \hat{d}\sqrt{r^2+H^2(r)}$
as the vector from the black hole to the surface element, $\beta$ as the
albedo, and $\cos\psi = -\hat{s}\cdot\hat{d}$ representing the angle between
the direction of the incident radiation and the disk surface normal vector. 
The effective temperature resulting from irradiation can be defined as
$$T_{Irr}^4 = F/\sigma = {1\over3} T_0^4 (r_i/d)^2 (1-\beta) \cos\psi$$ and the
total temperature for the disk surface element $$T_d^4 = T_a^4(r) + T_{Irr}^4$$
The X-ray irradiation of the secondary is treated in a similar fashion. 
With respect to the black hole, a stellar surface element at ($x,y,z$) raises
above the orbital plane with an angle of $\delta = \arcsin{z \over d}$ to the
black hole with $\vec{d} = d\hat{d} = (x-1,y,z)$, while the accretion disk edge
at the outer radius $r_o$ raises above the orbital plane with an angle of
$\Delta = \arctan{H(r_o)\over r_o}$.
The effective temperature from irradiation is defined as $$T_{Irr,*} =
{1\over3} Q T_0^4 (r_i/d)^2 \gamma \cos\psi$$ with $\cos\psi =
-\hat{d}\cdot\mbox{-}\hat{g}$, $\mbox{-}\hat{g}$ as the surface normal vector,
$Q$ as the transmission function, and $\gamma$ representing the fraction of
intercepted X-ray flux converted to thermal energy.
The stellar surface element with $\cos\psi < 0$ is blocked by the star itself
and thus not irradiated.
The transmission function describes how the X-ray is blocked by the accretion
disk, with $Q=0$ if $\delta < \Delta$ (i.e., completely blocked), and $Q=1 -
e^{\delta-\Delta \over \Delta}$ if $\delta > \Delta$ to account for the
photoelectric absorption by the extended disk atmosphere.
The total temperature at each element of the secondary surface is then computed
as $$T_s^4 = T_g^4 + T_{Irr,*}^4$$ 

\subsection{Viewing the binary}

Whether a surface element of the accretion disk or the secondary is visible to
the observer at an inclination of $\Theta$ is determined as follows.
For each binary phase $\Phi$ with the observing vector $\hat{v}(\Theta,\Phi)$
from the binary to the observer, treated first is the component closer to the
observer, which is the accretion disk at $\mbox{-}{\pi\over2} < \Phi <
{\pi\over2}$ and the secondary at ${\pi\over2} < \Phi < {3\pi\over2}$ under the
convention that $\Phi=0$ if the accretion disk is right in between the
secondary and the observer.
For the accretion disk, the disk surface may for some $\Theta$ be blocked by
the disk edge, which is treated as a cylinder with a radius of $r_o$, a half
height of $H(r_o)$, a surface temperature of $T_a(r_o)$, and a surface normal
vector $\hat{m}$. We thus project the visible parts of the disk edge (with
$\hat{m}\cdot\hat{v} \ge 0$) onto a plane normal to $\hat{v}$, and compute its
silhouette as a polygon. For the disk surface elements, visible are those with
$\hat{s}\cdot\hat{v} \ge 0$ and outside the edge's silhouette. A disk silhouette is
computed for the disk edge and surface as a whole.
For the secondary, we examine the surface elements from pole to pole, and
compute a silhouette polygon of all previously examined elements for each
latitude. A secondary surface element is visible if it is outside the previous
silhouette polygon and with $\mbox{-}\hat{g} \cdot \hat{v} \ge 0$. 
We then treat the component farther away from the observer in the same way,
except that a surface element is invisible if it is within the silhouette polygon
for the closer component.
The ray-tracing method developed by Haines (1994) is used to check whether a
surface element, after projecting to the plane normal to $\hat{v}$, is within a
silhouette polygon.

\subsection{Model output}

Both the spectrum and the wide-band photometry are computed for the emergent
emission from the X-ray irradiated binary model.
For the $i$-th visible surface element, the flux is computed as $$\Delta F_i(\lambda) =
I_i(\lambda) {A_i\cos\Psi_i\over D^2}$$ with $I_i(\lambda)$ as the specific
intensity, $A_i \cos\Psi_i$ as the surface element area projected along the
line of sight, and $D$ as the distance to the observer.
Here $\cos\Psi = \hat{v}\cdot\mbox{-}\hat{g}$ for the secondary, $\cos\Psi =
\hat{v}\cdot\hat{s}$ for the disk surface, and $\cos\Psi = \hat{v}\cdot\hat{m}$
for the disk edge.
The disk edge and surface are treated simply as a black body, and $I_i(\lambda)
= B_\lambda(T_d) $.
% = {2hc^2/\lambda^5 \over \exp(hc/\lambda kT_d)-1}$ .
%
The secondary is treated as Kurucz's stellar atmosphere model (Kurucz 1993), which gives the
emergent intensity spectrum $I(\lambda;\mu, T, g)$ for a grid of the incident
angle $\mu$, surface temperature $T$ and gravity $g$ (ref).
In our model, the intensity spectrum $I_i(\lambda) = I(\lambda; \mu_i, T_{s,i},
g_i)$ is interpolated from the model grid with $\mu =
\hat{v}\cdot\mbox{-}\hat{g}$.
Summing up all visible surface elements of the secondary and the accretion
disk, we obtain the total flux spectrum as $$F(\lambda) = \sum_i \Delta
F_i(\lambda)$$
The flux observed through a filter is computed as $$F = \int F(\lambda)
T(\lambda) d\lambda$$ and the magnitude is computed as $$m = -2.5\lg(F/F_0)$$
Here $T(\lambda)$ is the filter transmission, and $F_0$ is the zeroth magnitude
flux, which are taken from the {\tt stsdas.synphot} package.
In our model illustration, we calculate the magnitudes for Johnson $UBV$,
Cousins $RI$, and Bessel $JHK$, but any number of filters can be easily added
into the code.

Given the model parameters, the model first sets up the binary geometry, computes
the secondary and disk temperature profiles with and without X-ray irradiation,
then computes the emergent spectrum and the magnitudes for a binary phase
$\Phi$. 
For each binary phase, we put the model output in one figure as shown in
Figure~1 for a base model with $M_\bullet = 30M_\odot$, a B0V secondary ($M_* =
17.5 M_\odot, R_* = 7.4 R_\odot$, and $T_*$ = 30,000 K with solar abundance)
filling its Roche lobe ($f_R = 1$), $L_X = 10^{39}$ erg s$^{-1}$, $X_h = 1$,
$X_d = 0.5$, $\beta = 0.5$, $\gamma = 0.5$, and $\Theta = 60^\circ$.
In such a figure, the upper left panel shows the binary components projected on
the plane normal to the viewing vector $\hat{v}$, and the upper right panel
shows the $UBVRIJHK$ light curves for a whole period with the magnitudes marked
for this phase.
The lower panel shows the emergent spectra from the secondary with and without
X-ray irradiation, from the disk itself and the X-ray irradiation contribution,
and from the binary as a whole. 
As an illustrative product for a given model, we create an animated movie by
stacking up a sequence of such figures for binary phases over a period.

\subsection{Dependence on Key Parameters }

To illustrate how the model depends on the key parameters, we start from the
base model shown in Figure~1. In such a model, the star contributes 70\%-60\%
of the light in the U-I bands, the un-X-ray-irradiated disk contributes
15\%-20\%, and the X-ray irradiation of the disk contributes the rest
15\%-20\%.  
X-ray heating of the L1 side of the secondary boosts it to be the hottest side,
yielding a sinusoidal light curve instead of an ellipsoidal one.
We vary one parameter in its reasonable range with all other parameters fixed
at the base model, and investigate the according changes in the spectrum, the
magnitudes, the light curve shapes and variation amplitudes with respect to the
base model.

%% 3 %%
We first investigate the five parameters on X-ray irradiation.
The X-ray luminosity is varied from $10^{38}$ to $10^{41}$ erg s$^{-1}$.
With increasing X-ray luminosity, the X-ray heated disk and the L1 side of the
secondary become hotter and brighter.
Accordingly, the light curve shape changes from ellipsoidal to sinusoidal, with
brighter mean magnitudes and larger variation amplitudes.  The effects are
illustrated in Figure~B1.

%% 4 %%
The disk fattening factor is varied from 0.1 to 10. With
increasing fattening, the disk becomes thicker and intercepts more
X-ray radiation, thus becomes slightly brighter. The disk casts a larger shadow
on the secondary, and the secondary becomes less X-ray heated. Accordingly, the
light curve shape changes from sinusoidal to ellipsoidal, with slightly
brighter mean magnitudes but smaller variation amplitudes.
The effects are illustrated in Figure~B2.

%% 5 %%
The disk size factor is varied from 0.01 to 1. With a larger
radius and emitting area, the disk contributes slightly more in the optical, leading to 
slightly brighter mean magnitudes for the light curves.
The slightly larger disk shadow on the secondary reduces the X-ray heating
only slightly, and the light curve shape remains sinusoidal yet with smaller variation amplitudes. 
The effects are illustrated in Figure~B3.

%% 6 %%
The albedo for the disk surface is varied from 0.01 to 0.99. The increasing
albedo reduces the heating of the disk, leading to much smaller contribution
from the X-ray irradiation of the disk.
X-ray heating of the secondary remains the same, and the light curve shape
remains sinusoidal. 
The light curve mean magnitudes decrease slightly, and the variation amplitudes
increasing slightly, both due to the diminishing of the X-ray irradiation of
the disk.  The effects are illustrated in Figure~B4.

%% 7 %%
The X-ray to thermal conversion factor is varied from 0.01 to 0.99. We do not
link this factor to the disk albedo because the secondary albedo may be
different due to temperature and ionization differences, and some of the X-ray
energy may be converted to kinetic energy of the materials blown away from the
secondary surface.
With increasing conversion efficiency, the L1 side of the secondary becomes
hotter. The light curve shape changes from ellipsoidal to sinusoidal with
slightly brighter mean magnitudes and much larger variation amplitudes.
The effects are illustrated in Figure~B5.

%% 1 %%

Now we investigate the parameters that affect the binary configuration.
The black hole mass is varied from $3M_\odot$ to $3000M_\odot$. 
With the increasing black hole mass, the Roche lobe of the black hole gets
larger, the separation gets larger, and the secondary star is less distorted
and less gravitationally darkened.
The heating of the secondary becomes less and less significant so that the
light curves change from sinusoidal to ellipsoidal as illustrated in Figure~B6a.
Because the secondary star is less gravitationally darkened at higher black
hole mass, the variation amplitudes for the light curves become smaller and
smaller (Figure~B6b).
With a fixed disk size factor, the physical size of the disk gets larger at
higher black hole mass, and its contribution in the optical gets higher as
illustrated in Figure~B6a. This also leads to higher total optical light, and
brighter mean magnitudes of the light curves (Figure~B6b).
At black hole mass $\le10M_\odot$, the separation becomes so small that the
secondary star and the disk begin to block each other when they line up,
leading to extra shallow drops in the light curves around phases 0$^\circ$
and 180$^\circ$ (Figure~B6a).

The fraction the secondary fills its Roche lobe is varied from 0.01 to 1.
With a smaller filling fraction and a fixed secondary, the Roche lobe of the
secondary becomes larger. Subsequently, the separation, the Roche lobe of the
black hole, and the physical size of the accretion disk all become larger.
The larger disk contributes more in the optical, leading to brighter mean
magnitudes of the light curves, more so at longer wavelengths.
X-ray heating of the secondary becomes less significant, and the light curve
shape changes from sinusoidal to ellipsoidal with reduced variation amplitudes.
The tidal distortion and gravitational darkening of the secondary also become
less significant, adding to the decrease of the variation amplitudes.
The effects are illustrated in Figure~B7.

%% 8 %%
The inclination angle is varied from $0^\circ$ (face-on) to $90^\circ$
(edge-on). 
With the inclination angle increasing, the projected area of the disk becomes
smaller, leading to smaller disk contribution in the optical, and dimmer mean
magnitudes of the light curves.
The projected area of the L1 side of the secondary also becomes larger, leading
to larger variation amplitudes for the sinusoidal light curves.
When the inclination angle increases to $70^\circ$, the disk begins to block
parts of the secondary, which enhances the contrast between the heated side and
its antipodal side, and leads to much larger variation amplitudes.
When the inclination angle increases to above $80^\circ$, the disk begins to
block the hottest part of the L1 side of the secondary, which lowers the
contrast between the heated side and its antipodal side, and leads to reduced
variation amplitudes.
The effects are illustrated in Figure~B8.

\begin{table}
\begin{tabular}{|c|c|c|c|}
\hline
SpType & Mass & Radius & Temp \\
       & ($M_\odot$) & ($R_\odot$) & (K) \\
\hline
O5V  & 60 &  12 &  42000 \\
B5V  & 5.9 &  3.9 &  15200 \\
A5V  & 2.0 &  2.4 &  8180 \\
F5V  & 1.4 &  1.3 &  6650 \\
G5V  & 0.92 &  0.92 &  5560 \\
%K5V  & 0.67 &  0.72 &  4410 \\
\hline
%G5III  & 1.1 & 10 &  5050 \\
%K5III  & 1.2 & 25 &  4050 \\
%\hline
B5I  & 20 &  50 &  13600 \\
A5I  & 13 &  60 &  8610 \\
F5I  & 10 &  100 &  6370 \\
G5I  & 12 &  150 &  4930 \\
%K5I  & 13 &  400 &  3990 \\
\hline
\end{tabular}
\caption{Experimented secondary spectral types.}
\end{table}

%% 2 %%
We experiment with the secondary properties by varying spectral types and
luminosity classes as listed in Table~B1.
The quoted mass, radius and temperature are taken from the MK spectral type
calibration (Cox 2000) for stars in the solar neighborhood.
For main sequence stars, later spectral types are less massive and smaller in
size, with higher densities, lower temperatures and lower luminosities.
For an O star secondary, the separation is the largest, and X-ray heating of
the secondary is a minor effect, leading to an ellipsoidal light curve.
X-ray heating becomes significant for a B star secondary, and the light curve
becomes sinusoidal with much larger variation amplitude. 
For still later stellar types, the variation amplitudes become smaller despite the
smaller separations and more significant X-ray heating effects, because the
accretion disk contribution to the optical light exceeds the star contribution
by larger and larger factors. 
The effects are illustrated in Figure~B9.

For supergiant stars, later spectral types are less massive yet larger in size,
with lower densities and lower temperatures.
For a B5I supergiant with a high temperature,  X-ray heating of the L1 side of
the secondary is not enough to raise it to be the hottest side, and the light
curve is an ellipsoidal one.
For later spectral types with much lower temperatures, X-ray heating is enough
to raise the L1 side of the secondary to be the hottest side, and the light
curves become sinusoidal.
The physical disk size, along with the Roche lobe of the black hole, becomes
larger for later spectral types. 
Because of its low temperature, the expanded outer disk does not contribute
much to the optical at shorter wavelengths (U/B bands), but contributes
significantly at longer wavelengths.
In the U/B bands, the mean magnitudes of the light curves become dimmer for
later spectral types because the secondary becomes dimmer while the disk
contribution remains largely the same.
In the V--K bands, the mean magnitudes become brighter for later spectral
types, more so in the redder bands, because both the disk and the secondary
become brighter.
The effects are illustrated in Figure~B10.

To summarize, five parameters are required to set up the binary model without
X-ray irradiation: (1) the black hole mass $M_\bullet$, (2-3) the secondary
mass ($M_*$), radius ($R_*$) and temperature ($T_*$) as determined by its
initial mass and age, (4) the fraction $f_R$ the secondary fills its Roche
lobe,  and (5) the inclination angle $\Theta$. 
Five extra parameters are required to incorporate the X-ray irradiation: (6)
the X-ray luminosity $L_X$, (7) the disk fattening factor $X_h$, (8) the disk
size factor $X_d$, (9) the disk albedo $\beta$, and (10) the X-ray to thermal
conversion factor $\gamma$.
In total, 10 key parameters are required to fully describe an X-ray irradiated
binary model for ULXs.

\begin{figure}

\plottwo{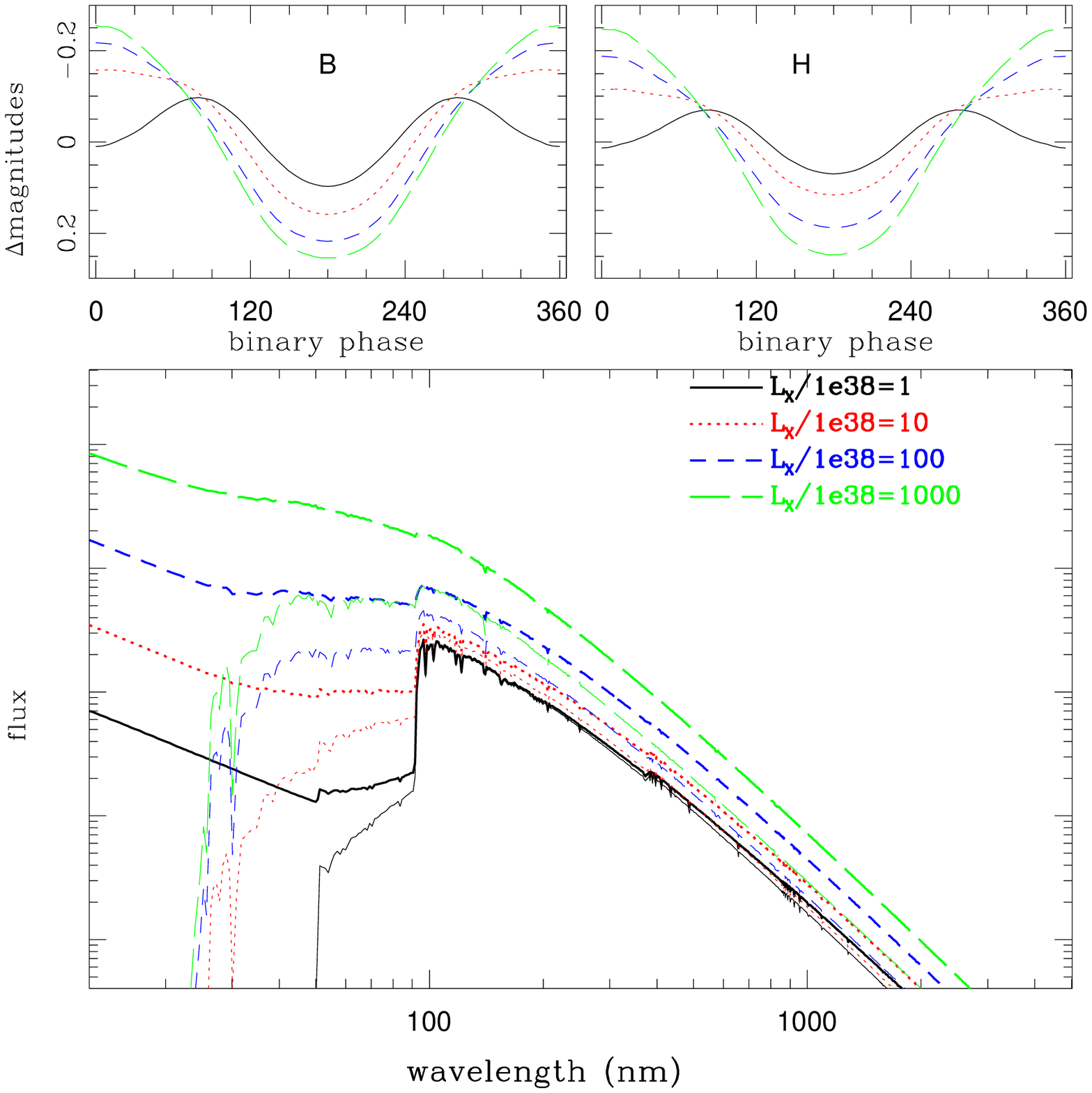}{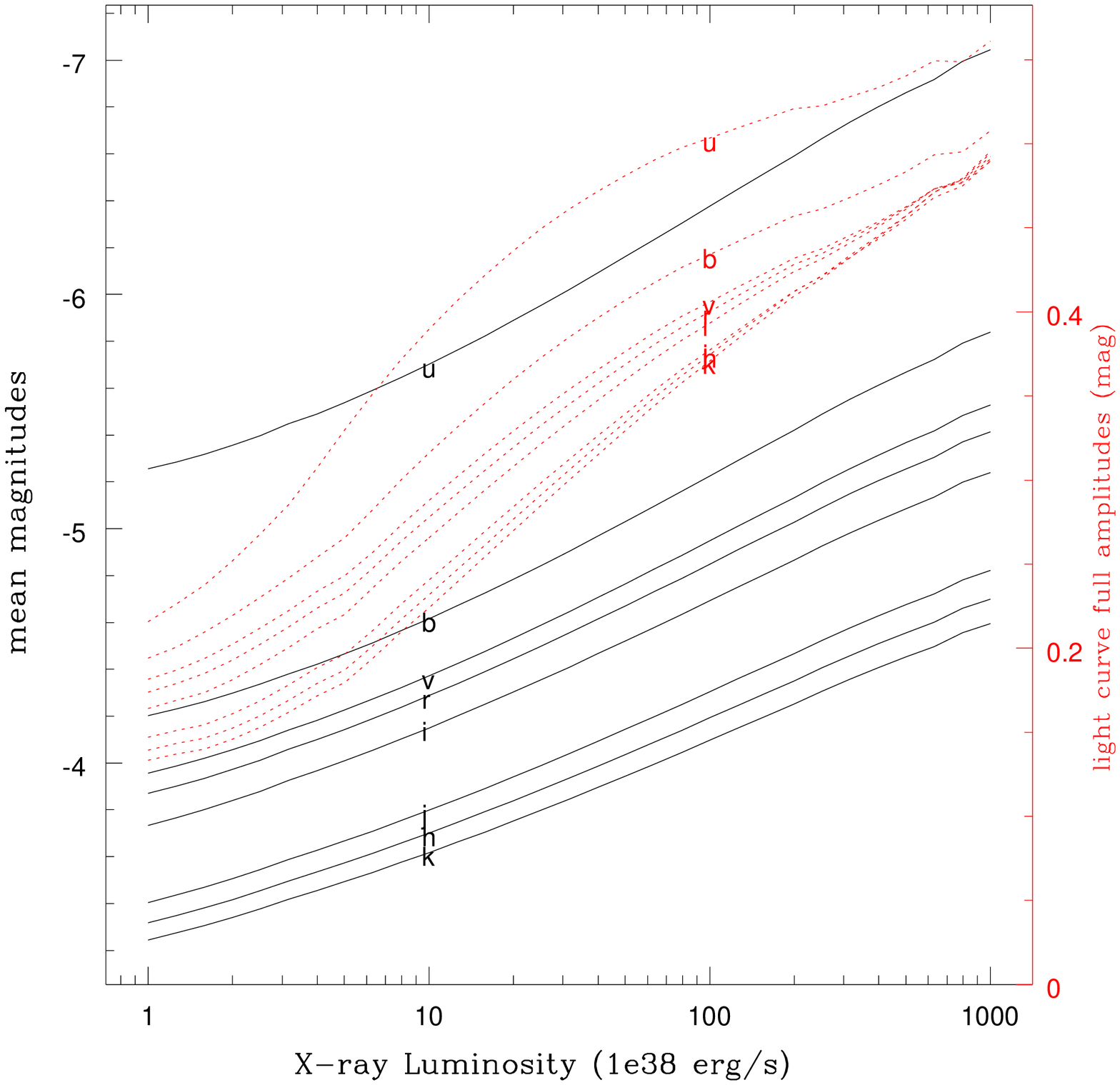}

\caption{({\it a}) The spectra (lower panel) and the B/H light curves (upper
panel) for models with different X-ray luminosities. The thin spectra are for
the X-ray  heated star, and the thick spectra are for the sum of the star and
the disk.  ({\it b}) The mean magnitudes (solid) and the variation
amplitudes (dotted) of the UBVRIJHK light curves for models with different
X-ray luminosities. }

\end{figure}

\begin{figure}

\plottwo{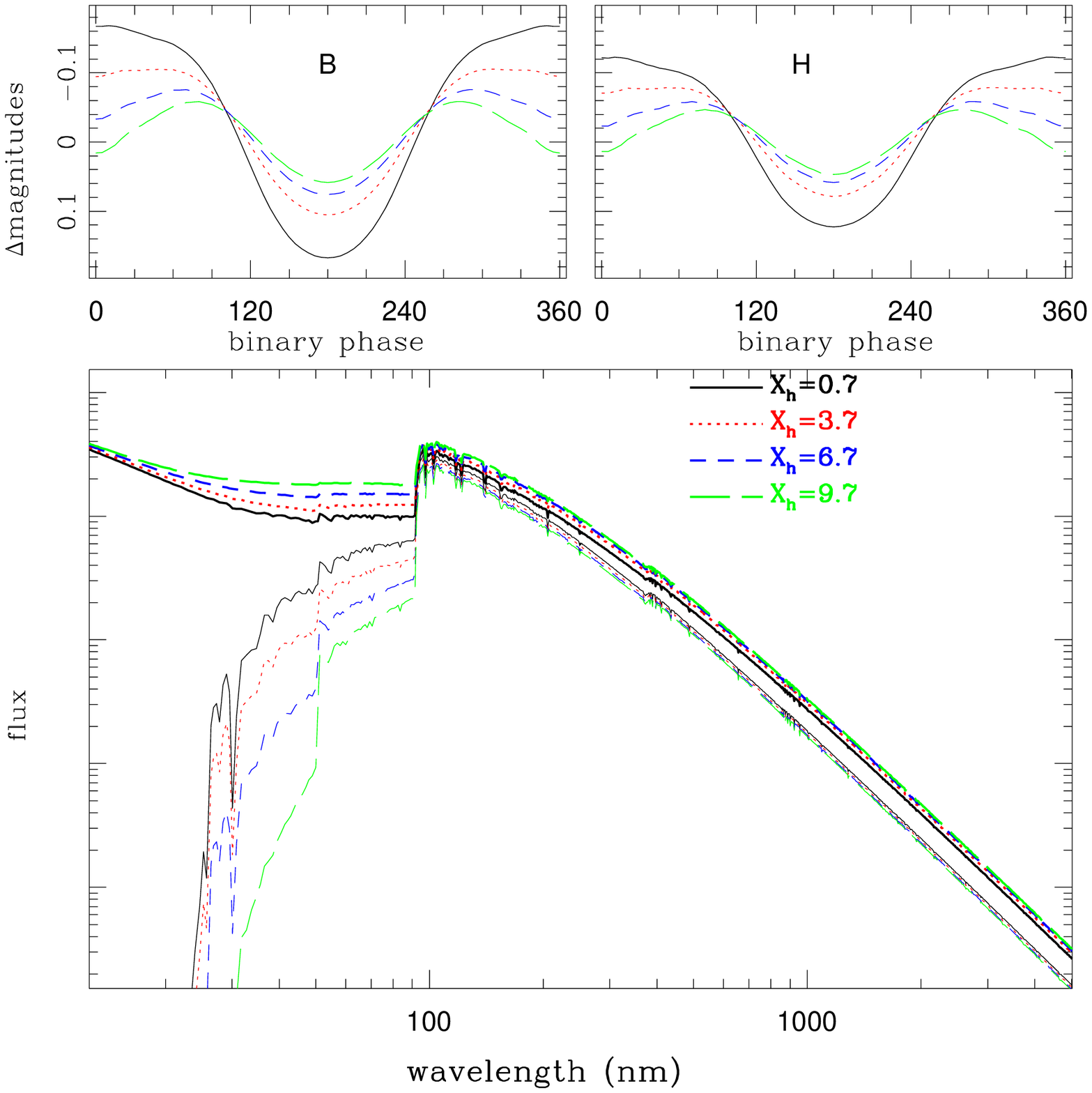}{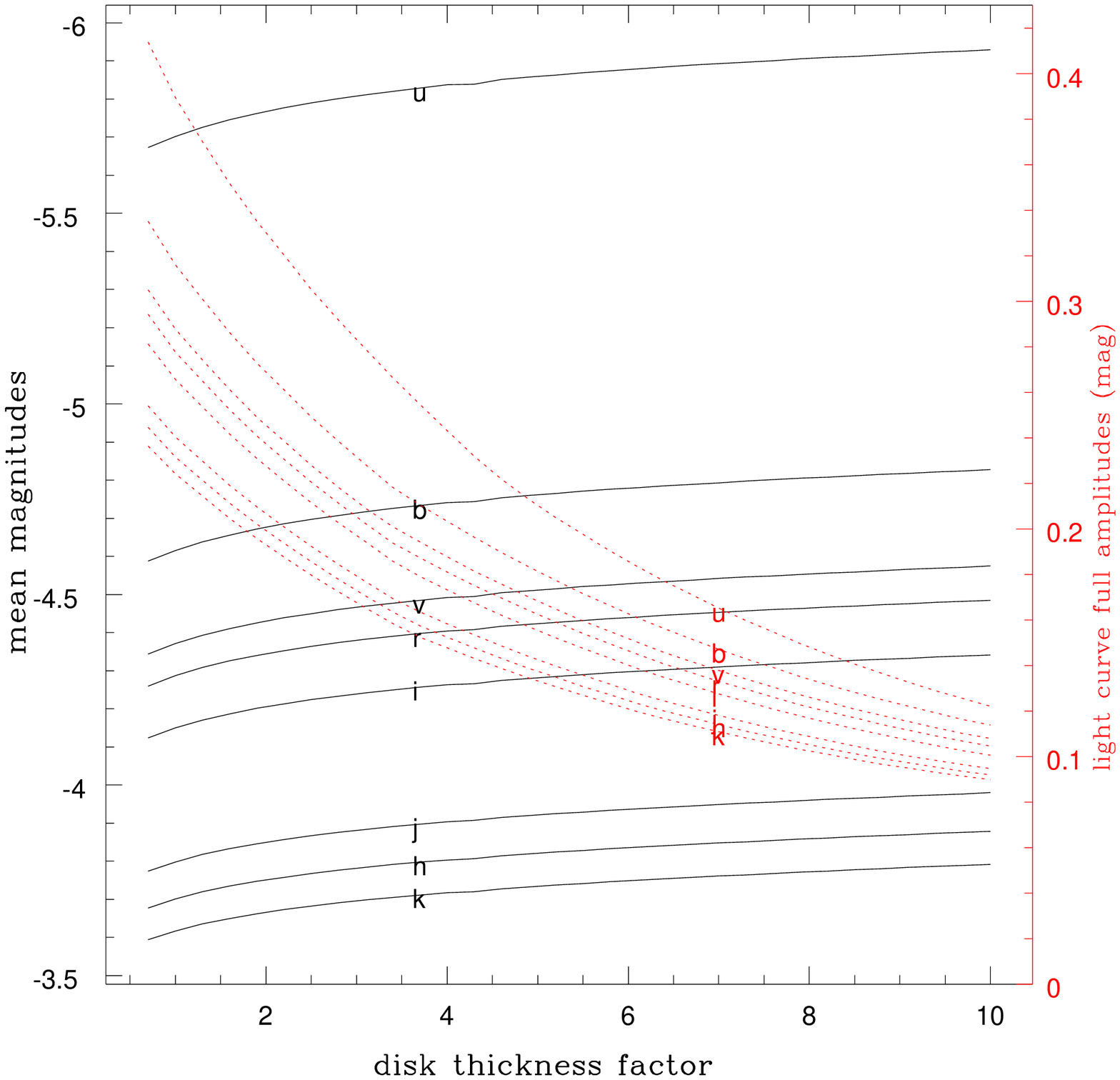}

\caption{({\it a}) The spectra (lower panel) and the B/H light curves (upper
panel) for models with different disk fattening factors. The thin spectra are
for the X-ray heated star, and the thick spectra are for the sum of the star
and the disk.  ({\it b}) The mean magnitudes (solid) and the variation
amplitudes (dotted) of the UBVRIJHK light curves for models with different disk
fattening factors. }

\end{figure}

\begin{figure}

\plottwo{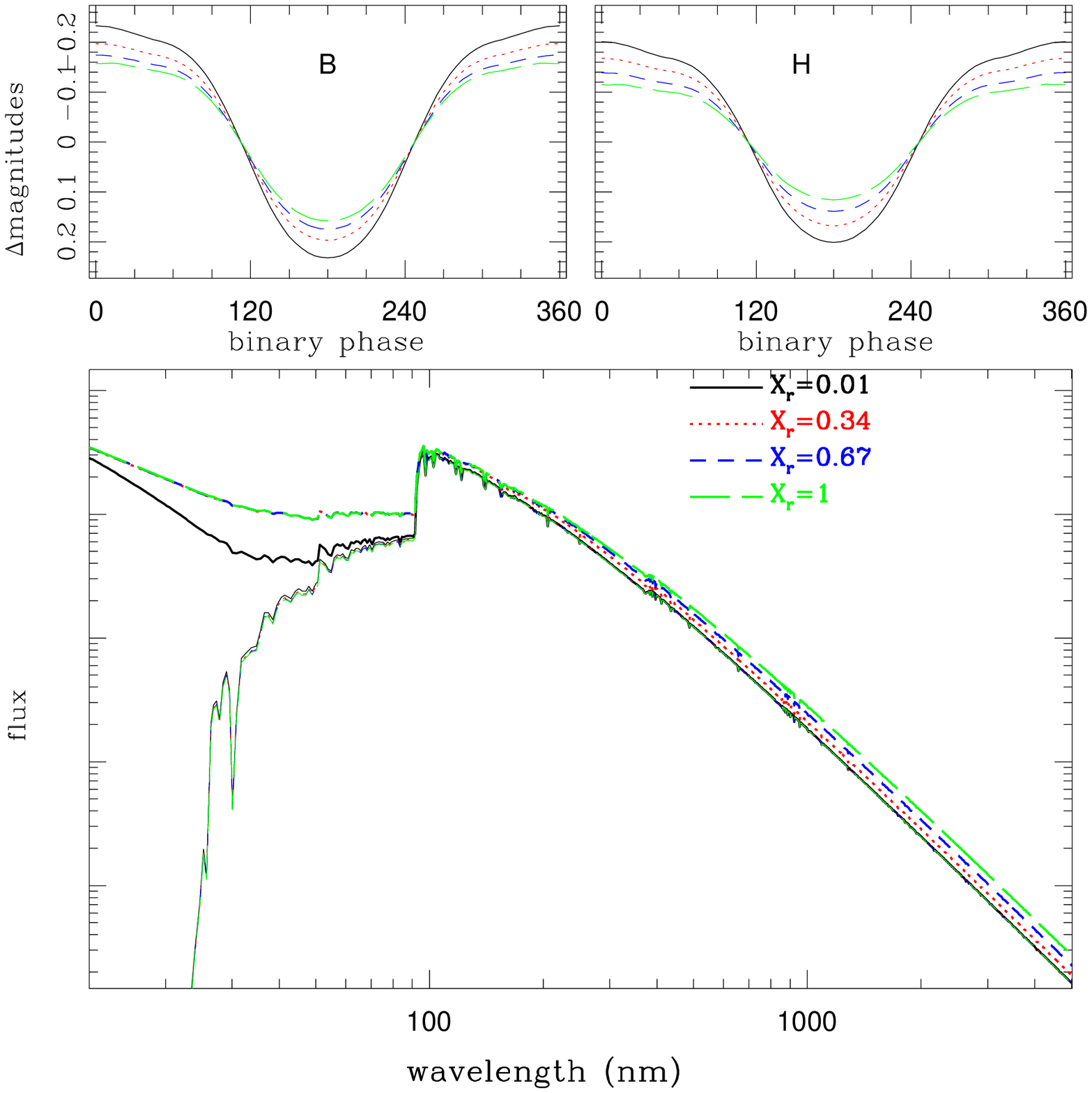}{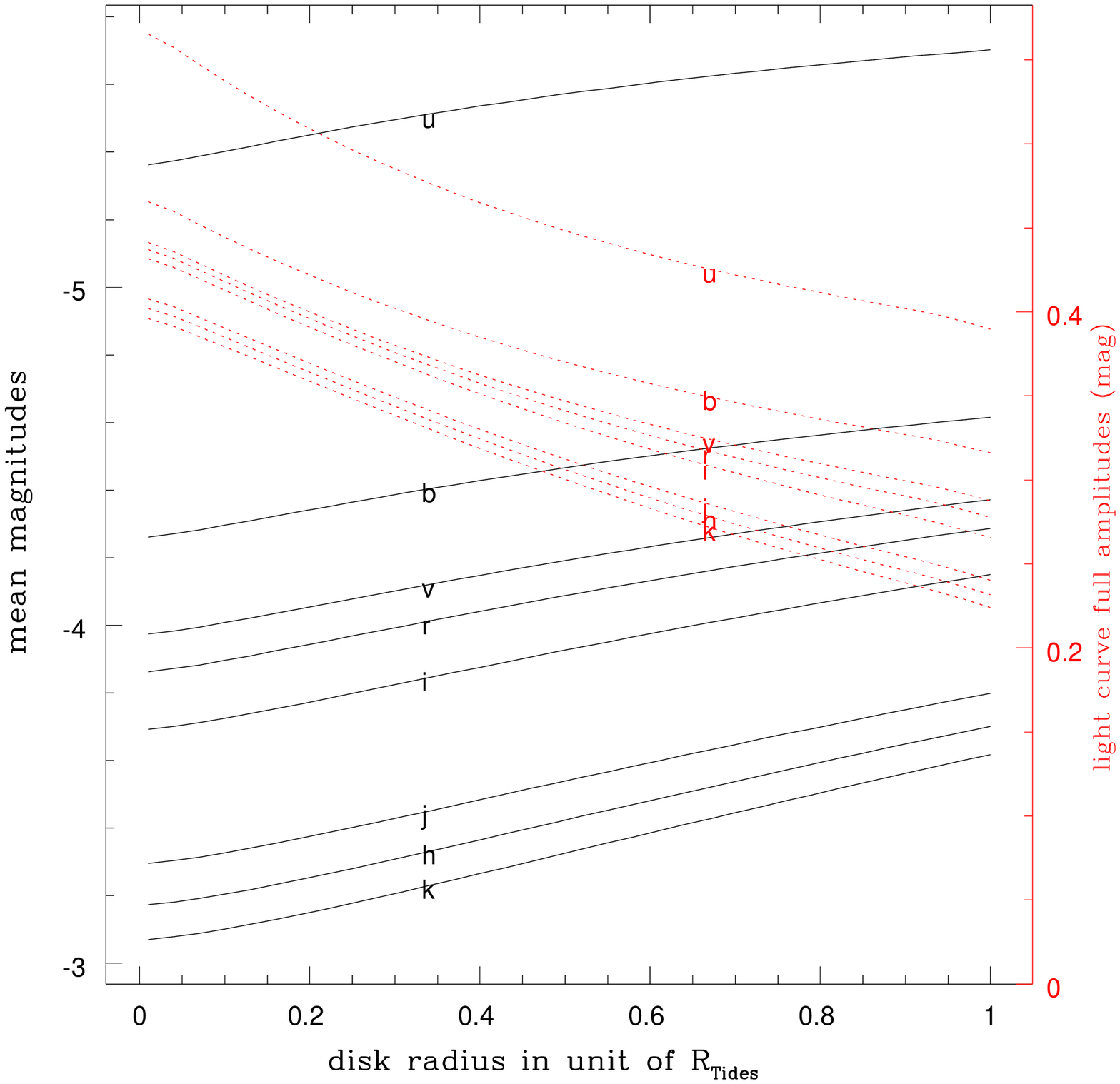}

\caption{({\it a}) The spectra (lower panel) and the B/H light curves (upper
panel) for models with different disk size factors. The thin spectra are for
the X-ray heated star, and the thick spectra are for the sum of the star and
the disk.  ({\it b}) The mean magnitudes (solid) and the variation
amplitudes (dotted) of the UBVRIJHK light curves for models with different disk
size factors. }

\end{figure}

\begin{figure}

\plottwo{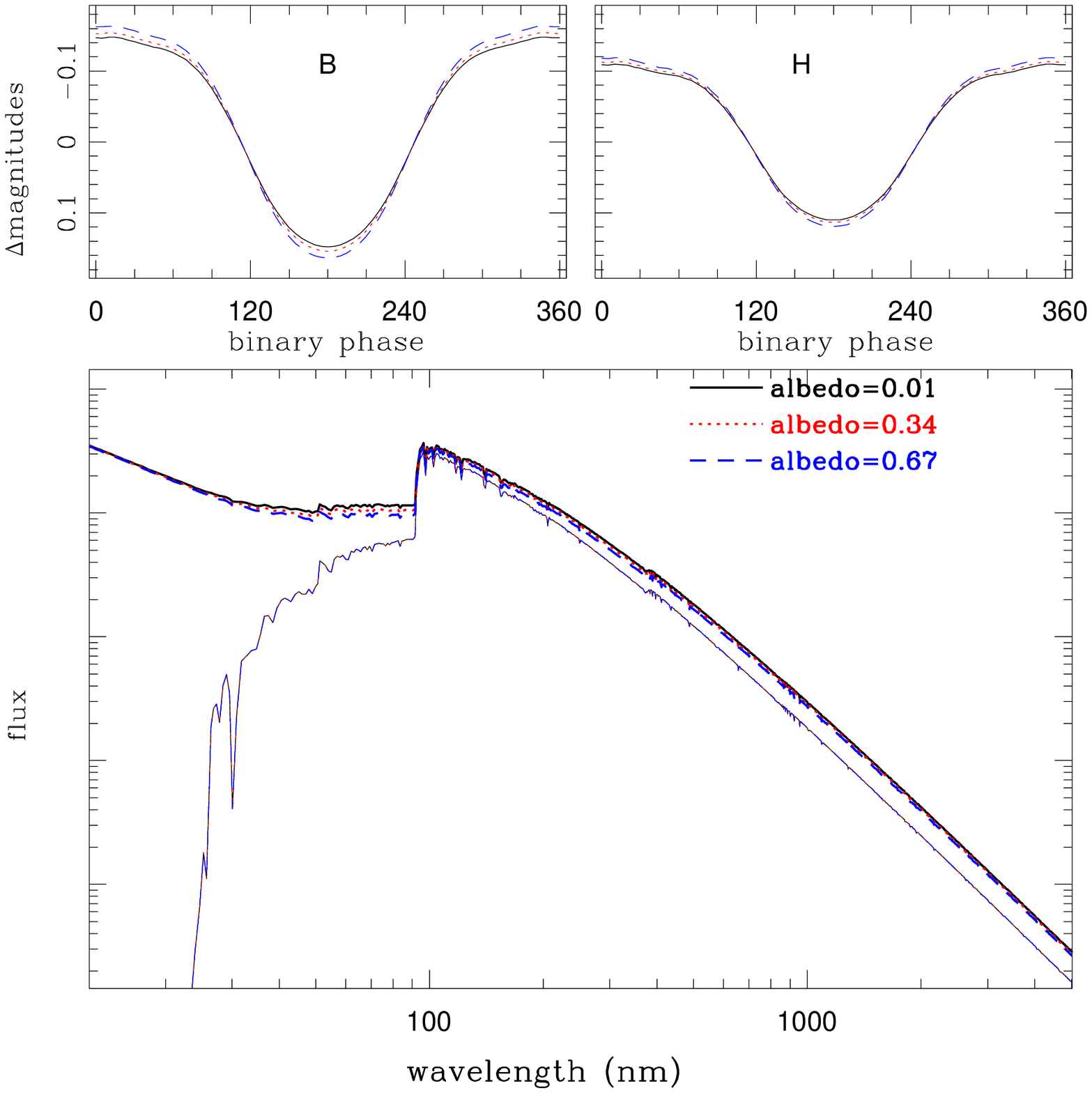}{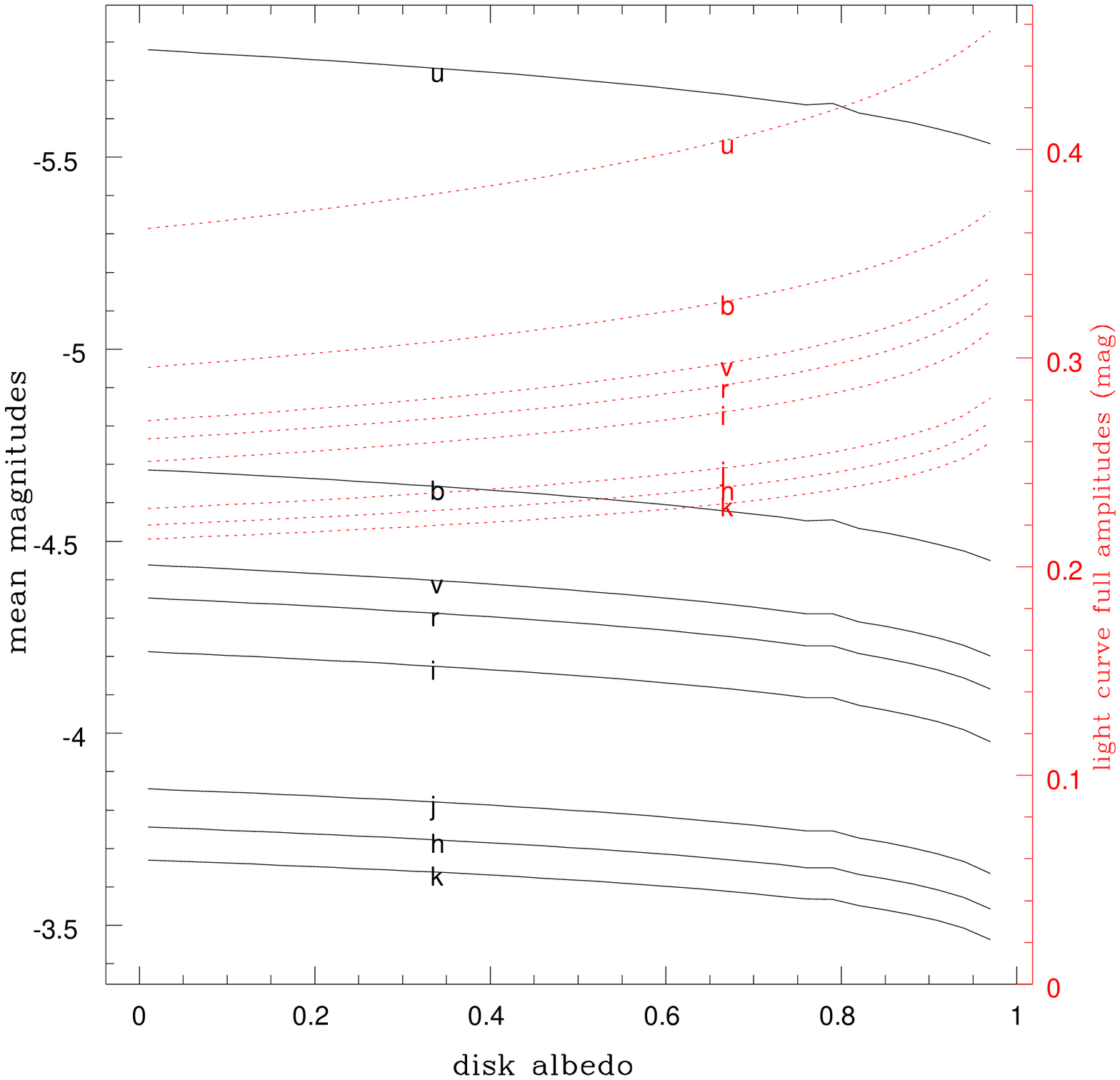}

\caption{({\it a}) The spectra (lower panel) and the B/H light curves (upper panel) for
models with different disk albedo. The thin spectra are for the X-ray
heated star, and the thick spectra are for the sum of the star and the disk.  
({\it b}) The mean magnitudes (solid) and the variation amplitudes (dotted) of
the UBVRIJHK light curves for models with different disk albedo. }

\end{figure}

\begin{figure}

\plottwo{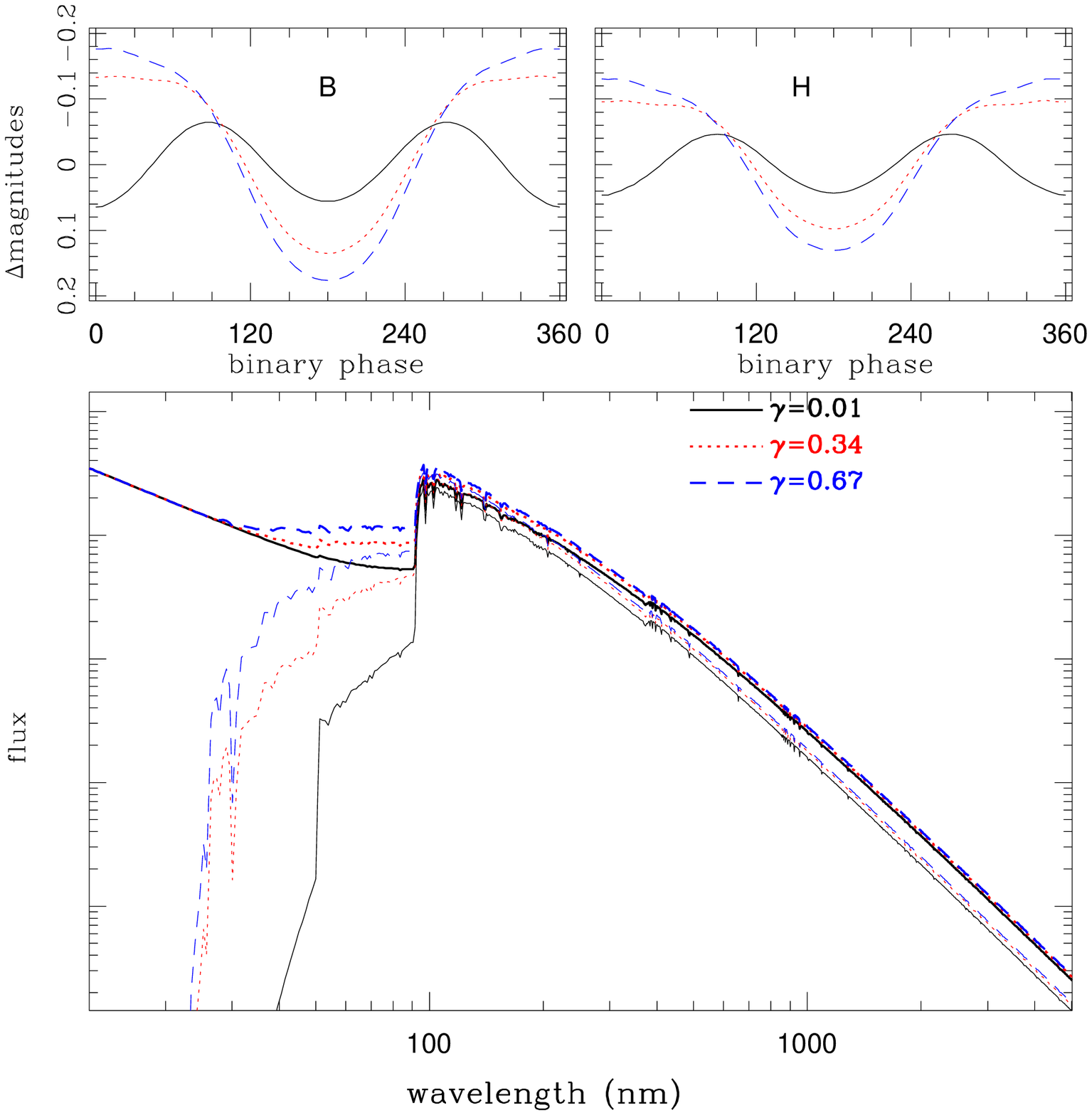}{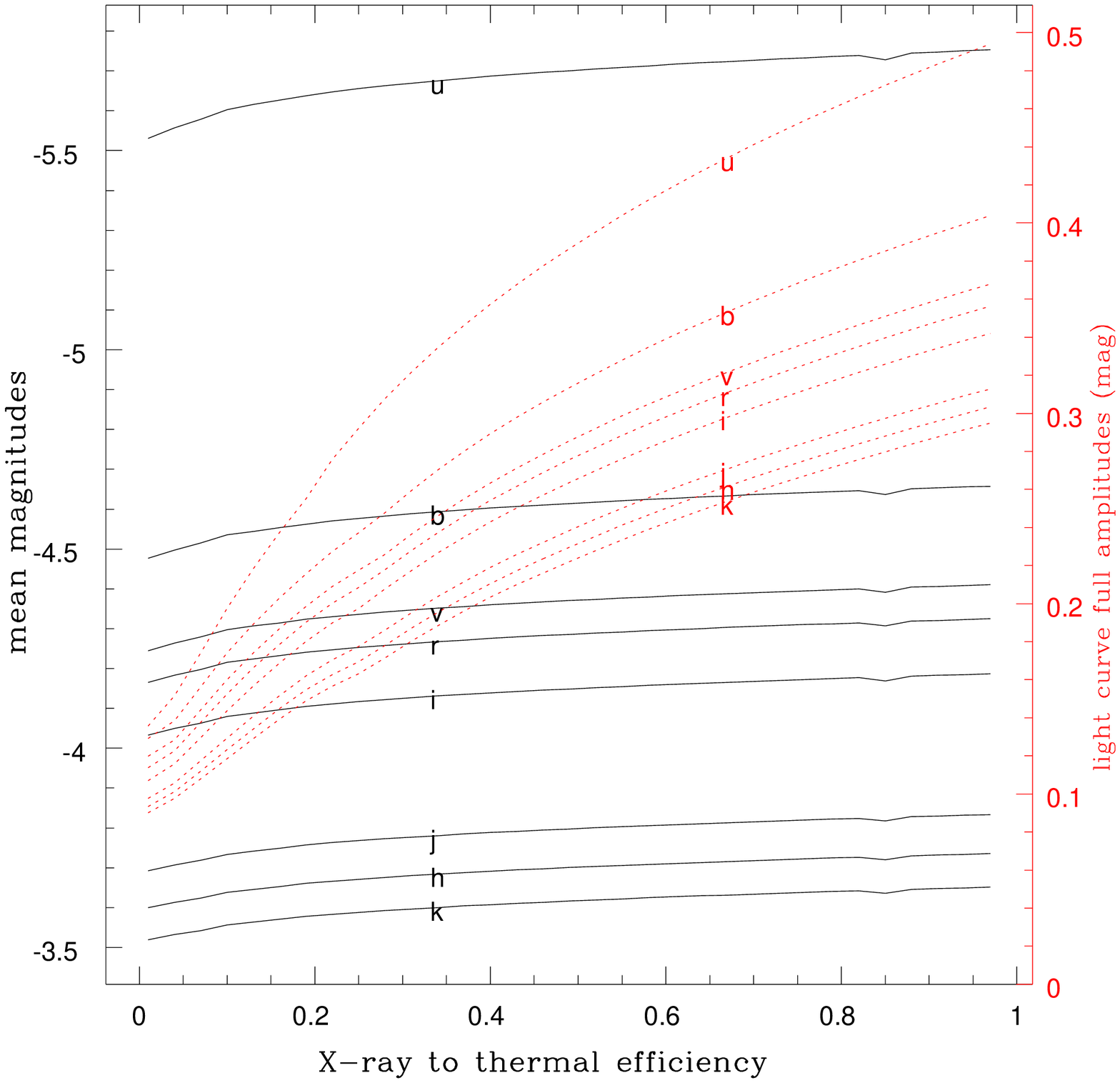}

\caption{({\it a}) The spectra (lower panel) and the B/H light curves (upper panel) for
models with different X-ray to thermal conversion factor on the secondary
surface. The thin spectra are for the X-ray heated star, and the thick spectra
are for the sum of the star and the disk.  ({\it b}) The mean magnitudes (solid) and the variation amplitudes (dotted) of
the UBVRIJHK light curves for models with different X-ray to thermal conversion
factor on the secondary surface. }

\end{figure}

\begin{figure}

\plottwo{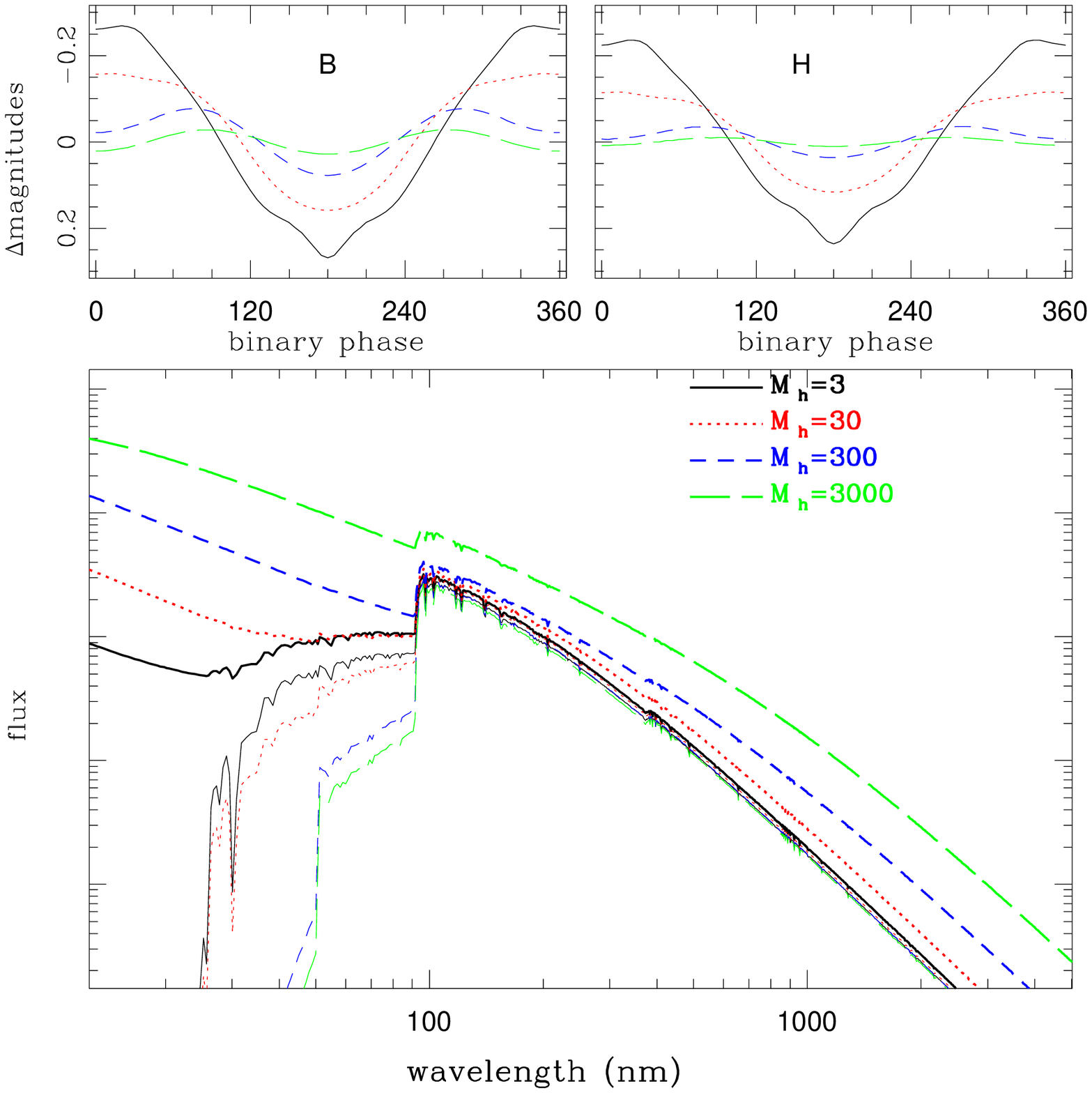}{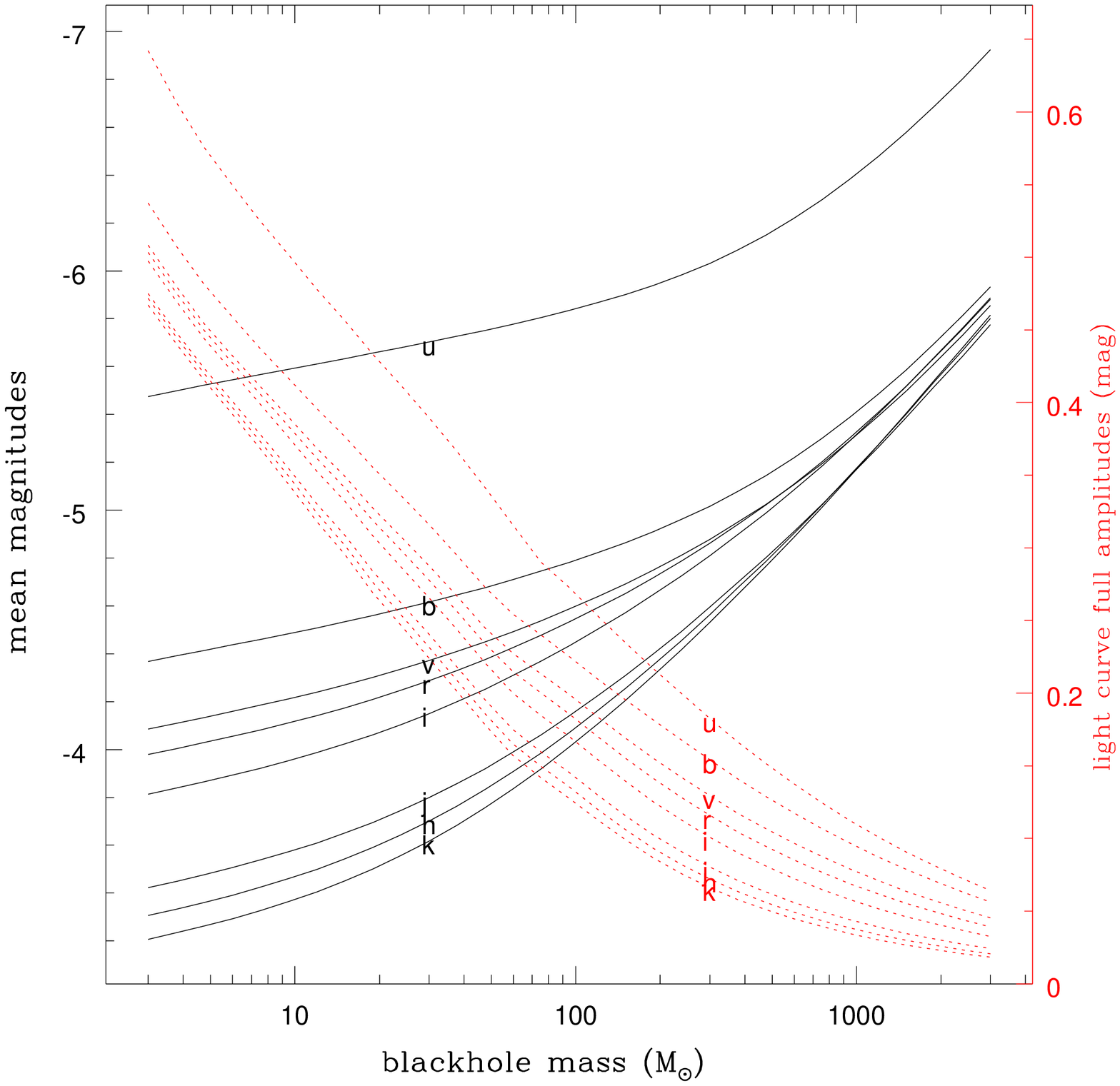}

\caption{({\it a})  The spectra (lower panel) and the B/H light curves (upper panel)
for models with different black hole masses. The thin spectra are for the X-ray
heated star, and the thick spectra are for the sum of the star and the disk.
({\it b}) The mean magnitudes (solid) and the variation amplitudes (dotted) of
the UBVRIJHK light curves for models with different black hole masses. }

\end{figure}

\begin{figure}

\plottwo{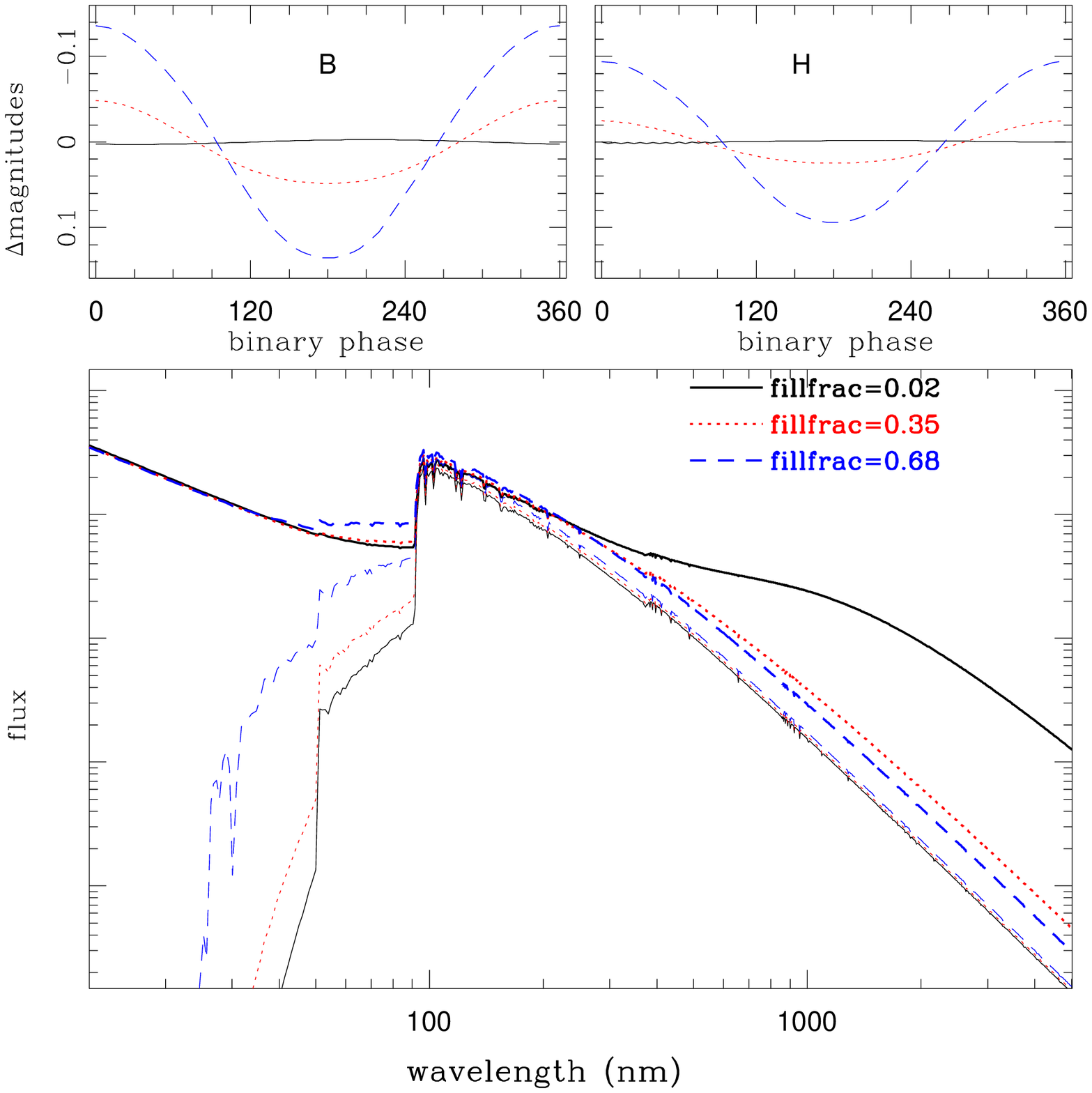}{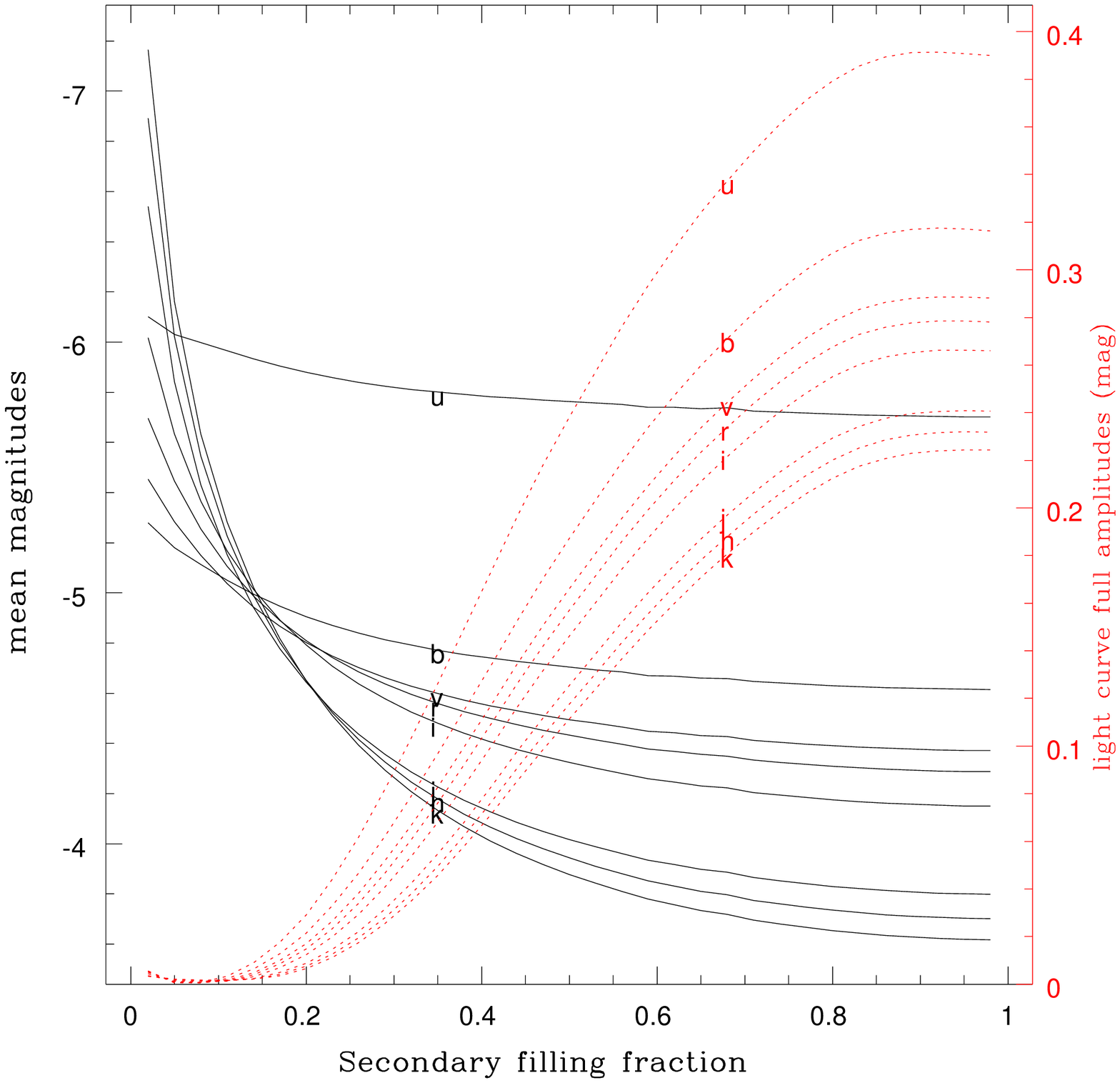}

\caption{({\it a}) The spectra (lower panel) and the B/H light curves (upper panel) for
models with different fractions the secondary fills its Roche lobe. The thin
spectra are for the X-ray heated star, and the thick spectra are for the sum of
the star and the disk. ({\it b})  The mean magnitudes (solid) and the variation amplitudes (dotted) of
the UBVRIJHK light curves for models with different fractions the secondary
fills its Roche lobe. }

\end{figure}

\begin{figure}

\plottwo{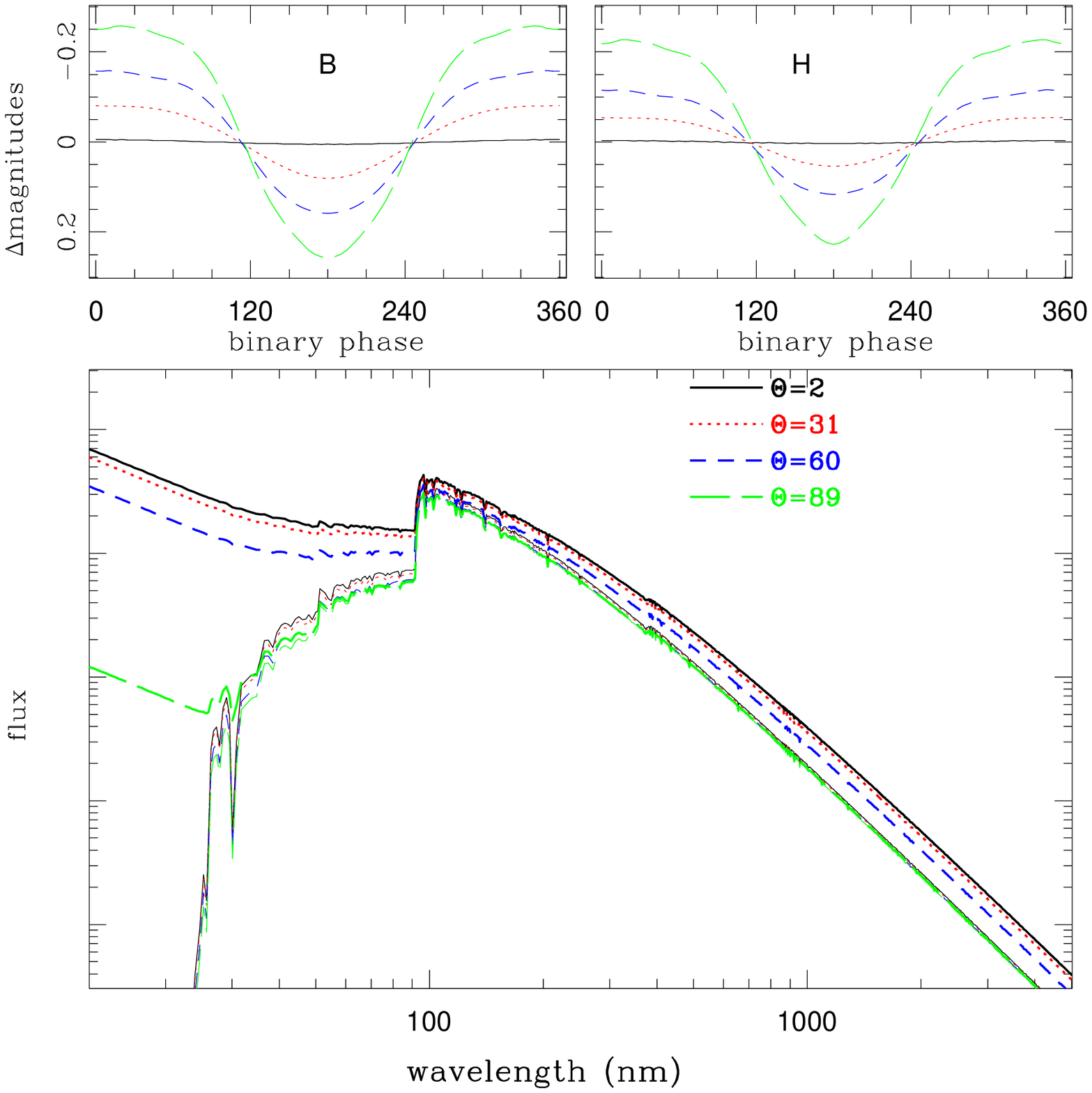}{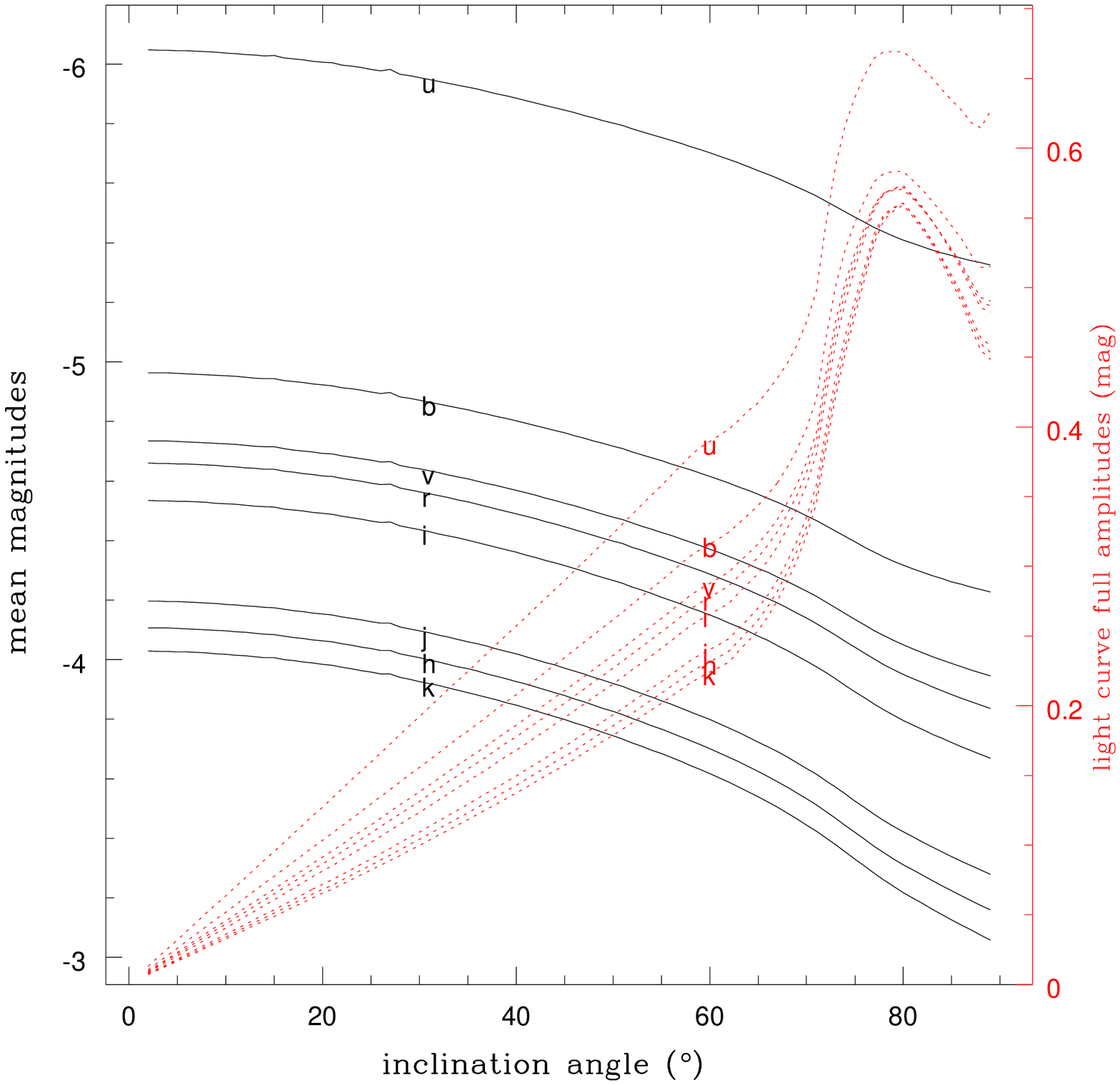}

\caption{({\it a}) The spectra (lower panel) and the B/H light curves (upper panel) for
models with different inclination angles. The thin spectra are for the X-ray
heated star, and the thick spectra are for the sum of the star and the disk.  
({\it b}) The mean magnitudes (solid) and the variation amplitudes (dotted) of
the UBVRIJHK light curves for models with different inclination angles. }

\end{figure}

\begin{figure}

\plottwo{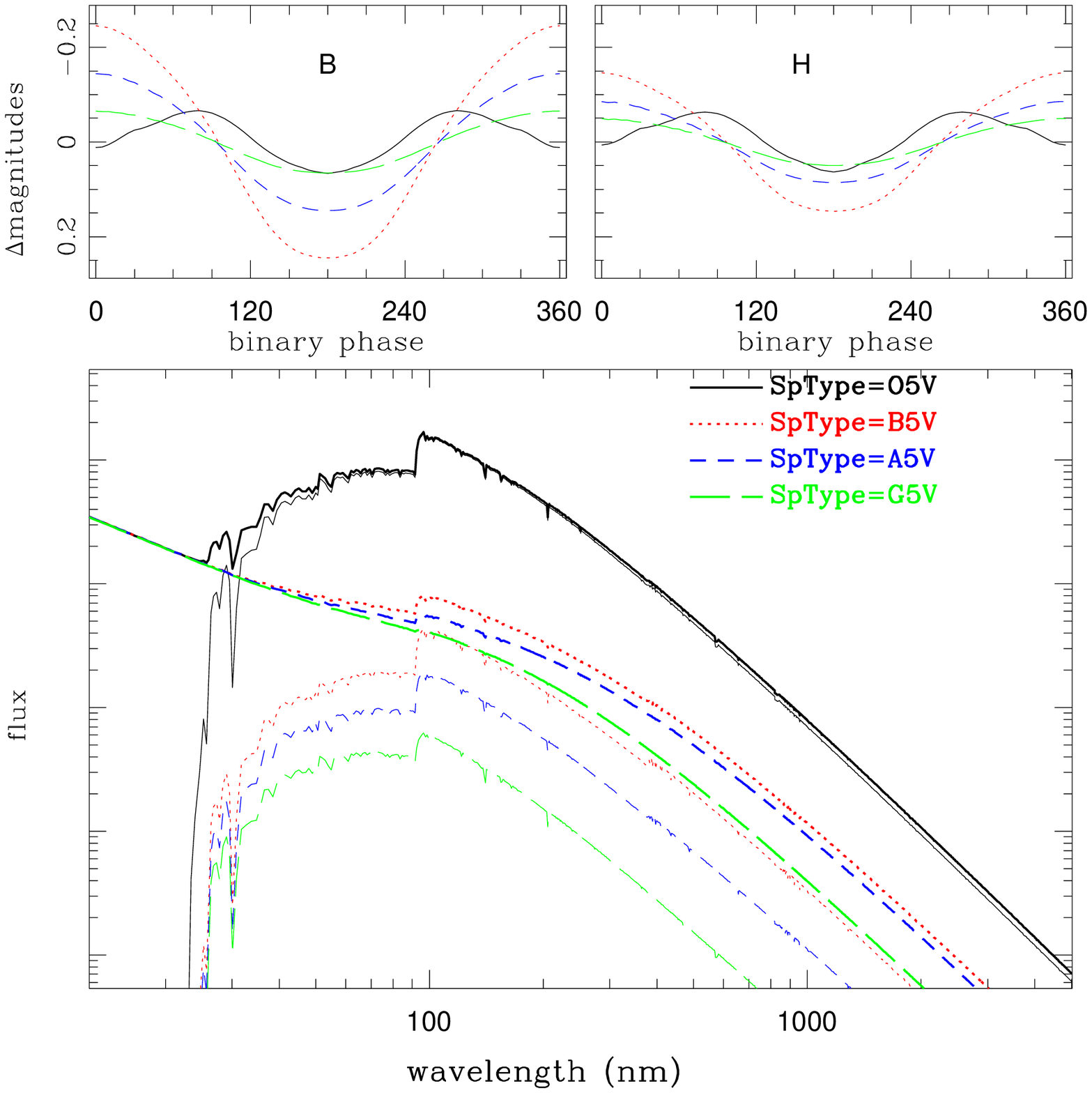}{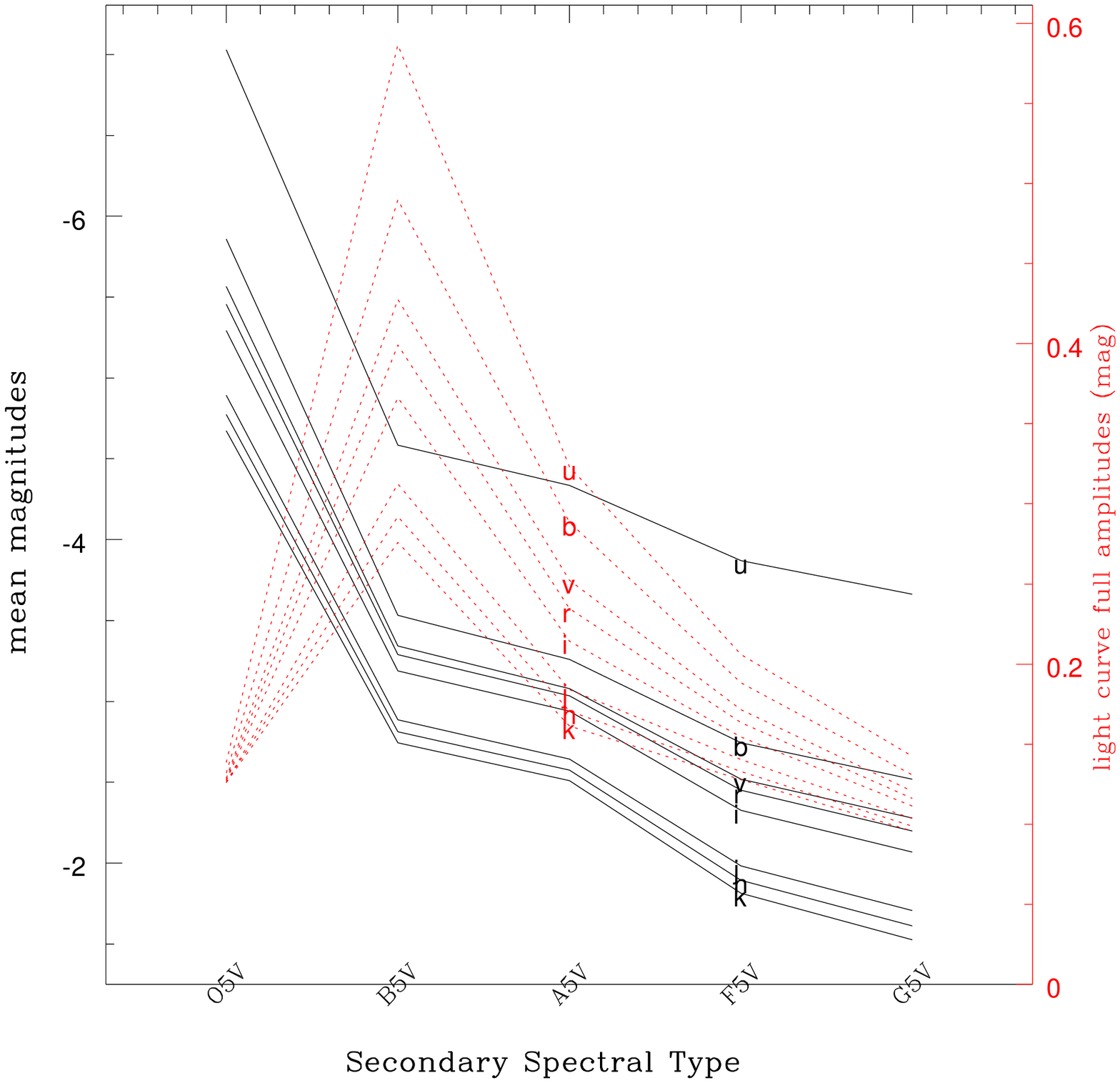}

\caption{({\it a}) The spectra (lower panel) and the B/H light curves (upper panel) for
models with dwarf secondaries of different spectral types. The thin spectra are for the X-ray
heated star, and the thick spectra are for the sum of the star and the disk.  
({\it b}) The mean magnitudes (solid) and the variation amplitudes (dotted) of
the UBVRIJHK light curves for models with dwarf secondaries of different spectral types.}

\end{figure}

\begin{figure}

\plottwo{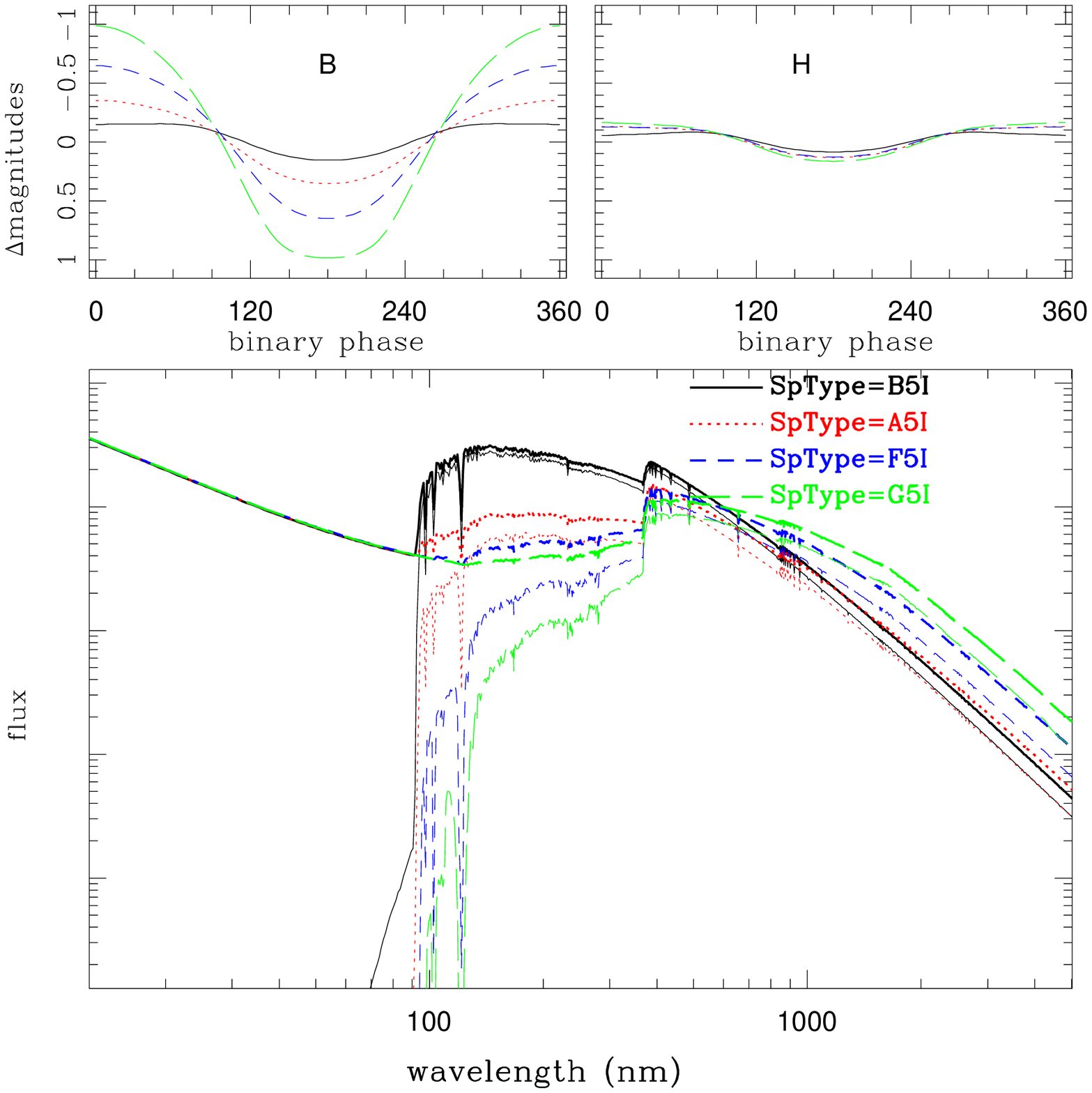}{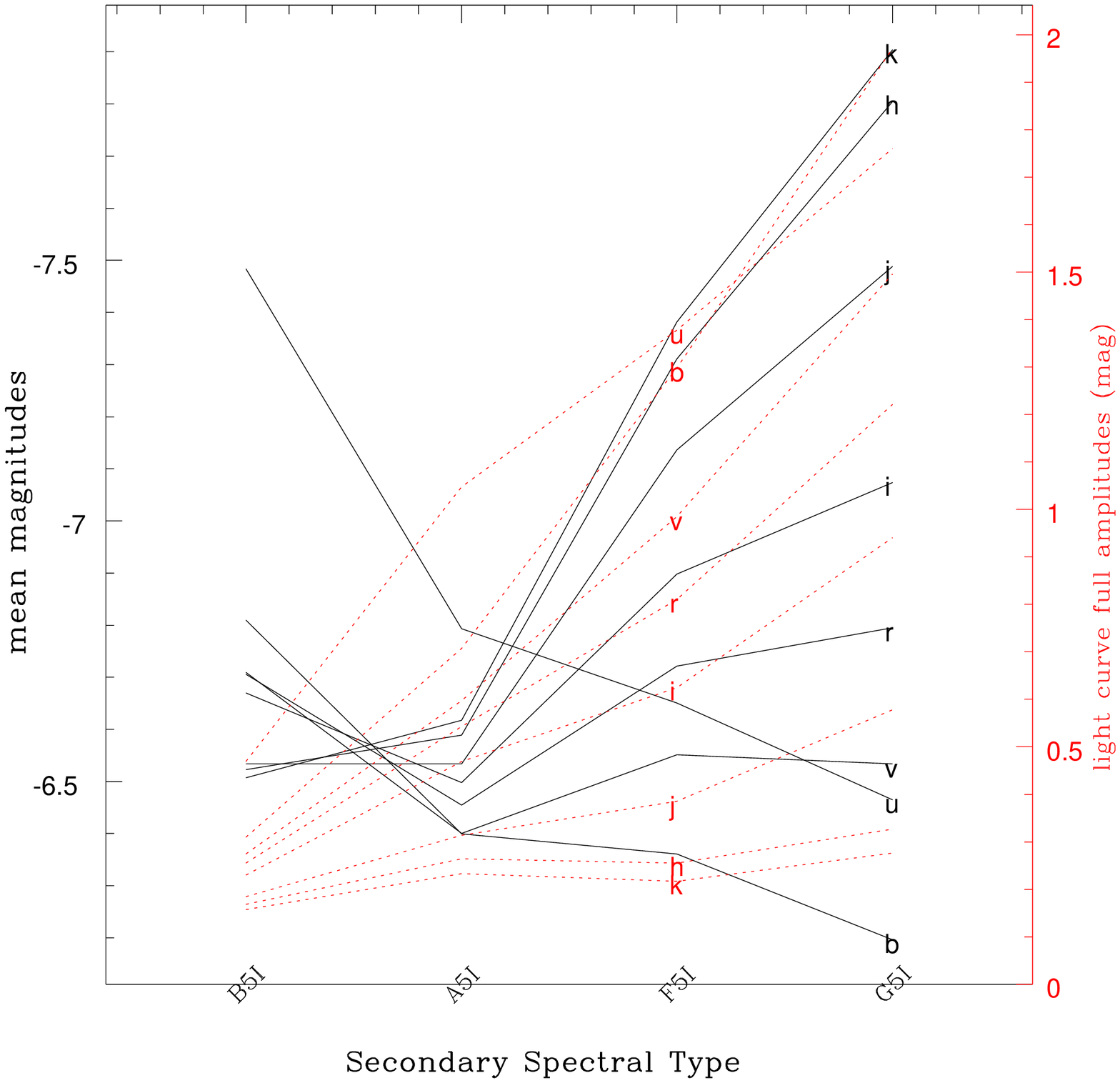}

\caption{({\it a}) The spectra (lower panel) and the B/H light curves (upper panel) for
models with supergiant secondaries of different spectral types. The thin spectra are for the X-ray
heated star, and the thick spectra are for the sum of the star and the disk.  
({\it b}) The mean magnitudes (solid) and the variation amplitudes (dotted) of 
the UBVRIJHK light curves for models with supergiant secondaries of different spectral types.}

\end{figure}


\begin{thebibliography}{}


\bibitem[Begelman (2002)]{} Begelman, M.C., 2002, \apj, 568, L97
\bibitem[Cantrell et al. (2007)]{} Cantrell, A. \& Bailyn, C., 2007, ApJ, 670, 727
\bibitem[Colbert et al. (1999)]{} Colbert, E. J. M. and Mushotzky, R. F. 1999, \apj, 519, 89
\bibitem[Copperwheat et al. (2005)]{} Copperwheat, C., Cropper, M., Soria, R., Wu, K., 2005, MNRAS, 362, 79
\bibitem[de Jager et al. (1988)]{} de Jager, C., Nieuwenhuijzen, H., van der Hucht, K. A. 1988, A\&AS, 72, 259
\bibitem[Fabbiano (1989)]{} Fabbiano, G. 1989, ARA\&A, 27, 87
\bibitem[Feng et al. (2006)]{} Feng, H. and Kaaret, P., 2006, ApJ, 650, 75L
\bibitem[Frankowski et al. (2001)]{} Frankowski, A. \& Tylenda, R., 2001, A\&A, 367, 513
\bibitem[Frank et al. (2002)]{} Frank, J., King, A., and Raine, D, 2002, {\it Accretion Power in Astrophysics}, \S5.10
\bibitem[Fryer et al. (2001)]{} Fryer, C.~L. \& Kalogera, V., 2001, ApJ, 554, 548-560
\bibitem[Georganopoulos et al. (2002)]{} Georganopoulos, M., Aharonian, F.A., and Kirk, J.G., 2002, A\&A, 388, L25
\bibitem[Girardi et al. (2002)]{} Girardi, L., Bertelli, G., Bressan, A., et al. 2002, A\&A, 391, 195.
\bibitem[Gladstone et al. (2009)]{} Gladstone, J.~C., Roberts, T.~P., Done, C.\ 2009.\ MNRAS, 397, 1836
\bibitem[Gladstone et al. (2011)]{} Gladstone, J.~C., Roberts, T.~P., et al. 2011, in preparation
\bibitem[Grise et al. (2008)]{} Gri{\'e}, F., Pakull, M. W., Soria, R., et al. 2008, A\&A, 486, 151
\bibitem[Gutierrez et al. (2006)]{} Guti{\'e}rrez, C.M., 2006, \apj, 640, 17L
\bibitem[Haines (1994)]{} Haines, E., 1994, Graphics Gems IV, p. 24-46
\bibitem[Kaaret et al. (2009)]{} Kaaret, P. \& Corbel, S. 2009, \apj, 697, 950
\bibitem[King et al. (2001)]{} King, A. R., Davies, M. B., Ward, M. J., Fabbiano, G. and Elvis, M.  2001, \apj, 552, L109
\bibitem[Kopal (1959)]{} Kopal, Z., {\it Close Binary Systems}, 1959, p136
\bibitem[Kurucz (1993)]{} Kurucz, R.L., 1993, Kurucz CD-ROM No. 13. Stellar Atmosphere Programs and 2 km/s Grid
\bibitem[Liu (2009)]{} Liu, J. 2009, ApJ, 704, 1628
\bibitem[Liu et al. (2002)]{} Liu, J., Bregman, J., and Seitzer, P., 2002, ApJL, 580, 31
\bibitem[Liu et al. (2004)]{} Liu, J., Bregman, J., and Seitzer, P., 2004, ApJ, 602, 249
\bibitem[Liu et al. (2009)]{} Liu J., Bregman J., \& McClintock, J., 2009, ApJL, 679, 37
\bibitem[Liu et al. (2007)]{} Liu, J.; Bregman, J.; Miller, J. \& Kaaret, P., 2007, ApJ, 661, 165
\bibitem[Liu et al. (2008)]{} Liu, J. \& Di Stefano, R. 2008, ApJL, 674, 73
\bibitem[Lucy (1967)]{} Lucy, L. B., 1967, ZA, 65, 89
\bibitem[Miller et al. (2002)]{} Miller, M. \& Hamilton, D., 2002, MNRAS, 330, 232
\bibitem[Mucciarelli et al. (2005)]{} Mucciarelli, P.,Zampieri, L., et al. 2005, ApJ, 633, L101
\bibitem[Orosz et al. (2000)]{} Orosz, J.A. \& Hauschildt, P.H., 2000, A\&A, 364, 265
\bibitem[Orosz et al. (2007)]{} Orosz, J.A., McClintock, J.E., et al. 2007, Nature, 449, 872
\bibitem[Pakull et al. (2006)]{} Pakull, M.W., Grise, F., and Motch, C., 2006, IAUS, 230, 293 (astro-ph/0603771)
\bibitem[Portegies et al. (2002)]{} Portegies Zwart, S., and McMillan, S., 2002, ApJ, 576, 899
\bibitem[Press et al. (2003)]{} Press, W., Flannery, B., Teukolsky, S., Vetterling, W., 2003, {\it Numerical Recipe in C}, Cambridge University Press
\bibitem[Ramsey et al. (2006)]{} Ramsey, C., Williams, R., et al. 2006, ApJ, 641, 241
\bibitem[Raymond (1993)]{} Raymond, J. 1993, ApJ, 412, 267 
\bibitem[Roberts et al. (2010)]{} Roberts, T.~P., Gladstone, J.C., Goulding, A.D., et al. 2010, to be published in Astronomische Nachrichten (astro-ph/1011.2155)
\bibitem[Roberts et al. (2008)]{} Roberts, T.P.; Levan, A. J.; Goad, M. R. 2008, MNRAS, 387, 73
\bibitem[Ryder et al. (1993)]{} Ryder, S. D. 1993, Ph.D. thesis, Australian National Univ.
\bibitem[Shakura et al. (1973)]{} Shakura, N., and Sunyaev, R., 1973, A\&A, 24, 337
\bibitem[Strohmayer et al. (2003)]{} Strohmayer, T., and Mushotzky, R., 2003, ApJL, 586, 61L
\bibitem[Tully et al. (1988)]{} Tully R.B, 1988, {\it Nearby Galaxies Catalogue}, Cambridge University Press.
\bibitem[van Paradijs et al. (1994)]{} van Paradijs, J., and McClintock, J., 1994, A\&A, 290, 133
\bibitem[von Zeipel (1924)]{} von Zeipel H., 1924, MNRAS, 84, 702
\bibitem[Vrtilek et al. (1990)]{} Vrtilek, S.D., Raymond, J.C., Garcia, M.R. et al. 1990, A\&A, 235, 162
\bibitem[Zampieri et al. (2004)]{} Zampieri, L., Mucciarelli, P. et al., 2004, ApJ, 603, 523
\bibitem[Zampieri et al. (2011)]{} Zampieri, L., Impiombato, D., Falomo, R. et al., 2011, MNRAS, in press

\end{thebibliography}
\end{document}